\documentclass[twocolumn,citeautoscript, prx, superscriptaddress,longbibliography,amsmath,amssymb]{revtex4-1}

\usepackage{graphicx}
\usepackage{color}
\usepackage{upgreek}
\usepackage{float}
\usepackage{lipsum}
\usepackage{mathrsfs}
\usepackage{booktabs}
\usepackage{url}
\usepackage{xcolor}
\usepackage{siunitx}
\usepackage{times}
\usepackage[colorlinks,citecolor=teal,linkcolor=teal, urlcolor=teal]{hyperref}
 
\usepackage[T1]{fontenc}

\usepackage{enumerate}
\usepackage{physics} 
\usepackage{stmaryrd} 
\usepackage{ragged2e} 
\usepackage{bbm} 
\usepackage{wasysym}

\newcommand{\diag}{\mathrm{diag}}
\newcommand{\Swap}{\textsc{Swap}}
\newcommand{\Cnot}{\textsc{Cnot}}
\newcommand{\CZ}{\text{CZ}}

\newcommand{\GL}{\mathrm{GL}}

\newcommand{\FF}{\mathbb{F}}

\newcommand{\RR}{\mathbb{R}}
\newcommand{\CC}{\mathbb{C}}

\newcommand{\wt}{\mathrm{wt}}
\newcommand{\boldalpha}{\boldsymbol{\alpha}}
\newcommand{\sing}{\text{sing}}
\newcommand{\rhotranspose}{\rho^{\! \top}}
\newcommand{\Psitranspose}{\Psi^{\! \top}}
\newcommand{\qtranspose}{Q^{\!      \top}}

\newcommand{\twocopyobservable}{{\add \Omega}}
\newcommand{\twocopyobservablegeneric}{{\add \mho}}

\renewcommand{\subset}{\subseteq}

\newcommand{\add}{}

\newcounter{mycounter}

\newcommand{\fu}{Dahlem Center for Complex Quantum Systems, Freie Universit{\"a}t Berlin, 14195 Berlin, Germany}
\newcommand{\fzj}{Institute for Theoretical Nanoelectronics (PGI-2), Forschungszentrum J\"ulich, 52428 J\"ulich, Germany}
\newcommand{\innsbruck}{Universit\"{a}t Innsbruck, Institut f\"{u}r Experimentalphysik, Technikerstrasse 25, 6020 Innsbruck, Austria}
\newcommand{\aqt}{Alpine Quantum Technologies GmbH, 6020 Innsbruck, Austria}
\newcommand{\iqoqi}{Institut f\"{u}r Quantenoptik und Quanteninformation, \"{O}sterreichische Akademie der Wissenschaften, Otto-Hittmair-Platz 1, 6020 Innsbruck, Austria}
\newcommand{\vt}{Department of Computer Science, Virginia Tech, Blacksburg, Virginia 24061, USA}
\newcommand{\vtq}{Virginia Tech Center for Quantum Information Science and Engineering, Blacksburg, Virginia 24061, USA}

\begin{document}

\preprint{APS/123-QED}

\newcommand{\coauthor}[1]{\author{\mbox{#1}}}

\author{Daniel Miller}
\email{d.miller@fu-berlin.de}
\affiliation{\fu} \affiliation{\fzj} 
\coauthor{Kyano Levi} \affiliation{\fu}
\coauthor{Lukas Postler}  \affiliation{\innsbruck}
\coauthor{Alex Steiner}   \affiliation{\innsbruck}
\coauthor{Lennart Bittel} \affiliation{\fu}
\coauthor{Gregory A.~L. White} \affiliation{\fu}
\coauthor{Yifan Tang}        \affiliation{\fu} 
\coauthor{Eric J. Kuehnke}   \affiliation{\fu}
\coauthor{Antonio A. Mele} \affiliation{\fu}
\coauthor{Sumeet Khatri} \affiliation{\fu} \affiliation{\vt} \affiliation{\vtq}
\coauthor{Lorenzo Leone} \affiliation{\fu}
\coauthor{Jose Carrasco} \affiliation{\fu}
\coauthor{Christian D. Marciniak} \affiliation{\innsbruck}
\coauthor{Ivan Pogorelov} \affiliation{\innsbruck}
\coauthor{Milena Guevara-Bertsch} \affiliation{\innsbruck}
\coauthor{Robert Freund} \affiliation{\innsbruck}
\coauthor{Rainer Blatt}  \affiliation{\innsbruck} \affiliation{\iqoqi} 
\coauthor{Philipp Schindler} \affiliation{\innsbruck}
\coauthor{Thomas Monz} \affiliation{\innsbruck} \affiliation{\aqt}
\coauthor{Martin Ringbauer}  \affiliation{\innsbruck}
\coauthor{Jens Eisert} \affiliation{\fu}

\title{\add Experimental measurement and a physical interpretation of quantum shadow enumerators}


\begin{abstract}
Throughout its history, the theory of quantum error correction has heavily benefited from translating classical concepts into the quantum setting.
In particular, classical notions of weight enumerators, which relate to the performance of an error-correcting code, and MacWilliams' identity, which {\add links} enumerators {\add of a code to the ones of its dual},
have been generalized to the quantum case. 
In this work, we establish a relationship between the theoretical machinery of quantum weight enumerators and a seemingly unrelated physics experiment:
we prove that Rains' quantum shadow enumerators---a powerful mathematical tool---arise as probabilities of observing fixed numbers of triplets in a 
{\add two-copy} Bell sampling experiment.
This insight allows us to develop here a rigorous framework for the direct measurement of quantum weight enumerators, thus enabling experimental and theoretical studies of the entanglement structure of any quantum error-correcting code or {\add quantum} state under investigation.
On top of that, we derive concrete sample complexity bounds and physically-motivated robustness guarantees against unavoidable experimental imperfections.
Finally,
we demonstrate the feasibility of {\add experimentally learning} weight enumerators on a trapped-ion quantum computer.
Our experimental findings are in good agreement with theoretical predictions and illuminate how entanglement theory and quantum error correction crossfertilize each other once 
{\add two-copy} Bell sampling experiments are combined with the theoretical machinery of quantum weight enumerators.

\end{abstract}

\maketitle


\section{Introduction}
The ultimate goal in quantum technology is to see high-fidelity quantum devices both actualized and justified.
This will require us to solve two major challenges that---each individually---have received a great deal of attention.
The first challenge (actualization) poses the question
\emph{``how can we protect fragile quantum information against noise?''}
and indeed, the theory of \emph{quantum error correction} (QEC) allows for the detection and correction of errors that arise over the course of a quantum computation~\cite{lidar_quantum_error_2013, terhal_quantum_error_2015, campbell_roads_towards_2017}. 
It is widely accepted that QEC will be essential to the realization of full-scale quantum computers~\cite{eisert_mind_the_2025}, 
and so substantial efforts are being made 
{\add that}
have led to {\add impressive} progress in hardware development~\cite{bermudez_assessing_the_2017, egan_fault_tolerant_2021, ryan_anderson_realization_of_2021, krinner_realizing_repeated_2022, google_suppressing_quantum_2023, gupta_encoding_a_2024, bluvstein_logical_quantum_2024, acharya_quantum_error_2025, chiu_continuous_operation_2025}.
Of equal fundamental importance to the development of use cases for quantum computers and, thus, 
to the second challenge (justification) is the question
\emph{``how can we efficiently extract properties of a carefully constructed quantum state?''}.  
On the one hand, {\add understanding} what {\add can be learned from experimental data} informs {characterization and benchmarking} techniques~\cite{emerson_scalable_noise_2005, eisert_quantum_certification_2020, kliesch_theory_of_2021, helsen_general_framework_2022}, 
such as 
{\add verifying} the ability to generate large-scale
entanglement~\cite{flammia_direct_fidelity_2011, mooney_generation_and_2021,moses_a_race_2023, kam_characterization_of_2024, javadiabhari_big_cats_2025}.
On the other hand, and more provocatively, answers to this second question should also foster our understanding of genuine quantum phenomena, which could unlock entirely new applications.
As quantum devices are moving closer and closer
towards fault tolerance,
the above questions become increasingly pertinent.
At the same time, 
{\add they} are seldom considered in tandem, leaving exciting opportunity to do so.

The relationship between each challenge is not obvious from the outset.
Concepts from 
{\add device}
characterization 
have found their way into QEC, but not often vice versa. 
For linear properties such as 
{\add stabilizer} violations and fidelities, 
the
{\add learnability question} has been largely resolved in the context of classical shadows~\cite{ohliger_efficient_and_2013, aaronson_shadow_tomography_2018, huang_predicting_many_2020, elben_the_randomized_2023, fischer_dual_frame_2024}.
But not all quantities relevant for QEC are linear. 
In the {\add non-linear} regime, leveraging parallel \emph{multi-copy measurements} will become important~\cite{montanaro_learning_stabilizer_2017, chen_exponential_separations_2022, huang_quantum_advantage_2022, aharonov_quantum_algorithmic_2022, chen_a_hierarchy_2021, king_triply_efficient_2025}.
The sample complexity of estimating non-linear properties typically features an exponential separation between the
access models
where multiple copies of a quantum state are measured either (i) individually or (ii) jointly in an entangled basis.
Most prominently, \emph{two-copy Bell sampling} enables efficient 
{\add learnability} of purities and squared Pauli expectation values~\cite{bluvstein_logical_quantum_2024}
with wide-ranging applications in quantum machine learning~\cite{huang_information_theoretic_2021, huang_quantum_advantage_2022}
and topological data analysis~\cite{scali_the_topology_2024}.
Which other quadratic properties can be efficiently learned via Bell sampling is 
subject of active research~\cite{hangleiter_bell_sampling_2024}.
In the contexts of error correction and mitigation, 
simultaneous access to multiple copies is essential for benchmarking so-called resources of \emph{magic}~\cite{gross_schur_weyl_2021, haug_scalable_measures_2023, leone_learning_t_2024, haug_efficient_quantum_2024, hangleiter_bell_sampling_2024, bittel_a_complete_2025, bittel_operational_interpretation_2025}
and for \emph{virtual distillation}~\cite{cotler_quantum_virtual_2019, koczor_exponential_error_2021, huggins_virtual_distillation_2021, hakoshima_localized_virtual_2024}.
Having said this, multi-copy measurements are challenging to implement and experiments have just reached the stage of 
{\add two-copy} 
Bell sampling~\cite{schmid_experimental_direct_2008, islam_measuring_entanglement_2015, bluvstein_a_quantum_2022, huang_quantum_advantage_2022, bluvstein_logical_quantum_2024, acharya_quantum_error_2025}.

Ever since its first steps were taken~\cite{shor_scheme_for_1995, laflamme_perfect_quantum_1996, steane_multiple_particle_1996, steane_error_correcting_1996, knill_theory_of_1997, shor_quantum_analog_1997, calderbank_quantum_error_1998, rains_quantum_weight_1998, rains_quantum_shadow_1999, ashikhmin_upper_bounds_1999, ashikhmin_quantum_error_2000},
QEC has been deeply intertwined with concepts 
of
classical error correction 
\cite{macwilliams_a_theorem_1962, gleason_weight_polynomials_1970, macwilliams_the_theory_1978, pless_handbook_of_1998, nebe_self_dual_2006}.
On the {\add combinatorial} 
side,
\emph{quantum weight enumerators} (QWEs)~\cite{shor_quantum_analog_1997,calderbank_quantum_error_1998, rains_quantum_weight_1998, rains_quantum_shadow_1999}
have generalized 
corresponding
classical notions, carrying results such as the \emph{MacWilliams identity} 
{\add(see Eq.~\eqref{eq:macwilliams_identity} below)}
into the quantum setting.
The trifecta of 
 QWEs
comprises 
\emph{Shor--Laflamme}, 
\emph{Rains' unitary}, 
and \emph{Rains' shadow QWEs}, 
which 
{\add present the same information from different angles.}
The Shor--Laflamme QWE distribution (SLD) is best understood for stabilizer \emph{quantum error-correcting codes} (QECCs), where it coincides with the Pauli weight distribution of the code's stabilizer group 
Strikingly, the SLD can be converted---via
{\add the quantum} MacWilliams transform---into what we call
{\add the}
\emph{dual SLD}, which is the weight distribution of the logical Pauli group.
Hence, from the SLD, one can infer the code distance $d$, i.e., the smallest weight of a 
logical Pauli operator.
From a complexity-theoretic point of view, this is interesting because the problem of computing the distance of a linear code is \textsf{NP}-hard~\cite{vardy_algorithmic_complexity_1997}.
In turn, 
{\add also} computing SLDs is \textsf{NP}-hard. 
Nevertheless, recent advances involving tensor-network methods are pushing the boundaries of 
{\add computing SLDs} 
in practice 
\cite{cao_quantum_lego_2024, cao_quantum_weight_2024, braccia_computing_exact_2024, pato_planqtn_a_2025},
{\add 
for example through \href{https://zenodo.org/records/16761072}{PlanqTN}~\cite{pato_planqtn_a_2025},
a user-friendly software implementation of the quantum LEGO framework~\cite{cao_quantum_lego_2024}.}
 
QWEs are also powerful from alternate perspectives. 
From the very beginning~\cite{shor_quantum_analog_1997, rains_quantum_shadow_1999},
{\add they} were leveraged to rule out the existence of QECCs with certain code parameters $\llbracket n,k,d\rrbracket$,
where $n$ and $k$ denote the numbers of physical and logical qubits, respectively. More recently, a set of criteria has been found from QWEs to bootstrap codes that correct errors from codes that merely detect them~\cite{kubischta_quantum_weight_2023}.
For modern efforts in designing \emph{quantum low-density parity-check} (qLDPC) codes~\cite{bravyi_homological_product_2014, breuckmann_quantum_low_2021, panteleev_asymptotically_good_2022, leverrier_decoding_quantum_2023, bravyi_high_threshold_2024},
Rains' shadow enumerators provide bounds on parameters of asymptotically good qLDPC code families
for which the distance $d$ scales like 
{\add $c\times  n$}
for a non-zero constant $c>0$~\cite{panteleev_asymptotically_good_2022, leverrier_decoding_quantum_2023}. 
In this context, it is shown that the constant is bounded as $c \le 1/3$~\cite{rains_quantum_shadow_1999}.
Despite these applications,  
however,
the physical interpretation of Rains' shadow enumerators---the 
{\add central}
topic of this work---has remained a mystery for almost three decades.

The physics behind Shor--Laflamme enumerators, on the other hand, is well understood.
Originally introduced for the purpose of capturing the relationship between the 
entanglement fidelity 
and average fidelity 
of codewords sent through the local depolarizing channel~\cite{shor_quantum_analog_1997},
they have since then found numerous other interpretations.
In the literature on the geometry of quantum states and entanglement,
SLDs are often called \emph{correlation tensor norms}~\cite{de_vincente_multipartite_entanglement_2011, laskowski_correlation_tensor_2011, kloeckl_characterizing_multipartite_2015, morelli_correlation_constraints_2024},
but they have also recently appeared under the name 
\emph{sector length distribution}~\cite{wyderka_characterizing_quantum_2020, eltschka_maximum_nbody_2020, miller_graphstatevis_interactive_2021},
which decomposes the purity $\Tr[\rho^2]$ of an $n$-qubit state $\rho$ into $n+1$ Pauli-weight sectors 
and conveniently shares its acronym with the term 
Shor--Laflamme distribution.
Moreover, SLDs in disguise have been employed 
in the study of random quantum circuits~\cite{quek_exponentially_tighter_2022, braccia_computing_exact_2024} 
and of quantum chaos~\cite{schuster_operator_growth_2023}.
But despite their physical connection to quantum entanglement, it is not clear under what circumstances SLDs can be measured efficiently.

Suspecting a two-copy protocol for 
{\add learning}
SLDs is not far-fetched. 
A first clue is that SLDs decompose purities, 
which can be efficiently measured 
{\add in a two-copy learning experiment}~\cite{chen_exponential_separations_2022}. 
The same is true for all (exponentially many) squared Pauli expectation values, which enter the definition of SLDs [Eq.~\eqref{def:sld}].
In this work, we develop a method to circumvent the need of evaluating sums over exponentially many terms by establishing a direct connection between Rains' quantum shadow enumerators and two-copy Bell sampling experiments.
The latter give rise to notions of \emph{singlet} and \emph{triplet probabilities},
i.e., to the frequencies of measuring either the single \emph{antisymmetric} or one of the three \emph{symmetric} Bell states, respectively. 
Our main theoretical result (Thm.~\ref{thm:main}) is that these triplet probability distributions are the same as Rains' quantum shadow enumerators.
This 
{\add surprising}
discovery has several notable implications: 
(i) it provides a physical interpretation for the theoretical machinery of QWEs;
(ii) it enables exponential speedups in classical postprocessing compared to brute-force approaches for computing QWEs from Bell samples; and (iii) it identifies the conditions under which some enumerators can be efficiently estimated, while others cannot.


{\add 
Building on this, we develop here a rigorous framework that incorporates a wide range of practical considerations for learning QWEs. 
We then demonstrate the utility of this framework through entanglement characterization experiments conducted on a trapped-ion quantum processor.
}

In a first two-copy Bell sampling experiment, 
we investigate a suite of six-qubit states and showcase how the QWE machinery can be leveraged to reveal detailed information about the multi-faceted entanglement structure that differentiates these states.
Specifically, from the measured triplet probability distributions, 
we compute the \emph{total $i$-body correlations} [Eq.~\eqref{def:sld}],
\emph{averaged {\add subsystem} purities} [Eq.~\eqref{def:apd}], and a lower bound on the \emph{concurrence} [Ineq.~\eqref{eq:lower_bound_concurrence}].
Of high practical value is that all of this is possible without expensive postprocessing or the need for updating readout circuits.
Finally, to remove errors from the Bell samples, 
we propose and implement a heuristic error mitigation strategy that heavily relies on the QWE machinery itself.

In a second experiment, 
we directly measure the QWEs of the $\llbracket 7,1,3\rrbracket$ color code by performing transversal Bell measurements on two copies of the maximally mixed logical state.
Our experiment simultaneously constitutes a Bell measurement on the logical level of this code.
Therefore, we can detect and correct errors on the Bell samples.
After discarding all outcomes with violated parity checks (ca.~$57\%$ of the data),
we can correctly (within error bars) infer the number of weight-$i$ stabilizers and logical Pauli operators for all $i\in\{0,\ldots, 7\}$.
This observation can be regarded as an experimental signature confirming that the probed QECC has distance $d=3$.

Our work is organized as follows. 
{\add  We review 
QWEs and the MacWilliams transform in Sec.~\ref{sec:qwe_macwilliams}.
Our main result that QWEs can be understood in terms of two-copy Bell sampling experiments is established in
Sec.~\ref{sec:bell_sampling}. }
All our experimental demonstrations are then reported in Sec.~\ref{sec:experiment}.
We prove performance guarantees concerning the scalability of our 
{\add two-copy learning}
protocol in Sec.~\ref{sec:complexity_and_robustness}.
Lastly, in Sec.~\ref{sec:dicke_entanglement_stability}, we leverage {\add our new insights} to analyze the noise resilience of entanglement for certain quantum states on up to $n=1000$ qubits.

\section{Quantum weight enumerators}
\label{sec:qwe_macwilliams}

In 1962, MacWilliams discovered that the Hamming weight distribution of a classical linear code 
uniquely determines that of its dual, and vice versa~\cite{macwilliams_a_theorem_1962}.
Her discovery sparked a research program about error-correcting codes~\cite{gleason_weight_polynomials_1970, macwilliams_the_theory_1978, pless_handbook_of_1998,  nebe_self_dual_2006}
that quickly found its way into the quantum realm~\cite{shor_quantum_analog_1997, calderbank_quantum_error_1998, rains_quantum_weight_1998, rains_quantum_shadow_1999, ashikhmin_upper_bounds_1999, ashikhmin_quantum_error_2000}
{\add once} the first \emph{quantum error-correcting code} (QECC) had been constructed~\cite{shor_scheme_for_1995}.
At the core of the quantum weight enumerator machinery lies the 
\emph{quantum MacWilliams transform}, 
{\add a linear involution, $M=(M_{i,j})_{i,j=0}^n\in\GL(\RR^{n+1})$,
which is traditionally formulated in terms of polynomial identities.
Its matrix entries
\begin{align}
    \label{eq:macwilliams_matrix}
  M_{i,j} =  \frac{1}{{2^n}} \sum_{l=0}^n \binom{n-j}{i-l} \binom{j}{l} (-1)^l 3^{i-l} \,.
\end{align}
also appear as the coefficients in front of $x^i$ in the \mbox{polynomials} $f_j(x)=2^{-n}(1+3x)^{n-j}(1-x)^j$~\cite{macwilliams_the_theory_1978}.}
In this section, we reformulate this machinery from a fresh perspective. 
{\add First, we avoid the abstract language of polynomial rings,
which allows us to clarify the role of Rains' shadow QWEs 
in the linear algebra behind
the MacWilliams transform.}
Second, following the literature about sector lengths~\cite{de_vincente_multipartite_entanglement_2011, laskowski_correlation_tensor_2011, kloeckl_characterizing_multipartite_2015, morelli_correlation_constraints_2024, wyderka_characterizing_quantum_2020, eltschka_maximum_nbody_2020, miller_graphstatevis_interactive_2021}, 
we define all QWEs in terms of quantum states $\rho$ rather than code space projectors $\Pi$.
The latter concept is recovered as a special case (after appropriate renormalization) when specializing to 
\begin{align} \label{eq:rho_qecc}
    \rho_\text{QECC} =\frac{ \Pi }{ \Tr[\Pi]} \,,
\end{align}
which carries the physical interpretation of the maximally mixed state inside the code space of the QECC.


\subsection{Shor and Laflamme's QWEs }
\label{sec:sld_intro}

The first notion of QWEs reviewed here 
{\add has been coined in a seminal paper}
by Shor and Laflamme~\cite{shor_quantum_analog_1997}.
Readers without any prior exposure 
{\add to this topic}
might benefit from consulting 
{\add the examples in}
App.~\ref{app:example} before proceeding.
For an $n$-qubit state $\rho$, we define the \emph{Shor--Laflamme QWE distribution} (SLD) as
$\mathbf{a}[\rho] = (a_i[\rho])_{i=0}^n 
 \in \RR^{n+1}$, 
{\add where }
\begin{align} \label{def:sld}
    a_i[\rho] = \frac{1}{2^n}\sum_{\substack{P\in \{I,X,Y,Z\}^{\otimes n} \\\wt(P)=i}} \Tr[\rho P]^2 \, .
\end{align}
Here, $\wt(P)$ is the \emph{weight} of an $n$-qubit Pauli operator $P$, i.e., its number of non-identity tensor factors.
{\add Similarly, we define the \emph{dual SLD}, $\mathbf{b}[\rho] = (b_i[\rho])_{i=0}^n \in \RR^{n+1}$, 
where
\begin{align} \label{def:sld_dual}
    b_i[\rho] = \frac{1}{2^n}\sum_{\substack{P\in \{I,X,Y,Z\}^{\otimes n} \\\wt(P)=i}} \Tr[\rho P \rho P] \, .
\end{align} 
The SLD and the dual SLD uniquely determine each other via
\begin{align} \label{eq:macwilliams_identity}
    M \mathbf{a}[\rho] = \mathbf{b}[\rho] \, ,
\end{align}
which is known as the \emph{quantum MacWilliams identity}~\cite{shor_quantum_analog_1997}.
Note that our normalizations of $M$, $\mathbf{a}[\rho]$, and $\mathbf{b}[\rho]$ differ from those in Ref.~\cite{shor_quantum_analog_1997},
which has the advantage of $M$ being self-inverse
and the purity of $\rho$ decomposing into
\begin{align} \label{eq:sld_and_purity}
    \Tr[\rho^2] = \sum_{i=0}^n a_i[\rho] \, .
\end{align}}

SLDs have 
{\add various} 
useful properties:
(i)
for every $n$-qubit state $\rho$ and 
{\add  every Pauli weight}
$i\in\{0,\ldots,n\}$,
the 
{\add QWE}
$a_i[\rho]$
is invariant under local  unitary operations~\cite{rains_quantum_weight_1998}, which implies
\begin{align} \label{eq:sld_pure_product}
a_i \left[\Psi_1\otimes \ldots\otimes \Psi_n \right] 
= a_i\left[\ket{0}\!\!\bra{0}^{\otimes n}\right ]
=  \frac{1}{2^n}\binom{n}{i} \, ,
\end{align}
for   {\add all pure} product states $\Psi_1\otimes \ldots\otimes \Psi_n$;
{\add and}
(ii) 
due to the triangle inequality, $a_i[\rho]$ is convex in $\rho$.
In other words, 
{\add Shor--Laflamme QWEs cannot increase}
under incoherent mixtures.
In combination, these two properties show  
\begin{align} \label{eq:nbody_criterion}
{\add \exists \ i: \ a_i    [\rho] > \frac{1}{2^n} \binom ni }
    \hspace{1mm}
    \Longrightarrow
    \hspace{1mm}
    \rho \, \text{ {\add is}   entangled}\,,
\end{align}
which we refer to as the 
\emph{{\add $i$-body} sector length criterion}~\cite{de_vincente_multipartite_entanglement_2011}.

Under the physically well-motivated local depolarizing noise channel, 
SLDs decay as 
\cite{ashikhmin_quantum_error_2000}
\begin{align}  \label{eq:sector_length_decay}
    a_i \left[\mathcal{E}_p^{\otimes n} [ \rho] \right] = (1-p)^{2i} a_i[\rho] \, ,
\end{align}
where $\mathcal{E}_p[\cdot] = (1-p)[\cdot]+p\tfrac{\mathbbm 1}{2}$ denotes the single-qubit depolarizing channel with error 
{\add probability}
$p\in[0,1]$.
Inserting Eq.~\eqref{eq:sector_length_decay} into Eq.~\eqref{eq:sld_and_purity}
shows that purities decay as 
\begin{align} \label{eq:purity_decay}
\Tr\left[(\mathcal{E}_p^{\otimes n} [ \rho])^2\right ]= \sum_{i=0}^n a_i[\rho] (1-p)^{2i}
\end{align}
under local depolarizing noise.
Our first technical contribution is an analog of Eq.~\eqref{eq:purity_decay} for fidelities rather than purities:

\bigskip
\refstepcounter{mycounter}
\noindent
\textbf{Proposition~\arabic{mycounter}\label{lem:overlap_decay}}  (Overlap decay under local depolarizing noise)
\emph{Let $\rho$ be an $n$-qubit state, $\mathbf{a}[\rho] = (a_i[\rho])_{i=0}^n \in \RR^{n+1}$ its SLD, and $p\in[0,1]$ a {\add depolarization  probability}.
Then, it holds}
\begin{align} \label{eq:overlap_decay}
    \Tr\left[ \,  \rho \  \mathcal{E}_p^{\otimes n} [ \rho]  \right ]= \sum_{i=0}^n a_i[\rho] (1-p)^{i} \, .
\end{align} \smallskip
\emph{Proof:} See App.~\ref{app:overlap_decay}. \hfill $\square$ 

\bigskip

Proposition~\ref{lem:overlap_decay} is especially useful for pure states {\add of the form} $\Psi = \ket{\psi}\!\bra{\psi}$, for which
the overlap $\Tr[\Psi \sigma]$ with any other state $\sigma$ is the same as the fidelity $\bra{\psi} \sigma \ket{\psi}$.
Furthermore, if $\Psi$ is a \emph{genuinely multipartite entangled} (GME) stabilizer state---which implies that the operator  $ {\add W=  \frac{\mathbbm 1}{2}- \Psi}$ is a GME witness~\cite{terhal_bell_inequalities_2000, guehne_detection_of_2002, bourennane_experimental_detection_2004, guehne_entanglement_detection_2009}---it
follows that the noisy state $\sigma = \mathcal{E}_p^{\otimes n} [\Psi] $ remains GME as long as
\begin{align} \label{eq:fidelity_criterion_stab}
     \sum_{i=0}^n a_i[\Psi] (1-p)^i > 0.5 \, .
\end{align}
This and other entanglement criteria [Eqs.~\eqref{eq:nbody_criterion}, \eqref{eq:purity_criterion}, \eqref{eq:lower_bound_concurrence}] that can be tested via SLDs motivate computing them,
which has been achieved for \emph{Greenberger-Horne-Zeilinger}  (GHZ)~\cite{aschauer_local_invariants_2004} 
and for 1-dimensional periodic \emph{cluster states}~\cite{miller_shor_laflamme_2023} 
by counting their weight-$i$ stabilizer operators.

Even in the more general case where 
\begin{align} \label{eq:rho_qecc_stab}
    \rho_\text{QECC}^\text{stab} = \frac{1}{2^n}\sum_{S \in\mathcal{S}} S
\end{align}
is the maximally mixed state [Eq.~\eqref{eq:rho_qecc}] within the code space of an $\llbracket n,k,d\rrbracket$ QECC with stabilizer group $\mathcal{S}$~\cite{calderbank_quantum_error_1998},
one can find the $i$-body {\add Shor--Laflamme enumerator}
\begin{align} \label{eq:sld_stabilizer_qecc}
    a_i[\rho_\text{QECC}^\text{stab}  ] = \frac{1}{2^n}
    \left| \left\{ S \in \mathcal{S} \ \big | \  \wt(S)=i \right \}\right| \, ,
\end{align}
by counting weight-$i$ stabilizer operators.
{\add Similarly, one can show (see App.~\ref{app:derivation_pwd_dual_and_shadow}) that
\begin{align}\label{eq:sld_dual_step3}
 b_i[    \rho_\text{QECC}^\text{stab}  ] 
 = \frac{1}{2^{n+k}}
    \left|\{ P \in \mathcal{S}^\perp \ | \  \wt(P)=i \}\right| \, , \text{ where}
\end{align}
\begin{align}  \label{def:normalizer}
\mathcal{S}^\perp  = \left\{ P\in \{I,X,Y,Z\}^{\otimes n} \ \vert \  \forall S \in \mathcal{S}: SP = PS \right\} 
\end{align} 
is the \emph{dual of $\mathcal{S}$} considered as an additive code over the field $\FF_4$~\cite{nebe_self_dual_2006}.
The latter can be understood as the set $\{I,X,Y,Z\}$ of single-qubit Pauli operators~\cite{calderbank_quantum_error_1998}.
Other interpretations of $\mathcal{S}^\perp$ are as the normalizer of $ \mathcal{S}$, and as  the logical Pauli group of the QECC (both modulo global phases). 
}

{\add
For an arbitrary $\llbracket n,k,d\rrbracket$ QECC with code state $\rho_\text{QECC}$ as in Eq.~\eqref{eq:rho_qecc},}
the discrepancy between 
$\mathbf{a}[\rho_\text{QECC}]$ and $\mathbf{b}[\rho_\text{QECC}]$ reveals the code's distance 
\begin{align} \label{eq:distance}
d= \min \left\{ i>0  \ \vert \ a_i[\rho_\text{QECC}] <  2^k b_i[\rho_\text{QECC}] \right\} \,.
\end{align}
This is clear for stabilizer QECCs, where 
{\add $A_i=2^{n}a_i[\rho_\text{QECC}^\text{stab}]$}
and 
{\add $B_i=2^{n+k} b_i[\rho_\text{QECC}^\text{stab}]$}
are equal to the number of weight-$i$ stabilizers [Eq.~\eqref{eq:sld_stabilizer_qecc}] and logicals [Eq.~\eqref{eq:sld_dual_step3}], respectively, and remains true for non-stabilizer QECCs~\cite{shor_quantum_analog_1997}.



\subsection{Rains' unitary QWEs}
\label{sec:apd_intro}

The second notion of QWEs 
{\add can be}
explained in terms of subsystem purities~\cite{rains_quantum_weight_1998}.
The reduced state  $\rho_S = \Tr_{S^\text{c}}[\rho]$ on a subset of qubits $S \subset \{1,\ldots, n\}$ is obtained from $\rho$ by tracing out the qubits in the complement $S^\text{c} = \{1,\ldots,n\}\setminus S$.
We denote the average of the subsystem purity $\Tr[\rho_S ^2]$ over all marginals of a fixed size $i \in \{0,\ldots, n\} $ by 
\begin{align}\label{def:apd}
    a'_i[\rho] = \frac{1}{\binom{n}{i}} \sum\limits_{\substack{S\subset \{1,\ldots,n\}\\ \vert S\vert = i}} \Tr[ \rho_S ^2  ] \,,
\end{align}
which is the definition of \emph{Rains' unitary QWEs} with a more physically-motivated 
{\add prefactor}.
Here, we will often refer to the vector $\mathbf{a'}[\rho] = (a'_i[\rho])_{i=0}^n  \in \RR^{n+1}$ 
of Rains' unitary QWEs as 
{\add the}
\emph{averaged purity distribution} (APD). 
{\add Following Rains, we dually define
$\mathbf{b}'[\rho]=(b'_i[\rho])_{i=0}^n$ via $b'_i[\rho] = a'_{n-i}[\rho]$.}

In the exact case without statistical errors,
APDs and SLDs carry exactly the same information about $\rho$ since they can be linearly converted into each other via 
$\mathbf{a'}[\rho] = T' \mathbf{a}[\rho]$ and 
$\mathbf{a}[\rho] = T'^{-1} \mathbf{a} '[\rho]$,
where
\begin{align} \label{eq:trafo_sld_to_apd}
    T'_{i,j} &= {2^{n-i}} {\binom{n}{i}}^{-1} \binom{n-j}{n-i} 
    \\
    \text{and} 
    \hspace{5mm} 
    \label{eq:trafo_apd_to_sld}
    T'^{-1}_{i,j} &= {2^{j-n}}{\binom{n}{j}}
     \binom{n-j}{n-i} (-1)^{i+j}\, .
\end{align}

{\add Consider an $n$-qubit state $\rho$ with  $a'_i[\rho]< a'_n[\rho]$  for some subsystem size  $i\in\{1,\ldots, n-1\}$.
Since $a'_n[\rho]= \Tr[\rho^2]$, 
this implies the existence of at least one subset $S$, with $\lvert S \rvert = i$, that is bipartite-entangled with its complement $S^\text{c}$ 
\cite{horodecki_method_for_2002, horodecki_quantum_entanglement_2009}.
In particular, this shows
\begin{align} \label{eq:purity_criterion}
   {\add \exists \ i: \ a'_{i}[\rho]} < a'_n[\rho]
    \hspace{2mm}
    \Longrightarrow
    \hspace{2mm}
    \rho \,  \text{ {\add is}  entangled} \, ,
\end{align}
which we refer to as the \emph{purity criterion}.}
Empirically~\cite{miller_shor_laflamme_2023}, the most noise-robust version of 
{\add this   criterion}
is attained for $i=n-1$, where Rains' unitary QWE takes the simple form
\begin{align}
    a'_{n-1}[\rho] &= \frac{1}{n} \sum_{i=1}^n \Tr\left[ (\Tr_{\{i\}}[\rho])^2 \right] \,.
\end{align}


\subsection{Rains' shadow QWEs}
\label{sec:qwe_shadow_intro}

When Rains introduced the notion of quantum shadow enumerators~\cite{rains_quantum_shadow_1999}, 
he took inspiration from the classical coding literature~\cite{conway_a_new_1990}. 
Here, however, we find it more 
{\add instructive} 
to provide a 
physical introduction 
{\add to this third notion of QWEs.}
Let $\rho$ be an $n$-qubit 
{\add density matrix}, 
$\rhotranspose$ its transpose in the computational basis, and  $\rho \mapsto \tilde \rho$ the  \emph{state inversion map},
where 
\begin{align} \label{eq:spin_flipped_state}
     \tilde \rho &= Y^{\otimes n} \rhotranspose Y^{\otimes n} 
\end{align}
is the \emph{spin-flipped} state~\cite{wootters_entanglement_of_1998, rungta_universal_state_2001, wong_potential_multipartite_2001, hall_multipartite_reduction_2005}.
The name of $\tilde{\rho}$ stems from the fact that in the case of a single qubit $\rho = \frac{I+\mathbf{r}\cdot \sigma}{2}$ with Bloch sphere vector $\mathbf{r}\in \RR^3$ (aka spin), the state inversion map is flipping the spin 
{\add as} 
$\mathbf{r}\mapsto - \mathbf{r}$, i.e., $\tilde \rho = \frac{I-\mathbf{r}\cdot \sigma}{2}$.
This should not be confused with the \emph{bit ``flip'' channel} $\rho\mapsto X\rho X$, where the spin is \emph{rotated} around the $x$-axis of the Bloch sphere.
{\add In contrast, the state inversion map is a \emph{reflection} and, therefore,} 
impossible to physically implement for an unknown quantum state $\rho$.
In technical terms,
the state inversion map 
$\rho \mapsto \tilde \rho$ 
is positive but not completely positive.
%
{\add Thus equipped,}
we are 
{\add now} 
in the position to {\add define} \emph{Rains' shadow enumerators}~\cite{rains_quantum_shadow_1999},
{\add $\mathbf{\tilde a}[\rho] = (\tilde a_i[\rho])_{i=0}^n\in \RR^{n+1}$, where}
\begin{align} \label{def:rains_qwe}
\tilde a_i[\rho] = \frac{1}{2^n} \sum\limits_{\substack{P\in \{I,X,Y,Z\}^{\otimes n} \\\wt(P)= i}} \Tr[ \rho P \tilde \rho P ]  \, .
\end{align}

{\add 
SLDs and APDs are not the only QWEs that can be transformed into each other [recall Eq.~\eqref{eq:trafo_sld_to_apd}].
To achieve the same for Rains' shadow QWEs, we need the matrix $\tilde T = (\tilde T _ {i,j})_{i,j=0}^n$ whose entries
\begin{align} \label{eq:trafo_sld_to_tpd_via_signed_mwt}
\tilde T _{i,j} = (-1)^j M_{i,j} 
\end{align}
arise from MacWilliams' transform [Eq.~\eqref{eq:macwilliams_matrix}] 
after alternating the sign of its columns.
Acting with this matrix on the SLD of an $n$-qubit state $\rho$ yields its shadow QWEs.
In other words,
\begin{align}
\tilde{\mathbf{a}}[\rho] = \tilde T \, \mathbf{a}[\rho] \, .
\end{align}
The inverse transformation (from shadow QWEs to the SLD) is similarly given by
\begin{align} \label{eq:trafo_tpd_to_sld_via_signed_mwt}
\tilde T^{-1}_{i,j} = (-1)^i M_{i,j} \, .
\end{align}
}

{\add  
In the special case where $\rho = \rho^\text{stab}_\text{QECC}$ [recall Eq.~\eqref{eq:rho_qecc_stab}] for some stabilizer group $\mathcal{S}$,
one can show (see App.~\ref{app:derivation_pwd_dual_and_shadow}) that
\begin{align}\label{eq:shadow_step4}
 \tilde a_i[    \rho_\text{QECC}^\text{stab}  ] 
 = \frac{1}{2^{n+k}}
    \big| \big\{ P \in \tilde{\mathcal{S}}  \ \big | \  \wt(P)=i \big \}\big| \, , 
\end{align}
where the subset $\tilde {\mathcal{S}}\subset \{I,X,Y,Z\}^{\otimes n}$ is defined as
\begin{align}
\label{def:shadow_coset}
    \tilde {\mathcal{S}} = \{P  \   \vert \ \forall S \in \mathcal{S} \!: PS = (-1)^{\wt(S)}SP\} \,,
\end{align}
in reminiscence of Eq.~\eqref{def:normalizer}.
While $\mathcal{S}^\perp$ is called the dual of $\mathcal{S}$,
the subset $\tilde{\mathcal{S}}$ known as its \emph{shadow}~\cite{nebe_self_dual_2006}.
This terminology is rooted in the following characterization:
}

\bigskip
\refstepcounter{mycounter}
\noindent
\textbf{Proposition~\arabic{mycounter}\label{lem:shadow_characterization}}  (Rains' shadow is a coset of the logical group)
\emph{Let $\mathcal{S}$ be the stabilizer group of an $\llbracket n,k,d\rrbracket$ QECC.
\add 
Then, it holds $\tilde{\mathcal{S}} \in \{I,X,Y,Z\}^{\otimes n}/\mathcal{S^\perp}$. 
In other words, we have}
\begin{align} \label{eq:shadow_characterization}
    \tilde{\mathcal{S}} =  P \mathcal{S}^\perp \,.
\end{align}
\emph{{\add for every $P\in  \tilde{\mathcal{S}}$.
Moreover, we have $\tilde{\mathcal{S}}=\mathcal{S}^\perp$} 
iff $\mathcal{S}$ admits generators $S_1,\ldots, S_{n-k}$ such that $\wt(S_i)$ is even for all $i$.  
}

\smallskip
\emph{Proof:} See App.~\ref{app:proof_shadow_characterization}. \hfill $\square$ 

\bigskip

{\add 
Note that  Prop.~\ref{lem:shadow_characterization}  naturally generalizes Thm.~3 in Chap.~3 of Ref.~\cite{pless_handbook_of_1998},
which establishes the analogous result for linear codes over $\FF_2$.
To our knowledge, Prop.~\ref{lem:shadow_characterization} has not been explicitly formulated in the existing literature.
As a corollary,
we obtain $|\tilde{\mathcal{S}}|=2^{n+k}$.
In this light, Eq.~\eqref{eq:shadow_step4} implies that $\tilde a_i[\rho^\text{stab}_\text{QECC}]$ can be regarded as the probability that a uniformly random Pauli operator $P\in \tilde{\mathcal{S}}$ has $\wt(P)=i$.

}

\subsection{\add The quantum weight enumerator machinery}
\label{sec:macwillams_linear_algebra_behind}

\begin{figure}
    \centering
    \includegraphics{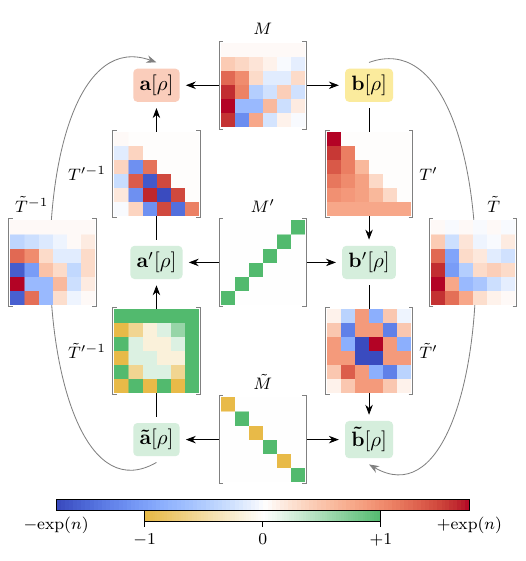}
    \caption{The self-inverse MacWilliams transforms $M$, $M'$, and $\tilde M$ 
    convert QWEs (left) into dual QWEs (right).
    They represent the same linear map in three different bases of $\RR^{n+1}$ that correspond to  Shor--Laflamme QWEs 
    (top), Rains' unitary QWEs 
    (center), and Rains' shadow QWEs 
    (bottom).
    The three transforms are related via basis change matrices $T'$, $\tilde T$, and $\tilde T'$, e.g., $M' = T' M T'^{-1} $. 
    Plotted are the matrix entries in the special case of $n=5$ qubits,
    where red and blue (green and yellow) colors show which matrices have entries that are exponentially diverging (bounded by $\pm 1$) in the general case of $n$ qubits.\footnote{\href{https://mc-zen.github.io/qsalto}{Digital feature: interactive version of this figure for arbitrary $n < 500$.}}
    In Thm.~\ref{thm:main}, we show that Rains' shadow QWEs $\mathbf{\tilde a}[\rho]$ are the same as the triplet probability distribution obtained in a Bell sampling experiment on $\rho\otimes \rho$, where $\rho $ is any $n$-qubit state. 
    Thus, one can efficiently {\add learn} $\mathbf{\tilde a}[\rho]$,
    which we {\add experimentally} demonstrate in Sec.~\ref{sec:experiment}.
    Note that statistical errors on $\mathbf{\tilde a}[\rho]$ never uncontrollably spread onto Rains' unitary QWEs $\mathbf{a'}[\rho]$ as the entries of ${\tilde T}'^{-1}$ are bounded by $\pm 1$, 
    {\add also} see Thm.~\ref{thm:sample_complexity_bounds}. 
    For some states $\rho$, however, errors can unfavorably amplify during the transform to the SLD $\mathbf{a}[\rho]$ 
    {\add since} some matrix entries of  $\tilde T^{-1}$ are exponentially large.
    }
    \label{fig:macwilliams}
\end{figure}

{\add
Now that we have reviewed the three notions of QWEs,
let us further illuminate their roles  in the linear algebra behind the quantum MacWilliams identity.
Figure~\ref{fig:macwilliams} visualizes this situation for the instructive case of $n=5$ qubits
by means of a commutative diagram.
From top to bottom, the three rows depict the MacWilliams identity
expressed in the bases of SLDs, APDs, and shadow enumerators.
First, the top-center box shows the matrix entries $M_{i,j}$ from Eq.~\eqref{eq:macwilliams_matrix}, with positive and negative values in red and blue, respectively,
where greater color intensity indicates larger magnitude.
The unfortunate fact that $\vert M_{i,j}\vert $ can attain exponentially large values (in $n$)
is highlighted by the broad color bar at the bottom of the figure.
In contrast, the matrices $M' = T'M T'^{-1}$ and $\tilde M = \tilde{T} M \tilde{T}^{-1}$,
which correspond to the MacWilliams transform in the bases of APDs and shadow enumerators, respectively,
only contain the entries $-1$ (yellow), $0$ (white), and $+1$ (green).
It is important that 
all matrices in Fig.~\ref{fig:macwilliams} depend only on the number of qubits $n$,  
not on the state  whose QWEs they act upon.
For a concise collection of all matrix definitions, see App.~\ref{app:notation}.

The original MacWilliams identity  [Eq.~\eqref{eq:macwilliams_identity}] 
connects the SLD, $\mathbf{a}[\rho]$, to the dual SLD, $\mathbf{b}[\rho]$,
which are shown in the upper corners of the diagram.
Applying $T'$ (top-right box)
converts them into the APD, $\mathbf{a}'[\rho] = T' \mathbf{a}[\rho]$, and to the dual APD, $\mathbf{b}'[\rho] = T'\mathbf{b}[\rho] $.
The reverse transformation, $T'^{-1}$, is shown in the top-left box.
Both $T'$ and $T'^{-1}$ are lower-triangular matrices.
While $T'$ only contains non-negative matrix elements, 
we can see that the signs of the entries of $T'^{-1}$ 
follow a checkerboard pattern.
Similarly, the transformation matrices $\tilde T$ and $\tilde{T}^{-1}$, 
shown in the center-right and  -left boxes, 
arise from $M$ by flipping the signs of its columns and rows, respectively.
The transformation matrix $\tilde{T}' = \tilde T \, T'^{-1}$,
which directly converts the APD  into Rains' shadow enumerators,
i.e., $\tilde{\mathbf{a}}[\rho] = \tilde{T}' \mathbf{a}'[\rho]$,
is shown in the bottom-right box. 
The color coding shows that its matrix entries,
\begin{align} \label{eq:trafo_apd_to_tpd}
     \tilde T ' _{i,j}  &= \frac{1}{2^n}
     \binom{n}{j} \sum_{l=0}^i
     \binom{n-j}{i-l}\binom{j}{l} (-1)^{j-l} \, ,
\end{align}
suffer from exponentially large values.
Fortunately, its inverse (bottom-left box) has matrix entries,
\begin{align}
 \label{eq:trafo_tpd_to_apd}
    \tilde T '^{-1}_{i,j} & = 
\frac{1} {\binom{n}{i}}
\sum_{l=0}^i
\binom{n-j}{i-l}
\binom{j}{l}
{\left(-1\right)^{i-l}} \, ,
\end{align} 
that are well-behaved and, thus, displayed in green and yellow.

We have now explained all QWEs in Fig.~\ref{fig:macwilliams},
except for the one in the bottom-right corner.
This quantity is introduced here for the first time 
(to our knowledge).
We define it as $\tilde{\mathbf{b}}[\rho] = \tilde M \tilde{\mathbf{a}}[\rho]$
and call it the \emph{dual shadow enumerator distribution}
because it arises from Rains' shadow QWEs by applying the MacWilliams transform in the appropriate basis.
Note that, in a similar spirit,
Rains defined the dual APD, $\mathbf{b}'[\rho]$,
by applying the anti\-diagonal matrix $M' =T ' M T'^{-1}$, 
where
\begin{align} \label{eq:macwilliams_antidiagonal}
     M'_{i,j} = \delta_{i,n-j} \,,
\end{align} 
as shown in the box at the very center of Fig.~\ref{fig:macwilliams}.
In this basis, it is clear (from the Schmidt decomposition) that a state $\rho $
is pure iff $\mathbf{a}'[\rho] = \mathbf{b}'[\rho]$,
which is equivalent to
$\mathbf{a}[\rho] = \mathbf{b}[\rho]$,
and to
$\tilde{\mathbf{a}}[\rho] = \tilde{\mathbf{b}}[\rho]$.
Also for SLDs this criterion is lucid since 
$\mathbf{a}[\rho] $ sums up to the purity 
while $\mathbf{b}[\rho] $ is always a normalized probability distribution. 

In linear-algebraic terms, this means that $\rho$ is pure iff its QWE distribution
(in any basis) 
lies in the $(+1)$-eigenspace of the MacWilliams transform 
(in the same basis).
Since $M$ is self-inverse, its only other eigenvalues is given by $- 1$.
From Eq.~\eqref{eq:macwilliams_antidiagonal}, we obtain $\Tr[M] = \delta_{n,\text{even}}$,
which implies that the $(+1)$-eigenspace of $M$ has dimension $\lceil \tfrac{n+1}{2}\rceil$ whereas its $(-1)$-eigenspace is $\lfloor \tfrac{n+1}{2}\rfloor$-fold degenerate.
We would like to point out (for the first time to our knowledge) 
that the quantum MacWilliams transform is diagonalized in the basis of Rains' shadow enumerators.
More precisely, we have
\begin{align} \label{eq:macwilliams_diagonal}
    \tilde{M}  = \tilde T M \tilde T ^{-1} 
    =\diag( (-1)^{n}, \ldots, -1, 1 , -1, 1) \,,
\end{align} 
as depicted in the bottom-center box of Fig.~\ref{fig:macwilliams}.

Let us summarize what we know  about Rains' shadow enumerators at this point:
(i) they have an obscure definition [Eq.~\eqref{def:rains_qwe}] that involves the spin-flipped state and an exponentially large sum;
(ii) they provide a vector space basis of $\RR^{n+1}$ in which the quantum MacWilliams transform is  diagonal;
(iii) they were used~\cite{rains_quantum_shadow_1999} to derive a powerful bound, $d\le \tfrac n3 + 2$, that holds for all $\llbracket n,k,d\rrbracket$ QECCs;
(iv) for stabilizer QECCs they carry the interpretation of  the Pauli weight distribution over the coset $\tilde{\mathcal{S}}$ of  $\mathcal{S}^\perp$ that is defined in Eq.~\eqref{eq:shadow_step4};
and
(v) for arbitrary states  they remain a normalized probability distribution.
The physical interpretation of this probability distribution in the general case, 
however,
was hitherto unknown.
In Thm.~\ref{thm:main} below,
we will establish such an interpretation 
that was lacking for almost three decades:
\emph{Rains' shadow enumerators are the probabilities of observing  fixed triplet numbers  in a two-copy Bell sampling experiment.}

In this light, we can finally understand the aforementioned criterion 
``\emph{$\rho$ is pure $\Leftrightarrow$ $\tilde{\mathbf{a}}[\rho]=\tilde{\mathbf{b}}[\rho]$}''.
By Eq.~\eqref{eq:purity_from_bell_samples} below, $\rho $ is pure iff odd-singlet events are never observed, i.e., iff $\tilde a_{n-j}[\rho] = 0$ whenever  $j$ is even.
This, in turn, is equivalent to $\tilde{\mathbf{a}}[\rho]$ lying in the $(+1)$-eigenspace of $\tilde M$.

More importantly, however, this interpretation settles open problems in quantum learning theory.
It is well known that all six QWE distributions in Fig.~\ref{fig:macwilliams} can be approximated to inverse-polynomial accuracy  
with polynomially many samples (both in $n$)
in the single-copy access model,
provided $\rho$ is a stabilizer state as in Eq.~\eqref{eq:rho_qecc_stab}.
Indeed,
one first determines an $\FF_2$-linear basis of $\mathcal{S}$ using the algorithm from Ref.~\cite{gottesman_talk_2008}.
Then one can sample from $\mathcal{S}$, $\mathcal{S}^\perp$, and $\tilde{\mathcal{S}}$ to approximate
$\mathbf{a}[\rho]$, $\mathbf{b}[\rho]$, and $\tilde{\mathbf{a}}[\rho]$, respectively.
From the shadow QWEs, one can infer the remaining QWEs by applying benign transforms in Fig.~\ref{fig:macwilliams}. 
In the general case, where $\rho$ is an arbitrary state,
Thm.~\ref{thm:main} implies that, given access to Bell samples from $\rho\otimes\rho$,
one can efficiently learn the triplet probability distribution, i.e., Rains' shadow enumerators, $\tilde{\mathbf{a}}[\rho]$.
As before, this also grants access to $\tilde{\mathbf{b}}[\rho]$,  $\mathbf{a}'[\rho]$, and $\mathbf{b}'[\rho]$, also 
see App.~\ref{app:operator_norms}.
The situation  changes drastically, if we are instead given access to Bell samples from $\rho \otimes \tilde\rho$.
In this case, our results imply that the resulting triplet probability distribution is equal to the dual SLD, $\mathbf{b}[\rho]$.
%
Finally, let us remark that the learnable quantities in the $\rho\otimes \rho$ and $\rho \otimes \tilde\rho$ access models, respectively,
are highlighted in Fig.~\ref{fig:macwilliams} with a green and yellow background.
}

{\add
\section{Learning enumerators via Bell sampling}
\label{sec:bell_sampling}
}

\begin{figure}
    \centering
    \includegraphics[width= .9\columnwidth]{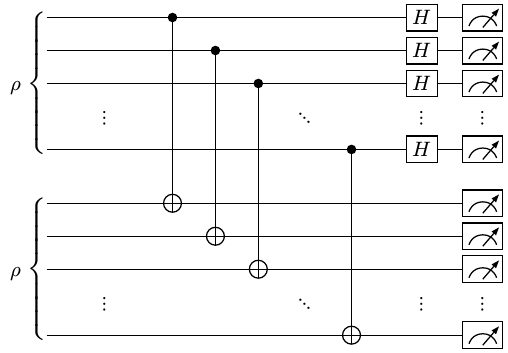}
    
    \vspace{1mm}
    \caption{Quantum circuit for a Bell sampling experiment with two copies of an $n$-qubit state $\rho$. 
    As discussed in the main text, this single-setting experiment allows us to measure the QWEs of $\rho$.
    In the special case of  $\rho = \rho_\text{QECC}$ [Eq.~\eqref{eq:rho_qecc}], this enables access to the QWEs of any QECC under investigation. 
    If the measurement outcomes of a qubit pair (connected via a $\Cnot$ gate) are both equal to $1$, the qubits are effectively projected into a singlet; otherwise into a triplet.
    By Thm.~\ref{thm:main}, the resulting triplet probability distribution [Eq.~\eqref{eq:tpd}] is the same as Rains' shadow QWEs [Eq.~\eqref{def:rains_qwe}],
    {\add which can be transformed into other QWEs as in Fig.~\ref{fig:macwilliams}.}
    }
    \label{fig:two_copy_bell_measurement}
\end{figure}

Two-copy Bell sampling is a technique for certifying and quantifying entanglement in many-body quantum systems that is well established 
both in theory~\cite{carvalho_decoherence_and_2004, mintert_concurrence_of_2005, mintert_observable_entanglement_2007, aolita_scalable_method_2008,
hangleiter_bell_sampling_2024}
and experiment~\cite{schmid_experimental_direct_2008, 
islam_measuring_entanglement_2015, bluvstein_a_quantum_2022, huang_quantum_advantage_2022,
bluvstein_logical_quantum_2024}.
It 
{\add crucially}
relies on the so-called $\Swap$ trick~\cite{hastings_measuring_renyi_2010, mele_introduction_to_2024}, which
allows us to write the purity of a single-qubit state $\rho$ as
\begin{align} \label{eq:swap_trick_1qubit}
     \Tr[\rho^2] =
     \Tr[ (\rho \otimes \rho) \Swap  ] \,.
\end{align}
$\Swap= \ket{0,0}\!\bra{0,0} + \ket{0,1}\!\bra{1,0} + \ket{1,0}\!\bra{0,1} + \ket{1,1}\!\bra{1,1}$
is the operator that decomposes the two-qubit Hilbert space
$\CC^2\otimes \CC^2$ into a direct sum of its symmetric and anti\-symmetric subspace~\footnote{Note that Eq.~\eqref{eq:swap_trick} can be visually understood in the calculus of tensor network rewirings~\cite{bridgeman_hand_waving_2017, mele_introduction_to_2024}.
This viewpoint leads to multi-copy generalizations of 
Bell sampling to estimate higher-order Rényi entropies~\cite{subasi_entanglement_spectroscopy_2019}.}.
The former is three-dimensional and spanned by the \emph{triplet} state vectors
\begin{align} \label{eq:triplet1}
    \ket{\Phi^+} &= \tfrac{1}{\sqrt{2}}(\ket{0,0} + \ket{1,1})\,, \\ 
\label{eq:triplet2}
\ket{\Psi^+} &= \tfrac{1}{\sqrt{2}}(\ket{0,1}+ \ket{1,0})\,,
\\
\label{eq:triplet3}
\text{and} \hspace{5mm} 
    \ket{\Phi^-} &= \tfrac{1}{\sqrt{2}}(\ket{0,0} - \ket{1,1})\,,
\end{align}
while the latter is one-dimensional and spanned by the \emph{singlet} state vector
\begin{align} \label{eq:singlet}
   \phantom{\text{and} \hspace{5mm} }
    \ket{\Psi^- } = \tfrac{1}{\sqrt{2}} (\ket{0,1} - \ket{1,0}) \,.
\end{align}
{\add The four vectors in Eqs.~\eqref{eq:triplet1}--\eqref{eq:singlet} form the \emph{Bell basis}.}
In practice, 
{\add measurements in the Bell basis} 
are usually realized by applying a two-qubit $\Cnot$ gate and reading out control and target qubits in the  $X$- and $Z$-basis, respectively, see Fig.~\ref{fig:two_copy_bell_measurement}.
For our purposes, it suffices to differentiate triplets from singlets (without fine-graining triplets), which amounts to estimating the probability 
$\bra{\Psi^-} (\rho \otimes \rho) \ket{\Psi^-}$ of projecting $\rho\otimes \rho $ into a singlet.
Indeed, inserting the expression $\Swap = \mathbbm 1 - 2 \ket{\Psi^-}\! \bra{\Psi^-} $
into Eq.~\eqref{eq:swap_trick_1qubit}, we can write
\begin{align}
    \Tr[\rho^2] = 1 - 2 \bra{\Psi^-} (\rho \otimes \rho) \ket{\Psi^-} \,.
\end{align}
In the more general case where we have access to two copies of an $n$-qubit state $\rho$, the situation is similar:
perform a Bell measurement on qubit pair $(s, n+s)$ 
for all $s \in \{1,\ldots, n\}$
and denote the outcome by ${\alpha_s} \in\{ {\Phi^\pm}, {\Psi^\pm } \}$.
Then, the overall outcome of the two-copy Bell sampling experiment is 
\begin{align}
    \label{eq:bell_pairs}
    \ket{\boldalpha} = \bigotimes _{s=1}^n \ket{\alpha_s }_{s, n+s}.
\end{align}
{\add
Note that such Bell samples coincidentally have been dubbed \emph{``circuit shadows''} in Ref.~\cite{hangleiter_bell_sampling_2024},
even though the relationship to Eq.~\eqref{def:shadow_coset} had not yet been discovered.}
We denote the number of singlets in $\ket{\boldalpha}$ by 
\begin{align}
    \sing(\boldalpha) 
= \left\vert\left\{ s \in \{1,\ldots, n\} \ \big \vert \ \ket{\alpha_s} = \ket{\Psi^-} \right\} \right\vert \,.
\end{align}
The purity of $\rho$ can then be estimated as the difference between probabilities of finding an even and odd number of singlets, respectively~\cite{bluvstein_a_quantum_2022}.
That is,
\begin{align} \label{eq:purity_from_bell_samples}
    \Tr[\rho ^2] = 1 - 2 
    \hspace{-8mm}
\sum_{\substack{\boldalpha \in \{\Phi^\pm, \Psi^\pm \}^n \\ \sing(\boldalpha) = 1 
\text { (mod 2)} }}
    \hspace{-6mm}
    \bra{\boldalpha} (\rho \otimes \rho) \ket{\boldalpha}\,.
\end{align}

{\add Learning} 
purities through two-copy Bell measurements is remarkably efficient:
only ${O}(\epsilon^{-2})$ samples are required for estimating $ \Tr[\rho ^2] $ up to additive error $\epsilon>0$.
For a similar result about QWEs,  see Thm.~\ref{thm:sample_complexity_bounds} below.
On the other hand, 
all \emph{single-copy} protocols for 
{\add learning} purities unavoidably suffer from an exponential (in $n$) sample complexity~\cite{chen_exponential_separations_2022}.
This example clearly illustrates the 
{\add learning-theoretical advantage} 
of the paradigm of so-called \emph{quantum memories} where part of the postprocessing (after preparing copies of $\rho$) 
is performed quantumly
\emph{before} the measurement~\cite{chen_exponential_separations_2022, huang_quantum_advantage_2022, aharonov_quantum_algorithmic_2022, chen_a_hierarchy_2021, king_triply_efficient_2025}.
Here we extend the power of quantum memories into the realm of QWEs.

{\add \vspace{5mm}

\subsection{How to learn sector length distributions \\ aka Shor--Laflamme QWEs}
\label{sec:measure_sld}
}

Consider the $2n$-qubit observable
\begin{align} \label{eq:sld_observable}
    \twocopyobservable_i 
    =  \frac{1}{2^n} \sum_{{\substack{P \in \{I,X,Y,Z\}^{\otimes n} \\ \wt(P) = i}}} P \otimes P \, .
\end{align}
Since $\Tr[(\rho\otimes \rho)(P \otimes P)] = \Tr[\rho P]^2$, we can reinterpret the entries in the 
SLD $\mathbf{a}[\rho]=(a_i[\rho])_{i=0}^n$ from Eq.~\eqref{def:sld} as two-copy expectation values 
\begin{align} \label{eq:sld_expectation_value}
a_i [\rho]  = \Tr[(\rho\otimes \rho)
\twocopyobservable_i 
] \,.
\end{align}
Inserting two identity operators,
$\mathbbm 1 = \sum_{\boldalpha} \ket{\boldalpha
}\! \bra{\boldalpha}$, in the basis of Bell pairs $\ket{\boldalpha}$ from Eq.~\eqref{eq:bell_pairs} yields
\begin{align} \label{eq:sld_bell_expansion}
          a_i [\rho] =  \sum_{\boldalpha, \boldalpha' \in \{ \Phi^\pm, \Psi^\pm \}^n}   \bra{\boldalpha} \twocopyobservable_i  \ket{\boldalpha'} \bra{\boldalpha'} (\rho\otimes\rho) \ket{\boldalpha} \, .
\end{align}
To compute the matrix entries $\bra{\boldalpha} \twocopyobservable_i \ket{\boldalpha'}$, we exploit that the $\Swap$ operator has the Pauli decomposition 
\begin{align}
    \Swap = \frac{1}{2} \left( I\otimes I \hspace{.5mm}+\hspace{.5mm} X\otimes X \hspace{.5mm}+\hspace{.5mm} Y\otimes Y \hspace{.5mm}+\hspace{.5mm} Z\otimes Z   \right) \, .
\end{align}
After a little algebra, we find
\begin{align} \label{eq:sld_observable_via_swaps}
    \twocopyobservable_i = \frac{1}{2^{n-i}}\sum_ {\substack{S  \subset \{1,\ldots, n\} \\ \vert S \vert = i}} \prod_{s \in S} \left( \Swap_{s, n+s} - \frac{1}{2} \right) \, .
\end{align}
Since all expressions of the form 
$\bra{\boldalpha} \Swap_{s, n+s} \ket{\boldalpha'} $ 
and $\bra{\boldalpha} -\frac{1}{2} \ket{\boldalpha'} $
with $\boldalpha\neq \boldalpha'$ vanish,
the operator $\twocopyobservable_i$ is diagonal in the basis of Bell pairs.
This already shows that it must be possible to infer the expectation value $a_i[\rho]= \Tr[ (\rho \otimes \rho) \twocopyobservable_i]$ via two-copy Bell sampling~\cite{hangleiter_bell_sampling_2024}.
To see how this can be achieved 
{\add with efficient computational complexity,
let us}
make use of elementary combinatorics: 
since we have $i$ factors in each of the $\binom{n}{i}$ terms in Eq.~\eqref{eq:sld_observable_via_swaps}, there is the possibility that exactly $l\in\{1,\ldots, i\}$ of the $\Swap$ gates hit some of the $\sing(\boldalpha)$ singlet pairs while the remaining ones hit exactly $i-l$ triplets.
Hereby, \emph{`hit'} means that a singlet and a $\Swap$ gate both have non-trivial support on a fixed subsystem $S$.
Because   
\begin{align} \label{eq:bell_symmetries}
\bra{\boldalpha}   \Swap_{s, n+s} -\tfrac{1}{2}  \ket{\boldalpha} 
= 
\begin{cases}
-\tfrac{3}{2}  & \text{if } \ket{\alpha_s} = \ket{ \Psi ^- }  \\
+\tfrac{1}{2}  & \text{ otherwise}, \\ 
\end{cases}
\end{align}
each such term contributes
\begin{align}
    \bra{\boldalpha}  \prod_{s \in S} \left( \Swap_{s, n+s} - \tfrac{1}{2} \right)  \ket{\boldalpha} = \left(\frac{-3}{2}\right)^l \left(\frac{+1}{2}\right)^{i-l} 
\end{align}
and since there are $\binom{\sing(\boldalpha)}{l} \binom{n-\sing(\boldalpha)}{i-l}$ possibilities for 
this to happen,
we find the eigenvalue
\begin{align}
\bra{\boldalpha} \twocopyobservable_i
\ket{\boldalpha} 
&=
\sum_{l=0}^i 
\binom{\sing(\boldalpha)}{l}
\binom{n-\sing(\boldalpha)}{i-l}
\frac{\left(-3\right)^l}{2^n} \,.
\label{eq:eigenvalues_sld}
\end{align}

{\add Therefore,} we can define an unbiased single-shot estimator
$\hat {\mathbf{a}}=(\hat{a}_i)_{i=0}^n$ for the SLD as follows:
perform a two-copy Bell measurement, 
record the outcome ${\boldalpha}\in \{\Phi^\pm, \Psi^\pm \}^n$,
and declare $\hat{a}_i = \bra{\boldalpha} \twocopyobservable_i \ket{\boldalpha}  $ 
for every $i\in\{0,\ldots, n\}$.
Further applying the transforms from Fig.~\ref{fig:macwilliams},
we obtain single-shot estimators for all six vectors of QWEs, 
e.g., $\hat{\mathbf{b}} = M \hat{\mathbf{a}} $ and $\hat{\mathbf{a}}' = T' \hat{\mathbf{a}}$ for the dual SLD $\mathbf{b}[\rho]$ and the APD $\mathbf{a'}[\rho]$, respectively.
Since $\bra{\boldalpha} \twocopyobservable_i \ket{\boldalpha}$ only depends on the number of singlets in $\ket{\boldalpha}$,
there are  $6(n+1)$ elements of $\RR^{n+1}$ that the vector-valued estimators ${\hat{\mathbf{a}}, \hat{\mathbf{b}}, \hat{\mathbf{a}}', \hat{\mathbf{b}}', \hat{\tilde{\mathbf{a}}}}$, and $\smash{\hat{\tilde{\mathbf{b}}}}$ may assume.
To support future experiments,
we implement open-access functionalities~\footnote{Our package is available under \url{https://github.com/Mc-Zen/qsalto} and via \texttt{pip install qsalto}.}.
{\add The computational complexity is $O(n^2)$, see App.~\ref{app:trafo_recursion}. }
Sample complexity, on the other hand, is an entirely different issue.
Since $\bra{\boldalpha} \twocopyobservable_i \ket{\boldalpha}$ takes exponentially large positive and negative values,
estimating SLDs by averaging over many realizations of $\hat{\mathbf{a}}$ 
{\add may suffer from cancellation effects (also see Sec.~\ref{sec:sample_complexity})}.
This challenge seems to be inherent to SLDs as all known protocols for measuring them have an exponential {\add worst-case} sample complexity,
{\add see App.~\ref{app:single_setting_benefits}.}


{\add
\vspace{5mm}

\subsection{How to learn averaged purity distributions \\ aka Rains' unitary QWEs}
\label{sec:measure_apd}
}

In the previous subsection, 
we have identified a single-shot estimator for the APD 
$\mathbf{a'}[\rho]$ of an $n$-qubit state $\rho$ that can be efficiently (but not yet accurately) computed.
This in itself is already surprising, as averaging all subsystem purities of a given size naively requires evaluating a sum over exponentially many terms.
As we show next, 
for APDs, the accuracy issue can be resolved, thus paving the way for 
{\add learning} 
them in large-scale experiments.
The idea is to directly compute the single-shot estimator $\hat{\mathbf{a}}'$ from the two-copy Bell measurement outcome  ${\boldalpha}\in \{\Phi^\pm, \Psi^\pm \}^n$ instead of detouring through the SLD, where errors could get amplified.
To this end, we introduce a second $2n$-qubit observable 
\begin{align}\label{eq:apd_observable}
    \twocopyobservable'_i = \frac{1}{\binom{n}{i}} \sum_{{\substack{S \subset \{1,\ldots, n\} \\ \vert S \vert   = i}}} \prod_{s \in S} \Swap_ {s, n+s}
\end{align}
for all $i\in\{0,\ldots, n\}$.
Leveraging the $\Swap$ trick, we find
\begin{align}\label{eq:swap_trick_subsystem}
    \Tr[\rho_S^2] = \Tr\left[(\rho\otimes  \rho) \prod_{s \in S} \Swap_{s,n+s} \right]
\end{align}
for every $S\subset \{1,\ldots, n\}$, and averaging over all such subsystems of a given size $|S|=i$ allows us to write the APD of $\rho$  [Eq.~\eqref{def:apd}] as
\begin{align}
    a'_i[\rho] = \Tr\left [(\rho\otimes \rho) \twocopyobservable'_i  \right] \,.
\end{align}
As in the case of SLDs [Eq.~\eqref{eq:eigenvalues_sld}], $\twocopyobservable'_i$ is diagonal  in the basis of Bell pairs 
and its eigenvalues are given by
\begin{align}
\label{eq:eigenvalues_apd}
\bra{\boldalpha} \twocopyobservable'_i
\ket{\boldalpha} 
&=
\sum_{l=0}^i
\binom{\sing(\boldalpha)}{l}
\binom{n-\sing(\boldalpha)}{i-l}
\frac{\left(-1\right)^l} {\binom{n}{i}} \,.
\end{align}
Hence, we can repeat the previous strategy and compute the already-defined single-shot estimator $\hat{\mathbf{a}}'=T' \hat{\mathbf{a}}$  as follows:
perform a Bell measurement on $\rho\otimes \rho$ and compute $\hat{a}'_i = \bra{\boldalpha} \twocopyobservable' _i \ket{\boldalpha}$ from the  ${\boldalpha}\in \{\Phi^\pm, \Psi^\pm \}^n$.
By Vandermonde's identity~\footnote{Vandermonde's identity, $\binom{n}{i}= \sum_l 
\binom{n-j}{i-l}\binom{j}{l}$, is a well-known result from combinatorics.},
Eq.~\eqref{eq:eigenvalues_apd} only takes 
values in the interval $[-1,1]$.
Thus,
statistical errors are no longer unfavorably amplified.
We defer formal proofs on efficiency and robustness of 
{\add learning} APDs to Sec.~\ref{sec:complexity_and_robustness}. 
In particular, we will show in Thm.~\ref{thm:sample_complexity_bounds} that the sample complexity of measuring any 
{\add given} entry of the APD is independent of $n$.

{\add \vspace{5mm} }

\subsection{Rains' shadow QWEs are triplet probabilities}
\label{sec:rains_shadow_enumerators_interpretation}

To complete the analogy between SLDs, APDs, and Rains' shadow QWEs, 
we should look for a third observable, $\tilde{\twocopyobservable}_i$.
The solution is surprisingly simple and constitutes our main theoretical discovery.

\bigskip
\refstepcounter{mycounter}
\noindent
\textbf{Theorem~\arabic{mycounter}\label{thm:main}}  (Physical interpretation of Rains' shadow QWEs)
\emph{Let $\rho$ be an $n$-qubit state and $\mathbf{\tilde a}[\rho] = (\tilde{a}_i[\rho])_{i=0}^n $ its vector of Rains' shadow QWEs from Eq.~\eqref{def:rains_qwe}.
For all $i\in\{0,\ldots, n\}$, we have }
\begin{align} 
    \tilde{a}_i[\rho] = \Tr\left[ (\rho \otimes \rho)  \tilde{\twocopyobservable}_i \right] \, ,
\end{align}
\emph{where we denote the projector onto the $i$-triplet subspace 
by }
\begin{align} \label{eq:tpd_observable}
    \tilde{\twocopyobservable}_i  = \sum_{\substack{\boldalpha \in \{\Phi^\pm, \Psi^\pm\}^{n}\\ \sing(\boldalpha)=n-i}}  \ket{\boldalpha}\! \bra{\boldalpha}.
\end{align}
\emph{In other words, the $i$-th Rains' shadow QWE of $\rho$ carries the interpretation of the probability for observing a fixed number of $i$ triplets in a Bell sampling experiment on $\rho\otimes \rho$.}

\smallskip

\emph{Proof:} 
{\add Let} $\mathbf{t}[\rho ] = (t_j[\rho ])_{j=0}^n $,
{\add where}
\begin{align}
    t_j[\rho] = \sum_{\substack{\boldalpha \in \{\Phi^\pm, \Psi^\pm\}^{n}\\ \sing(\boldalpha)=n-j}}  \bra{\boldalpha} (\rho\otimes \rho) \ket{\boldalpha}\, .
\end{align}
We have to show $\mathbf{\tilde a}[\rho] = \mathbf{t}[\rho]$,
{\add which}
is equivalent to the claim
$\mathbf{a}[\rho] = \tilde T ^{-1} \mathbf{t}[\rho] $.
{\add 
By Eq.~\eqref{eq:trafo_tpd_to_sld_via_signed_mwt}, 
we can write}
\begin{align} \label{eq:trafo_tpd_to_sld}
     \tilde T ^{-1}_{i,j} = \frac{1}{2^n}\sum_{l=0}^n \binom{n-j}{i-l} \binom{j}{l} (-3)^{i-l}\,.
\end{align}
{\add Let us}
express $\mathbf{a}[\rho]$ in terms of 
triplet probabilities.
The term $\bra{\boldalpha} \twocopyobservable_i \ket{\boldalpha}$ in Eq.~\eqref{eq:eigenvalues_sld} only depends on 
  $j = n-\sing(\boldalpha)$, the number of triplets in $\boldalpha$. 
This means that we can split up the sum and separate 
terms.
Expanding Eq.~\eqref{eq:sld_bell_expansion} yields
\begin{align}
   \begin{split}
   a_i[\rho] &= \sum_{j=0}^n\sum_{\substack{\boldalpha \in \{ \Phi^\pm, \Psi^\pm \}^n \\ \sing(\boldalpha)=n-j}}   \bra{\boldalpha} \twocopyobservable_i  \ket{\boldalpha} \!\bra{\boldalpha} (\rho\otimes\rho) \ket{\boldalpha} \\
   &= \sum_{j,l = 0}^n  \binom{n-j}{l} \binom{j}{i-l} \frac{(-3)^l}{2^n} t_j[\rho] \, .
   \end{split}
    \label{eq:main_thrm_step_2}
\end{align}
Since the summand in Eq.~\eqref{eq:main_thrm_step_2} is zero for $l>i$, we can exchange the order and write instead $l\mapsto i-l$, obtaining
\begin{align}
    a_i[\rho] &= \frac{1}{2^n} \sum_{j,l = 0}^n   \binom{n-j}{i-l} \binom{j}{l} (-3)^{i-l} t_j[\rho] \, .
\end{align}
{\add This shows} $\mathbf{a}[\rho] = \tilde T ^{-1}\mathbf{t}[\rho]$, 
which completes the proof. \hfill $\square$

\bigskip

{\add Having} established a novel connection between the physics of two-copy Bell sampling experiments and the powerful machinery of QWEs, 
we are 
{\add now}
in the   position to transfer valuable ideas from one paradigm to the other, see App.~\ref{app:moments_and_bounds_of_tpds}.

\bigskip
\refstepcounter{mycounter}
\noindent
\textbf{Corollary~\arabic{mycounter}\label{cor:shadow_inequalities}} (Physical origin of Rains' shadow inequalities)
\emph{For every $n$-qubit state  $\rho$ and all $i \in\{0,\ldots, n\}$, we have} 
\begin{align}
    \tilde a_i[\rho] \ge 0\,,
\end{align}
\emph{and the bound $d\le \tfrac{n}{3}+2$ for arbitrary  $\llbracket n,k,d\rrbracket $ QECCs follows from Thm.~15 in Ref.~\cite{rains_quantum_shadow_1999}.}

\smallskip
\emph{Proof:} Probabilities cannot be negative. \hfill $\square$

\bigskip
\refstepcounter{mycounter}
\noindent
\textbf{Corollary~\arabic{mycounter}\label{cor:triplet_mean_bound}}  (Lower bound on the average number of triplets)
\emph{Let $\rho$ be an $n$-qubit state and $(   \tilde a_x[\rho])_{x=0}^n$ its shadow QWEs. The average number of triplets obtained in a two-copy Bell sampling experiment can be bounded as} 
\begin{align}
 \frac{3n}{ 4}   \     \le   \  
 \sum_{x=0}^n      \tilde a_x[\rho] x
 \     \le \   n\,,
\end{align}
\emph{and these bounds are tight.}

\smallskip
\emph{Proof:} See App.~\ref{app:attainable_tpds}. \hfill $\square$

\bigskip


In light of Sec.~\ref{sec:qwe_macwilliams} and Thm.~\ref{thm:main}, it now becomes clear how one can learn QWEs of QECCs in experiments: first,
prepare two copies of the maximally mixed logical state $\rho_\text{QECC}$ [Eq.~\eqref{eq:rho_qecc}] and perform pairwise Bell measurements as in Fig.~\ref{fig:two_copy_bell_measurement} above.
Then, either (i) estimate the QWEs directly from the Bell samples using the single-shot estimators from Secs.~\ref{sec:measure_sld} and~\ref{sec:measure_apd} or, equivalently, (ii) estimate Rains' shadow enumerators $\tilde{\mathbf{a}}[\rho_\text{QECC}]$ and then apply the transforms from
{\add App.~\ref{app:notation}}.

\subsection{QWEs and entanglement}
\label{sec:concurrence_and_ntangle}

The idea {\add of quantifying} entanglement via QWEs is not new.
Already in Ref.~\cite{scott_multipartite_entanglement_2004}, it was shown that from SLDs it is possible to compute generalized Meyer-Wallach entanglement measures (which are conceptually {\add equivalent to} APDs).
In this section, 
we discuss important entanglement measures that {\add have been} defined in terms of triplet probabilities.
{\add By applying Thm.~\ref{thm:main}, 
it is now possible to efficiently compute these measures
whenever the SLD or the APD is analytically available.
Later, in Sec.~\ref{sec:dicke_entanglement_stability},
we will showcase theoretical applications of the connections established in this section.
}

A direct consequence of Thm.~\ref{thm:main} is that Rains' shadow QWEs [Eq.~\eqref{def:rains_qwe}] of an $n$-qubit state $\rho$ can be written as
  \begin{align} \label{eq:tpd}
    \tilde a_i[\rho] = \sum_{\substack{ \boldalpha \in \{\Phi^\pm, \Psi^\pm\}^n \\ \sing(\boldalpha)=n-i}}
    \bra{\boldalpha} (\rho\otimes\rho) \ket{\boldalpha} \,
    .
\end{align}
Hence, it makes sense to refer to $\mathbf{\tilde a}[\rho] = (\tilde a_i[\rho])_{i=0}^n$ as the \emph{triplet probability distribution} (TPD) of $\rho$.
By applying the linear maps from Fig.~\ref{fig:macwilliams},
it is possible to extract useful properties from TPDs.
For example, the final row of the vector equation  $ \mathbf{a'}[\rho] = \tilde T '^{-1} \mathbf{\tilde a}[\rho]$ reads 
\begin{align} \label{eq:purity_from_tpd}
    \Tr[\rho ^2] = a'_n[\rho] = \sum_{j=0}^n (-1)^{j} \tilde a_{n-j}[\rho] \,,
\end{align}
which is a condensed version of Eq.~\eqref{eq:purity_from_bell_samples}.
Conversely, the final row of 
$\mathbf{\tilde a}[\rho] = \tilde T ' \mathbf{a'}[\rho]$ is equivalent to
\begin{align} 
\label{eq:zero_singlet_from_apd}
    \tilde a_n[\rho] 
    &= {\frac{1}{2^n} \sum_{i=0}^n \binom{n}{i} a'_i[\rho]} 
    = \frac{1}{2^n} \sum_{S\subset \{1,\ldots, n\}} \Tr[\rho_S ^2]\,  .
\end{align}
In other words, 
the probability for obtaining $n$ triplets (zero singlets) is equal to the average over all subsystem purities.
Thus,  $\tilde a_n[\rho] < 1$ implies that at least one of the marginals of $\rho$ (including $\rho$ itself) cannot be pure.
Conversely, every pure state $\Psi$ with
\begin{align}\label{eq:concurrence_criterion}
    1- \tilde a_n[\Psi] > 0
\end{align}
must be entangled.
Therefore, the quantity
\begin{align} \label{eq:concurrence}
    C[\Psi] =  1- \tilde a_n[\Psi]
\end{align}
is of central importance.

A few historical comments are in order.
Originally, $C[\Psi]$ was studied 
under the name \emph{concurrence} in the modified form $2\sqrt{C[\Psi]}$~\cite{carvalho_decoherence_and_2004, mintert_concurrence_of_2005, mintert_observable_entanglement_2007, aolita_scalable_method_2008}.
Recently, concurrences 
have regained some attention~\cite{beckey_computable_and_2021, cullen_calculating_concentratable_2022,
beckey_multipartite_entanglement_2023,
schatzki_hierarchy_of_2024} in the form of Eq.~\eqref{eq:concurrence}.
We also prefer this version of the concurrence as it avoids superfluous notation and, thus, can be read off directly from the TPD.
For these reasons, in this work, the term concurrence will henceforth always refer to Eq.~\eqref{eq:concurrence}.
 

Via convex roof extension, it is possible to generalize this concept to the case where $\rho$  is an arbitrary mixed state~\cite{aolita_scalable_method_2008}.
While the resulting concurrence $C[\rho]$ is in general difficult to compute, 
it can always be lower bounded via~\cite{aolita_scalable_method_2008, beckey_multipartite_entanglement_2023}
\begin{align} \label{eq:lower_bound_concurrence}
C[\rho]  \ge  \tfrac{1}{2^n} +  \left(1-\tfrac{1}{2^n}\right) \Tr[\rho^2]  -\tilde a_n[\rho] \,.  
\end{align}
In combination with the QWE machinery, this bound is extremely powerful.
Indeed, provided we know the SLD $\mathbf{a}[\Psi]$ of a pure state $\Psi$, Eq.~\eqref{eq:sector_length_decay} yields the SLD of the locally-depolarized state $\rho = \mathcal{E}^{\otimes n}_p [\Psi]$.
This, in turn, determines $\Tr[\rho^2]=\sum_i a_i[\rho]$
as well as the all-triplet (zero-singlet) probability 
\begin{align} \label{eq:all_triplet_via_sld}
    \tilde a_n [\rho] = \frac{1}{2^n} \sum_{i=0}^n 3^{n-i}\,  a_i[\rho] \,,
\end{align}
which is all we need to evaluate the bound in Ineq.~\eqref{eq:lower_bound_concurrence}.

Note that Eq.~\eqref{eq:all_triplet_via_sld} is the final entry of $\mathbf{\tilde a}[\rho] = \tilde T  \mathbf{a}[\rho]$.
The first row of the same equation allows us to compute the zero-triplet (all-singlet) probability
\begin{align} \label{eq:zero_triplet_from_sld}
    \tilde a_0 [\rho] = \frac{1}{2^n} \sum_{i=0}^n (-1)^i a_i[\rho]
\end{align}
from the SLD of $\rho$.
In terms of subsystem purities, it can be expressed as
\begin{align} \label{eq:zero_triplet_from_apd}
     \tilde a_0 [\rho] 
     &
     = \frac{1}{2^n} \sum_{i=0}^n   (-1)^i  \binom{n}{i}  a'_i[\rho] 
     \\
     &= \frac{1}{2^n} \sum_{ S \subset \{1,\ldots, n\}} (-1)^{\vert S\vert }\Tr[\rho_S ^2] \,.
     \nonumber
\end{align}
{\add Yet another} expression is obtained by inserting $i=0$ into the original definition of Rains' shadow QWEs  [Eq.~\eqref{def:rains_qwe}],
\begin{align} \label{eq:ntangle}
\add    \tilde a_0[\rho] = \tfrac{1}{2^n} \Tr[\rho \tilde \rho] \,.
\end{align}
{\add For a pure state $\Psi=\ket\psi\!\bra\psi$, the quantity}  $\text{Tr}[ \Psi  \tilde\Psi ] = 2^n \tilde a_0[\Psi]$ is an entanglement measure
{\add that is}
known as 
{\add its}
\emph{$n$-tangle}~\cite{wong_potential_multipartite_2001}.

To summarize, the two endpoints of the TPD $\mathbf{\tilde a}[\Psi]$ of a pure state $\Psi$ directly correspond to two important entanglement measures: 
concurrence $C[\Psi]$ and $n$-tangle $\text{Tr}[ \Psi  \tilde\Psi ]$.
In particular, 
both of them follow from the SLD  $\mathbf{a}[\Psi]$.
While the concurrence lends itself to an entanglement detection protocol with an efficient sample complexity~\cite{aolita_scalable_method_2008},
the opposite is true about the $n$-tangle.
Being an overlap between two pure states, $\text{Tr}[ \Psi  \tilde\Psi ]$ is upper bounded by $1$.
Therefore, the zero-triplet probability $\tilde a_0[\Psi]$ is exponentially small by Eq.~\eqref{eq:ntangle}, 
which implies that exponentially many samples are required to resolve $\text{Tr}[ \Psi  \tilde\Psi ]$ up to constant additive error~\footnote{This redresses the claim made in Prop.~2 of Ref.~\cite{beckey_multipartite_entanglement_2023}. 
The chief problem is that the estimator for the $n$-tangle assumes the value $2^n$ in the rare event that none of the Bell measurements results in a triplet.
In all other events, the same estimator assumes the value 0.
Therefore, the denominator in the exponent of Höffding's inequality blows up exponentially, which renders measuring $\text{Tr}[\Psi \tilde \Psi]$ sample-inefficient
{\add for the protocol discussed in Ref.~\cite{beckey_multipartite_entanglement_2023}.
However, the situation is fully remedied if the access model is $\Psi\otimes \Psitranspose$ rather than $\Psi\otimes \Psi$, since in this case the $n$-tangle can be learned by applying the $\Swap${} test to $\Psi\otimes \tilde \Psi = \Psi\otimes Y^{\otimes n} \Psitranspose Y^{\otimes n}$}
.
}.
Nevertheless, concurrence and $n$-tangle are useful notions as they quantify \emph{how much} a given state is entangled, 
thus complementing the 
{\add $i$-body}
sector length criterion [Eq.~\eqref{eq:nbody_criterion}] and the purity criterion [Eq.~\eqref{eq:purity_criterion}],
which  certify \emph{that} a given state is entangled.
{\add Finally, note that, in the $\Psi\otimes \tilde{\Psi}$ access model,
$\text{Tr}[\Psi \tilde{\Psi}]$ is learnable  via Bell sampling.}

\section{Experimental demonstrations}
\label{sec:experiment}

We have now reviewed the theoretical machinery of QWEs in Sec.~\ref{sec:qwe_macwilliams} and established a strong connection to Bell sampling in Sec.~\ref{sec:bell_sampling} by identifying collections of $2n$-qubit observables (and showing they are diagonal in the transversal Bell basis) for measuring QWEs.
Having available these definitions is what will enable us in Sec.~\ref{sec:complexity_and_robustness} to derive rigorous performance guarantees.
Before that, however, we want to put the theory built so far to the ultimate test and implement experiments on a state-of-the-art quantum processor.
{\add 
Besides scientific curiosity, the main motivation for the presented experiments is to demonstrate that the proposed protocol does not rely on idealized assumptions. In particular, we aim to verify that the protocol remains viable in the presence of real-world noise sources and experimental imperfections.
}

Two-copy Bell sampling  has the following experimental requirements:
(i)
since we need to prepare two approximate (see Thm.~\ref{thm:robustness} below) copies of   
the same state in two quantum registers, 
it is beneficial to work with an experimental platform for which physical parameters exhibit little to no variation between the qubits;
(ii)
considering that the two registers eventually need to be read out via pairwise Bell measurements,
we also require a certain degree of connectivity.
For example, 
if we wanted $n$ linearly connected qubits at the preparation stage,
we would overall require at least a $2\times n$ rectangular connectivity.
Both requirements are met 
by all-to-all connected trapped-ion quantum computers.

In this section, we report experimental demonstrations of single-setting QWE measurements carried out on our trapped-ion quantum processor, as described in Ref.~\cite{pogorelov_compact_ion_2021}.
We trap sixteen $^{40}\text{Ca}^+$ ions in a macroscopic linear Paul trap,
where the electronic states are controlled via laser pulses.
Each ion encodes one qubit into the electronic Zeeman levels
$\ket{0} = \ket{4^2\text{S}_{1/2}, m_J=
-1/2}$ and
$\ket{1} = \ket{3^2\text{D}_{5/2}, m_J=
-1/2}$, 
which are connected via an optical quadrupole transition at a wavelength of $\SI{729}{nm}$. 
Coulomb interaction between the ions gives rise to collective motional modes, which are used to mediate entangling operations between any desired pair of qubits. Single ion addressing with $\SI{729}{nm}$ light allows for arbitrary single- and two-qubit operations.
The native gate set consists of arbitrary physical rotations around the $x$- and $y$-axis of the Bloch sphere and virtual rotations \cite{mckay_efficient_z_2017} around the $z$-axis for single-qubit operations, as well as the maximally entangling two-qubit M\o{}lmer-S\o{}rensen gate
$\exp ( \text{i}\frac{\pi}{4}X\otimes X)$~\cite{sorensen_quantum_computation_1999}.
The latter is equivalent to a $\Cnot$ gate  up to local rotations~\cite{maslov_basic_circuit_2017}. 

We begin by presenting experimental results on a variety of six-qubit states in  Sec.~\ref{sec:experiment_states}, 
before reporting a proof-of-principle demonstration of the direct measurement of QWEs of QECCs in Sec.~\ref{sec:experiment_qeccs}.

\subsection{Direct measurement of QWEs for states}
\label{sec:experiment_states}

\begin{figure*}
    \centering
    \includegraphics{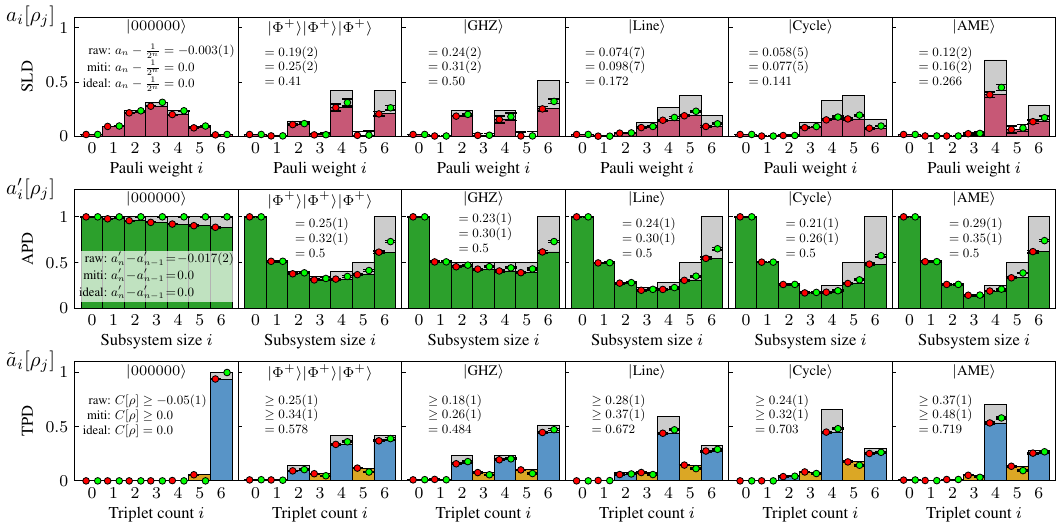}
    \caption{\justifying Experimentally measured QWEs for 
    six different states $\rho_j$ on $n=6$ trapped-ion qubits per copy.
    From 10,000 two-copy Bell samples per state, we simultaneously 
    estimate the SLD (top), 
    APD (center), and TPD (bottom).
    Respectively, these carry the interpretation of contribution to the purity [Eq.~\eqref{eq:sld_and_purity}], the averaged {\add subsystem} purity [Eq.~\eqref{def:apd}], and the observed triplet probability estimates [Eq.~\eqref{eq:tpd}].
    In the TPD picture, we distinguish bins of even (blue) and odd (yellow) singlet numbers to highlight their positive and negative impact, respectively, on the global purity $\Tr[\rho_j^2]$.
    Bootstrapped error bars show 95\% confidence intervals.
    Insets show quantities whose positivity certifies entanglement by the $n$-body sector length criterion (top), purity criterion (center), and concurrence criterion (bottom),
    where ``raw'', ``miti'', and ``ideal'' refer to raw experimental (dark red dots),
    {\add 
    experimental results with error mitigation as described in App.~\ref{app:error_mitigation_qwes},} 
    and ideal theoretical values (gray background), respectively. 
    For all states (except for the first) and all criteria, entanglement is experimentally verified.
    }
    \label{fig:experiment_states}
\end{figure*}

For our first experiment, we separate the sixteen ions into 
three groups with different purposes. On each end of the linear trap,
we devote six ions to host one copy of a quantum state $\rho$ that is our approximate experimental realization of a pure target state $\Psi $ on $n=6$ qubits. The remaining four ions in the center of the chain serve as a buffer zone to suppress the effect of laser fields leaking from one copy to the other.
We prepare six different states  
that exhibit various features of entanglement,
and directly measure their {triplet probability distribution} (TPD) in the two-copy Bell sampling experiment from Fig.~\ref{fig:two_copy_bell_measurement}.
The measured TPDs (blue and yellow) are plotted in the bottom panel of Fig.~\ref{fig:experiment_states}.
Recall from Thm.~\ref{thm:main} that the TPD is equal to the distribution $\mathbf{\tilde a}[\rho ]$ of Rains' shadow QWEs.
Hence (recall Fig.~\ref{fig:macwilliams}), we can transform the measured TPDs into Shor--Laflamme QWEs $\mathbf{a}[\rho ] = \tilde T ^{-1} \mathbf{\tilde a}[\rho ]$ (pink) and into Rains' unitary QWEs $\mathbf{a'}[\rho ] = \tilde T '^{-1} \mathbf{\tilde a}[\rho ]$ (green),
which are plotted in the top and central panel of Fig.~\ref{fig:experiment_states}, respectively.
In addition to the raw experimental data (red dots), 
we also present in Fig.~\ref{fig:experiment_states} error-mitigated results (green dots),
{\add 
obtained through a novel heuristic strategy described in App.~\ref{app:error_mitigation_qwes}.
}

Let us {\add introduce} the six states whose measured QWEs are displayed in Fig.~\ref{fig:experiment_states}.
As a digital feature, we  additionally link the definitions below to \texttt{GraphStateVis}~\cite{miller_graphstatevis_interactive_2021}, where the corresponding states and their SLDs are visualized~\footnote{For clarity, we point out the two-qubit gate count $N_j$ in our circuit for preparing $\Psi_j$:
$N_1=0$, $N_2=3$, $N_3=5$, $N_4=5$, $N_5=6$, $N_6=7$.
Except for $\Psi_6$ (App.~\ref{app:ame_circuit}),
preparation circuits are the obvious ones.
Two-qubit M\o{}lmer-S\o{}rensen gates were applied sequentially.
}.
{\add
Our first target state,}
$\href{https://graphvis.uber.space/?graph=6_0000}
{\Psi_1 = \ket{0}\!\bra{0}^{\otimes 6}}$ 
is a pure product state without any entanglement.
Second, $\href{https://graphvis.uber.space/?graph=6_4021}{\Psi_2 = \ket{\Phi^+}\!\bra{\Phi^+}^{\otimes 3}}$
is the state of three maximally entangled two-qubit Bell pairs without entanglement between the pairs.
All other target states are \emph{genuinely multipartite entangled} (GME).
First,  
the projector  
$\href{https://graphvis.uber.space/?graph=6_7C00}{\Psi_3}$ onto
$\href{https://graphvis.uber.space/?graph=6_7C00}{\ket{\text{GHZ}} = \tfrac{1}{\sqrt{2}}(\ket{0}^{\otimes 6} + \ket{1}^{\otimes 6}) }$,
followed by 
$\href{https://graphvis.uber.space/?graph=6_4225}{\Psi_4}$ and
$\href{https://graphvis.uber.space/?graph=6_4625}{\Psi_5}$ 
which project onto the \emph{graph state}~\cite{hein_multiparticle_entanglement_2004} vectors
\begin{align} 
    \href{https://graphvis.uber.space/?graph=6_4225}{\ket{\text{Line}}} 
    &
    \href{https://graphvis.uber.space/?graph=6_4225}{= \CZ_{1,2}\CZ_{2,3}\CZ_{3,4}\CZ_{4,5}\CZ_{5,6} \ket{+}^{\otimes 6} }
    \text{ and } \\
    \href{https://graphvis.uber.space/?graph=6_4625}{\ket{\text{Cycle}}}  
    &
    \href{https://graphvis.uber.space/?graph=6_4625}{= \CZ_{1,2}\CZ_{2,3}\CZ_{3,4}\CZ_{4,5}\CZ_{5,6}\CZ_{6,1} \ket{+}^{\otimes 6}},
\end{align}
respectively, where $\ket{+} = \tfrac{1}{\sqrt{2}}(\ket{0}+\ket{1})$,
and $\CZ_{i,j}$  is the controlled-$Z$ gate between qubit $i$ and $j$.
Finally, we construct a new circuit (App.~\ref{app:ame_circuit}) requiring only seven $\Cnot$ gates 
for preparing an \emph{absolutely maximally entangled} (AME) six-qubit state $\href{https://graphvis.uber.space/?graph=6_4E6F}{\Psi_6 = \ket{\text{AME}}\!\bra{\text{AME}}}$~\cite{cabello_optimal_preparation_2011}.
%
%

All states from above are stabilizer states.
Therefore, their $i$-body sector length [Eq.~\eqref{def:sld}] not only is interpretable as a term contributing to the purity [Eq.~\eqref{eq:sld_and_purity}],
but also as the (normalized) number of weight-$i$ stabilizer operators~[Eq.~\eqref{eq:sld_stabilizer_qecc}].
For $\href{https://graphvis.uber.space/?graph=6_0000}{\Psi_1}$, whose stabilizer group is $\{Z^{r_1}\otimes\ldots\otimes Z^{r_n}\ | \ \mathbf{r}\in\FF_2^n\}$, we thus expect a symmetrical binomial distribution [Eq.~\eqref{eq:sld_pure_product}], 
which is in 
{\add excellent} 
agreement with raw experimental observations in Fig.~\ref{fig:experiment_states} (top left panel).
For the remaining five states, we find experimental SLDs (pink) revealing various expected entanglement features despite noticeable amounts of noise.
For all measured states, the $n$-body sector length criterion [Eq.~\eqref{eq:nbody_criterion}] certifies the presence of entanglement.
More precisely, the difference $a_6[\rho_j]-2^{-6}$ (SLD insets) 
is strictly positive. 
The theoretical maximum,
$a_6[\href{https://graphvis.uber.space/?graph=6_7C00}{\Psi_3}]-2^{-6} = 0.5$, 
is attained by the ideal GHZ state~\cite{eltschka_maximum_nbody_2020}.
Also in our experiment,
we observe that the GHZ state has the largest $n$-body sector length $a_6^\text{miti}[\rho_3] = 0.32(2)$ among all measured states. 
{\add 
Since this value exceeds $\tfrac{1}{4} + 2^{-n}\delta_{n,\text{odd}}$,
we can conclude that $\rho_3$ is not semiseparable~\cite[Eq.~(185)]{miller_small_quantum_2019}.}
We find the GHZ state is closely followed by the three Bell pairs for which $a_6^\text{miti}[\rho_2] = 0.26(2)$. 
{\add Concerning the other sector lengths, 
we can see that the $i$-body criterion certifies entanglement in $\rho_4$ and $\rho_5$ for $i=5$,
as well as in $\rho_2$ and $\rho_6$ for $i=4$. }

As $\href{https://graphvis.uber.space/?graph=6_4021}{\Psi_2}$ is clearly triseparable, care should be taken when relating $a_n[\rho]$ to multipartite entanglement~\footnote{For example, 
$\rho = \tfrac{1}{2}\ket{D^n_1}\!\bra{D^n_1}+\tfrac{1}{2}\ket{D^n_{n-1}}\!\bra{D^n_{n-1}}$ in the notation of Eq.~\eqref{eq:dicke_state_definition} is a state that---despite being GME---has a vanishing full-body sector length $a_n[\rho]$, assuming $n$ is odd~\cite{kaszlikowski_quantum_correlation_2008}.}.
A more holistic analysis of QWE features can help.
A pure state $\Psi$ is called \emph{$m$-uniform} iff $a_i[\Psi]=0$ for all $1\le i \le m$,
{\add see App.~\ref{app:ame_circuit}.}
Such is impossible if $\Psi$ has a pure marginal of size $m$ (which would be necessarily unentangled with its complement, e.g., 
$\href{https://graphvis.uber.space/?graph=6_4021}{\Psi_2}$ is not $2$-uniform) and negating this statement shows in which sense $m$-uniformity can be interpreted as an extremal form of multipartite quantum entanglement.
In the design of our experiment, we have selected states of increasing levels of multipartite entanglement as measured by $m$-uniformity: 
$\href{https://graphvis.uber.space/?graph=6_0000}{\Psi_1}$
is $0$-uniform, 
$\href{https://graphvis.uber.space/?graph=6_4021}{\Psi_2}$,
$\href{https://graphvis.uber.space/?graph=6_7C00}{\Psi_3}$,  
$\href{https://graphvis.uber.space/?graph=6_4225}{\Psi_4}$ are $1$-uniform,
$\href{https://graphvis.uber.space/?graph=6_4625}{\Psi_5}$ is  $2$-uniform, and 
$\href{https://graphvis.uber.space/?graph=6_66A7}{\Psi_6}$ is  $3$-uniform.
Our experimental results confirm that $a_i[\rho_j]$ almost vanishes for the predicted combinations of $i$ {\add and} $j$.
Therefore, we can largely rule out systematic errors that generate few-body Pauli correlations, which are expected to be absent.
Moreover, we generally observe $a_i[\rho_j] \le a_i[\Psi_j]$ in Fig.~\ref{fig:experiment_states}, also when $ a_i[\Psi_j]$ is non-zero.
This provides further diagnostic information about the error mechanisms in the reported experiment as coherent overrotations in the preparation circuit and 
{\add non-unital}
noise, e.g., the amplitude damping channel~\cite{nielsen_quantum_computation_2000}, could increase $a_i[\rho]$.
Since we do not see any such effects, we attribute the noise affecting our states to entropy-increasing errors primarily.

Let us segue to a discussion about averaged purity distributions (APDs), $\mathbf{a}'[\rho]$, [Eq.~\eqref{def:apd}],
by pointing out again that the global purity $\Tr[\rho_j^2]$ arises in all three pictures:
it is the sum  [Eq.~\eqref{eq:sld_and_purity}] of all sector lengths (pink),
the rightmost entry  [Eq.~\eqref{def:apd}] of the APD (green),
and the alternating sum [Eq.~\eqref{eq:purity_from_tpd}] of triplet probabilities (blue for positive, yellow for negative).
In the APD picture, 
the hallmark of entanglement [Eq.~\eqref{eq:purity_criterion}] is an increase of $a'_i[\rho_j]$ with $i$, which is clearly observed at the right of the APD for all entangled states.
The averaged purity $a'_i[\rho_1]$ of the unentangled product state,
{\add on the other hand,}
monotonically \emph{decreases} with the subsystem size $i$, as expected.
The purity criterion [Eq.~\eqref{eq:purity_criterion}]
can be quantified as well:
in the APD inset of Fig.~\ref{fig:experiment_states},
we show the measured values of the purity difference $a'_6[\rho_j]- a'_5[\rho_j]$.
As with $a_n[\rho]-2^{-n}$ 
{\add before}, 
here a positive value of the purity difference certifies entanglement.
In all cases (except for 
$\href{https://graphvis.uber.space/?graph=6_0000}{\Psi_1}$ of course),
we would ideally expect
$a'_6[\Psi_j]- a'_5[\Psi_j]=0.5$ (which is simultaneously the theoretical maximum), however, we observe (raw) values between $0.211(6)$ for  
$\href{https://graphvis.uber.space/?graph=6_4625}{\Psi_5}$
and 
$0.285(5)$ for 
$\href{https://graphvis.uber.space/?graph=6_66A7}{\Psi_6}$.
{\add This implies that} 
the purity criterion clearly detects entanglement in \mbox{$\rho_2$, \ldots, $\rho_6$} despite significant noise levels.
{\add Bell}
readout error mitigation further boosts these purity differences by about 22\%--32\% as shown in the APD inset of Fig.~\ref{fig:experiment_states}.
We can also see that---on average---the marginals of the AME state are (as expected) more mixed than for lesser-entangled states such as the GHZ state,  e.g., by looking at $a'_3[\rho_j]$.
{\add More generally, we find $a'_i[\rho_j]< a'_n[\rho_j]$ for all $i\in\{3,4,5\}$ (and all $j\neq1$), which implies the existence of at least one entangled bipartition of size $i$ vs. $n-i$ of every state $\rho_j$ under investigation.}
Note that measuring APDs (in contrast to SLDs) is always scalable, see Thm.~\ref{thm:robustness} and Thm.~\ref{thm:sample_complexity_bounds} below. 

Finally, consider the TPDs [Eq.~\eqref{eq:tpd}] that are shown in the bottom panel of Fig.~\ref{fig:experiment_states}.
Among the three  QWEs, TPDs can be most straightforwardly extracted from  Bell samples simply by counting triplets. 
By Thm.~\ref{thm:main}, TPDs are the same as Rain's shadow QWEs [Eq.~\eqref{def:rains_qwe}] and,
in accordance with Cor.~\ref{cor:triplet_mean_bound},
we observe in Fig.~\ref{fig:experiment_states} that all measured TPDs concentrate towards larger triplet numbers.
Overall, we observe fair experimental agreement with theoretical predictions 
(gray/yellow bars for values that are smaller/larger than expected).

Let us start our discussion of the TPDs with the all-triplet probability $\tilde{a}_n[\rho]$.
For pure product states, we would never find any singlets, i.e.,
$\tilde a_6[\href{https://graphvis.uber.space/?graph=6_0000}{\Psi_1}]=1$.
This prediction is closely matched by our experimental observation
$\tilde a_6^\text{raw}[\rho_1]=0.94(1)$.
Moreover, Fig.~\ref{fig:experiment_states} reveals that almost all of what is missing from $\tilde a_6^\text{raw}[\rho_1]$ has leaked into $\tilde a_5^\text{raw}[\rho_1]=0.06(1)$, 
where just one singlet was measured.
The difference between even-singlet (blue) and odd-singlet (yellow) probabilities [Eq.~\eqref{eq:purity_from_tpd}] equals the global purity $\Tr[\rho^2] = a'_n[\rho]$.
In particular, we find $a'_6{}^\text{raw}[\rho_1] = 0.89(1)$,
where ``raw''
refers to the fact that readout errors (mainly due to $\Cnot$s) are mathematically included as noise on the state $\rho_1$,
{\add see App.~\ref{app:error_mitigation_qwes} for details. }

For the other (entangled) states, we see that the all-triplet probability $\tilde a_6 ^\text{miti}[\rho_j]$ remains slightly below each respective value of $\tilde a_6{}[\Psi_j]$.
In general, any value of $\tilde a_n[\rho]$ below unity is intimately linked to a loss of purity of a state $\rho$, its marginals, or both [Eq.~\eqref{eq:zero_singlet_from_apd}].
Then, the probability missing in $\tilde a_n[\rho]$ must reappear in some of the $\tilde a_i[\rho]$'s, which correspond to either an even or odd number of singlets.
Whenever the number of singlets is even (and positive),
we can conclude the existence of mixed marginals [Eq.~\eqref{eq:zero_singlet_from_apd}] 
that cannot be traced back to losses of global purity [Eq.~\eqref{eq:purity_from_tpd}].
In other words, Bell samples with an even (non-zero) number of singlets (blue) are the hallmark of entanglement in the TPD picture.
On the other hand, measurement outcomes with an odd number of singlets (yellow) always stem from mixedness [Eq.~\eqref{eq:purity_from_tpd}].
Both effects are simultaneously captured in the lower bound [Ineq.~\eqref{eq:lower_bound_concurrence}] on the concurrence $C[\rho]$,
see TPD inset of Fig.~\ref{fig:experiment_states}.\par
Among the entangled states studied here, the GHZ state has the smallest ideal concurrence, 
$C[\href{https://graphvis.uber.space/?graph=6_7C00}{\Psi_3}] = 0.484(1)$.
This is because the concurrence $C[\Psi] = 1-\tilde a_n[\Psi]$ {\add of a pure state $\Psi$} 
equals one minus the weighted sum over the APD [Eq.~\eqref{eq:zero_singlet_from_apd}]
and---for the GHZ state $\href{https://graphvis.uber.space/?graph=6_7C00}{\Psi_3}$---the APD has comparatively large entries (because all non-trivial marginals are mixtures of 
{\add just two computational basis} states).
At the other extreme, the ideal AME state $\href{https://graphvis.uber.space/?graph=6_66A7}{\Psi_6}$ has maximally mixed marginals and, 
{\add accordingly,}
the largest possible concurrence, 
$C[\href{https://graphvis.uber.space/?graph=6_66A7}{\Psi_6}] = 0.719(1)$.
Our experimental observations qualitatively match these theoretical predictions:
among the entangled states $\rho_j$, the experimental GHZ (AME) state minimizes (maximizes) the lower bound on $C[\rho_j]$.
In particular, this shows that also the concurrence criterion is robust enough to detect entanglement in our experiment---even without error mitigation.
We stress the fact that (in contrast to $a_n[\rho]-2^{-n}$ and $a'_n[\rho] - a'_{n-1}[\rho]$) the concurrence $C[\rho]$ is an entanglement measure.
In all cases, we find that Bell readout error mitigation significantly increases the measured lower bound on the concurrence.

We also see in Fig.~\ref{fig:experiment_states} that Bell samples with zero triplets are almost never observed.
This is easily explained as the corresponding probability $\tilde a_0[\rho]$ is exponentially suppressed in the number of qubits $n$, 
which relates to the fact that $\tilde a_0[\rho]$ (leftmost blue bar) equals the alternating weighted sum [Eq.~\eqref{eq:zero_triplet_from_apd}] over the APD of $\rho$ (green).
Finally, let us point out that, if our experimental states $\rho_j$ were pure, then $\tilde a_0[\rho_j]$ would be (exponentially)  proportional to yet another entanglement measure known as the $n$-tangle [Eq.~\eqref{eq:ntangle}].

The above discussion   illustrates how various entanglement properties can be read off from the three QWE distributions.
For supplementary theoretical examples on how other distinct quantum features manifest themselves in QWEs, we refer the interested reader to Apps.~\ref{app:superposition_vs_mixtures} 
{\add and}
\ref{app:manybody_examples}.
In App.~\ref{app:superposition_vs_mixtures}, we elaborate on the subtle differences between coherent superpositions and incoherent mixtures whereas App.~\ref{app:manybody_examples} exemplifies characteristics of various many-body states.

{\add Finally, note that two-copy Bell sampling has also been performed in complementary experiments using reconfigurable atoms~\cite{bluvstein_a_quantum_2022, bluvstein_logical_quantum_2024}. 
From the raw data of those experiments it is possible to compute QWEs.
}

\subsection{Direct measurement of QWEs for QECCs}
\label{sec:experiment_qeccs}

In the previous subsection, we have experimentally investigated QWEs of pure target states.
Let us next turn our attention to QECCs.
As we established in Sec.~\ref{sec:rains_shadow_enumerators_interpretation},
the QWEs of a QECC can be directly measured via Bell sampling from  $\rho_\text{QECC}^{\otimes 2}$,
where  $\rho_\text{QECC}$ [Eq.~\eqref{eq:rho_qecc}] is the maximally mixed state within the code space.
In principle, this protocol (Fig.~\ref{fig:two_copy_bell_measurement}) has the desirable single-setting property, see App.~\ref{app:single_setting_benefits},
provided one can deterministically prepare $\rho_\text{QECC}$.
Alternatively,  one could update the state preparation circuit on a shot-by-shot basis as  $\rho_\text{QECC}^{\otimes 2}$ is physically indistinguishable from a uniformly at random (and independently for each copy) prepared logical computational basis state.
Fortunately, this second protocol can be derandomized when the number of qubits is small.
This is best explained for an $\llbracket n, 1 ,d \rrbracket$ code with logical state vectors $\ket{0}_\text{L}$ and $\ket{1}_\text{L}$.
Assume that the logical bit flip gate $X_\text{L}$ admits a transversal implementation,
which is the case for all stabilizer codes~\cite{lidar_quantum_error_2013} and some non-stabilizer codes~\cite{kubischta_family_of_2023}.
Then, it suffices to perform Bell sampling from just two different input states, namely $\ket{0,0}_\text{L}$ and $\ket{0,1}_\text{L}$.
Indeed, since QWEs are invariant~\cite{rains_quantum_weight_1998} under local unitary gates such as $X_\text{L} \otimes X_\text{L}$, 
the TPDs arising from $\ket{1,1}_\text{L}$ and $\ket{1,0}_\text{L}$, respectively,
 ideally coincide with those from  $\ket{0,0}_\text{L}$ and $\ket{0,1}_\text{L}$.
Hence, there is no need for taking Bell samples from 
$\ket{1,1}_\text{L}$ and $\ket{1,0}_\text{L}$~\footnote{For codes where $X_\text{L}$ is not transversal,  it is sufficient to sample from $\ket{0,0}_\text{L}$, $\ket{0,1}_\text{L}$, and $\ket{1,1}_\text{L}$ because Bell states are symmetric  under particle exchange (up to a global phase that is irrelevant here).}.

\begin{figure}
    \centering
    \includegraphics{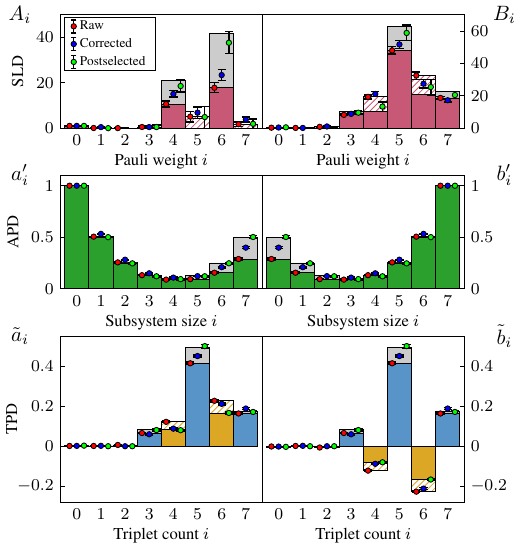}
    \caption{Experimentally measured QWEs (left) and dual QWEs (right) of the seven-qubit color code.
    In total, 20,000 Bell samples are averaged over two input settings: $\ket{0,0}_\text{L} $  and $\ket{0,1}_\text{L}$. 
    From this, we estimate SLD (top), APD (center), and TPD (bottom).
    For TPDs, we use blue (yellow) color to highlight that even (odd) singlet counts have a positive (negative) impact on the global purity.
    Shaded (gray) segments correspond to raw QWE values that are larger (smaller) in magnitude than in the ideal case.
    Besides raw data (left red dots), we also show error-corrected (centered blue dots) and post\-selected (right green dots) QWEs, respectively, for which Bell samples with violated parity checks are either corrected or simply discarded.
    Bootstrapped error bars show $95\%$ confidence intervals. 
    Note that, for some $\llbracket n,k,d\rrbracket$ codes, measurements of $A_i = 2^n a_i[\rho_\text{QECC}]$ and $B_i= 2^{n+k} b_i[\rho_\text{QECC}]$ QWEs are limited to intermediate system sizes,
    but TPD and APD measurements are always scalable, see main text Sec.~\ref{sec:complexity_and_robustness}.
    The organizational similarity to Fig.~\ref{fig:macwilliams} is intentional.
    }
    \label{fig:experiment_qecc}
\end{figure}

We perform the described two-setting experiment for the seven-qubit color code~\cite{steane_error_correcting_1996, bombin_topological_quantum_2006a, bombin_topological_quantum_2006b}
and present the measured QWEs and their dual QWEs in Fig.~\ref{fig:experiment_qecc}.
The seven-qubit color code is a $\llbracket 7,1,3 \rrbracket$ stabilizer QECC that is well explored in experiments~\cite{
ryan_anderson_implementing_fault_2022,
bluvstein_logical_quantum_2024,
butt_fault_tolerant_2024,
da_silva_demonstration_of_2024, 
mayer_benchmarking_logical_2024,
ryan_anderson_high_fidelity_2024, 
valentini_demonstration_of_2024,
zhou_low_overhead_2025,
ryan_anderson_realization_of_2021,
postler_demonstration_of_2022,
heussen_strategies_for_2023,
postler_demonstration_of_2024,
pogorelov_experimental_fault_2025}.
For our purposes, 
it suffices to directly prepare 
logical states
using a circuit with eight $\Cnot$ gates per copy~\cite{postler_demonstration_of_2022}
without resorting to sophisticated fault-tolerant preparation circuits~\cite{ryan_anderson_realization_of_2021, postler_demonstration_of_2022, heussen_strategies_for_2023, postler_demonstration_of_2024,  
zen_quantum_circuit_2024, peham_automated_synthesis_2025}.
This time, we use a buffer zone of only two ions to separate the two seven-qubit copies on our trapped-ion quantum device.
In all three QWE pictures, we can see that---in contrast to Fig.~\ref{fig:experiment_states}---the target state $\rho^\text{ideal}_\text{QECC}$ is not a pure state and that our experimental realization $\rho^\text{exp}_\text{QECC}$ is even more mixed.
Indeed, for a pure state, each of the three QWE distributions would have coincided---by MacWilliams' identity 
{\add [Eq.~\eqref{eq:macwilliams_identity}]}---with its respective dual.
This is clearly not the case in Fig.~\ref{fig:experiment_qecc} as the distributions in the left panels do not coincide with those on the right.
In quantitative terms, 
we can most easily read off the global purities 
$a'_7[\rho^\text{ideal}_\text{QECC}] = 0.5 $
and 
$a'_7[\rho^\text{exp}_\text{QECC}] = 0.228(1) $
from the APD (green), however, 
the same information is contained in the SLD (pink) and TPD (blue and yellow), whose sum [Eq.~\eqref{eq:sld_and_purity}] and alternating sum [Eq.~\eqref{eq:purity_from_tpd}], respectively, also yield the global purity.
Since {\add the MacWilliams transform} is diagonal in the TPD basis [Eq.~\eqref{eq:macwilliams_diagonal}],
the odd-singlet  (yellow) bins are  negated when passing to the dual TPD
$\mathbf{\tilde b}[\rho]= \tilde M \mathbf{\tilde a}[\rho]$ (bottom right in Fig.~\ref{fig:experiment_qecc}).
The fact that these bins are occupied, i.e., $\mathbf{\tilde a}[\rho] \neq \mathbf{\tilde b}[\rho]$, thus causes the aforementioned loss in  $\Tr[\rho^2] = \sum_{i=0}^n \tilde{b}_i [\rho]$.
Similarly, since $M'$ is the antidiagonal matrix [Eq.~\eqref{eq:macwilliams_antidiagonal}],
the dual APD, $\mathbf{b'}[\rho] = M' \mathbf{a'}[\rho]$, is just the reversed APD.
Therefore, 
$\mathbf{a'}[\rho] \neq \mathbf{b'}[\rho]$ implies the existence of marginals whose purity distinguishes them from their complementary marginal, which---by virtue of the Schmidt decomposition---again implies that $\rho $ is a mixed state.

Despite being mixed (both by construction and by noise), our state $\rho = \rho^\text{exp}_\text{QECC}$ is nevertheless entangled as certified by the averaged subsystem purities $a'_3[\rho],\ldots, a'_6[\rho]$, 
which are all clearly smaller than $a'_7[\rho]$. 
The maximal difference is $a'_7[\rho] - a'_4[\rho] = 0.20(2)>0$.
Also in the TPD picture, entanglement is observed as the all-triplet probability 
$\tilde{a}_7[\rho] = 0.163(1)$ is sufficiently small to grant a non-trivial lower bound [Ineq.~\eqref{eq:lower_bound_concurrence}] on the concurrence $C[\rho] \ge 0.13(1)$~\footnote{The respective ideal values  are $\tilde{a}_7[\rho^\text{ideal}_\text{QECC}] = 0.176(1)$ and  $C[\rho^\text{ideal}_\text{QECC}] \ge 0.33(1)$. }.
{\add Finally, the $i$-body sector length criterion [Eq.~\eqref{eq:nbody_criterion}] for $i=n-1$ certifies the presence of entanglement in $\rho^\text{exp}_\text{QECC}$ as we observe $a_6^\text{raw}[\rho^\text{exp}_\text{QECC}] = 0.14(3)> 0.06 \gtrsim a_6[\ket{0}\! \bra{0}^{\otimes 7}]$. 
}

The $n$-body sector length criterion, 
{\add however,} 
does not reveal that $\rho$ is entangled as $a_7^\text{\add raw}[\rho^\text{\add exp}_\text{\add QECC}] -2^{-7} = 0.006(11)$ is not undoubtedly positive within error bars.
This is to be expected as $A_i = 2^n a_i[\rho^\text{ideal}_\text{QECC}]$ (top left panel in Fig.~\ref{fig:experiment_qecc}) is the number of weight-$i$ stabilizer operators---and the $\llbracket 7,1,3\rrbracket$ code does not have any full-weight stabilizers.
In the ideal case, the Pauli weight distributions of the stabilizer group $\mathcal{S}$ and its normalizer $\mathcal{S}^\perp$  [Eq.~\eqref{def:normalizer}] are given by
\begin{align} \label{eq:sld_steane}
   \mathbf{A} &= (1, 0, 0, \hspace{0.85mm} 0\hspace{.85mm} , 21, \hspace{1.75mm} 0 \hspace{1.75mm}, 42, \hspace{0.85mm} 0\hspace{.85mm} )
   \hspace{3mm}
   \text{and}\\ 
\label{eq:dual_sld_steane}
   \mathbf{B} &= (1, 0, 0, 21, 21, 126, 42, 45),
\end{align}
respectively, from which we can read off the distance $d=3$ of the code [Eq.~\eqref{eq:distance}] as the smallest index $i>0$ for which $A_i< B_i$.
Moreover, $A_1 = A_2 = 0$ is equivalent~\cite{shor_quantum_analog_1997} to the fact that the seven-qubit color code (in contrast to its larger cousins) is a non-degenerate QECC.

In Fig.~\ref{fig:experiment_qecc}, we see that the SLD results are severely impacted by errors.
On the one hand, the 21 \mbox{weight-$4$} stabilizer operators (together with all other \mbox{weight-$4$}  Pauli operators, whose expectation values ideally vanish) 
only actualize a total of
\begin{align}
 2^7a_4^\text{\add raw}[\rho^\text{exp}_\text{QECC}] = 10.6^{+2.0}_{-1.8}    < 21 =
 2^7a_4[\rho^\text{ideal}_\text{QECC}]
\end{align}
squared Pauli expectation values. 
Similarly, we only find
\begin{align}
 2^7a_6^\text{\add raw}[\rho^\text{exp}_\text{QECC}] =  17.8^{+3.4}_{-3.0}   < 42 =
 2^7a_6[\rho^\text{ideal}_\text{QECC}],
\end{align}
which we mainly attribute to the $42$ \mbox{weight-$6$} stabilizer operators. 
On the other hand,  as we explain below, we observe that 
$ 2^7a_5^\text{\add raw}[\rho^\text{exp}_\text{QECC}]= 5.0^{+3.4}_{-3.0}$ and
$ 2^7a_7^\text{\add raw}[\rho^\text{exp}_\text{QECC}]= 1.7^{+1.4}_{-1.6}$ are higher than their ideal value zero. 
Despite these systematic errors due to experimental imperfections---which only amplified over the transform of (dual) TPDs to (dual) SLDs---we can still correctly infer the distance $d=3$ as well as the fact that the investigated code is non-degenerate from the raw data.

Let us point out an interesting noise characteristic that is observed 
in both the SLD and the dual SLD in Fig.~\ref{fig:experiment_qecc}:
the experiment underestimates (gray background) the number of Pauli operators $P\in\mathcal{S}$ and $P\in \mathcal{S}^\perp$ when $n-\wt(P)$ is even, 
whereas that of the opposite case is overestimated (pink shaded bars).
We explain this as follows.
First, we have $A_i=0$ in Eq.~\eqref{eq:sld_steane} for all odd $i$, which means that $\mathcal{S}$ only contains even-weight stabilizers.
Hence, Prop.~\ref{lem:shadow_characterization} implies that---for the investigated code---Rains' shadow $\tilde{\mathcal{S}}$ [Eq.~\eqref{def:shadow_coset}] coincides with the normalizer $\mathcal{S}^\perp$ [Eq.~\eqref{def:normalizer}].
For enumerators, this implies that $\tilde{\mathbf{a}}[\rho^\text{ideal}_\text{QECC}]  = \mathbf{b}[\rho^\text{ideal}_\text{QECC}] $ is a $(+1)$-eigenvector of the TPD-to-dual-SLD transform $\tilde{T}'^{-1} \tilde{M}$ from Fig.~\ref{fig:macwilliams}.
The impact of noise is easily understood in the TPD picture, where it manifests itself in the increased occurrence of odd-singlet events (within the constraints of Cor.~\ref{cor:triplet_mean_bound}).
In our case, we see that the largest deviation from the ideal case is that some of the probability $\tilde{a}_5[\rho]$ (blue) has leaked into $\tilde{a}_4[\rho]$ and $\tilde{a}_6[\rho]$ (yellow),
which correspond to two, three, and one singlet, respectively.
As the measured vector $\tilde{\mathbf{a}}[\rho]$ still has a fairly large overlap with its ideal version, it is unsurprising that 
$\mathbf{b}[\rho] = \tilde{T}'^{-1} \tilde{M}\tilde{\mathbf{a}}[\rho]$ is fairly close to $\tilde{\mathbf{a}}[\rho]$,
which is indeed what we observe in Fig.~\ref{fig:experiment_qecc}.
{\add Note that the top right panel displays $B_i = 2^{n+k}b_i[\rho] = 256 \times b_i[\rho]$, which enumerates the weight-$i$ logical Pauli operators of the code.}

Besides the raw data (red dots), we also show in Fig.~\ref{fig:experiment_qecc} the QWEs that are obtained when error correction (blue dots) or detection (green dots) is performed before the QWEs are estimated.
Here, it is important to note that---for color codes---all Clifford gates and, therefore, also Bell measurements can be implemented transversally. 
Thus, we can regard the here-reported experiment as \emph{logical} Bell sampling.
Leveraging a lookup table decoder for the $\llbracket 7,1,3 \rrbracket$ code, we can effectively (in classical postprocessing)  project $\rho^\text{exp}_\text{QECC}$ onto the logical code space without discarding any data.
Alternatively, we can postselect for events without detected errors by discarding all Bell samples where at least one parity check is violated.
Here, we can keep $4,067$ and $4,458$ Bell samples (from $10,000$ each) for $\ket{0,0}_\text{L}$ and $\ket{0,1}_\text{L}$, respectively, where no errors are detected.
Averaging the resulting two TPDs then results in the displayed error-detected, postselected QWEs.

We can see in Fig.~\ref{fig:experiment_qecc} that the error-corrected QWEs (blue dots) are generally closer to their ideal values than in the raw case (red dots).
Therefore our experiment demonstrates the hallmark of QEC: noise removal.
Furthermore, Fig.~\ref{fig:experiment_qecc} reveals that the postselected data (green dots) are in 
good
agreement with ideal predictions.
However, this {\add comes} at the expense of 
{\add increasing the} statistical uncertainty.
Even when accounting for error bars, the error detection results surpass those of error correction.
This {\add showcases} the practical advantage of error detection over correction
{\add for small qubit numbers.}

Before wrapping up this section,
we should mention that, although we limited ourselves to experiments with stabilizer states and codes, the general case can be treated analogously,
{\add see App.~\ref{app:qwes_of_cft_code} for an example.}
{\add Another}
interesting testbed for studying non-stabilizer codes is the one from Ref.~\cite{kubischta_family_of_2023} whose QWEs played an important role in their discovery.
The logical basis states of these exotic codes are superpositions of Dicke states [Eq.~\eqref{eq:dicke_state_definition}],
which can be prepared via \emph{entanglement carving}~\cite{skornia_nonclassical_interference_2001, sorensen_probabilistic_generation_2003, thiel_generation_of_2007, chen_carving_complex_2015, davis_painting_nonclassical_2018, ramette_carving_entangled_2025}.
Proof-of-principle entanglement carving experiments have been reported for neutral atoms trapped in optical lattices~\cite{welte_cavity_carving_2017, welte_photon_mediated_2018} and optical tweezers~\cite{dordevic_entanglement_transport_2021}
as well as for ions in a segmented Paul trap~\cite{richter_collective_photon_2023}.
As such, we anticipate that the outlined experiment will be within reach in the near future.

\section{Scalability considerations}
\label{sec:complexity_and_robustness}

Now that we have experimentally demonstrated that QWEs can be measured via two-copy Bell sampling,
we turn to the pressing question of scalability.
Specifically, we analyze two potential scalability bottlenecks.
For one, we assess how state preparation and measurement errors propagate into errors on the measured quantities in Sec.~\ref{sec:robustness_results}.
Second, in Sec.~\ref{sec:sample_complexity}, we investigate the effect of statistical errors arising from finite sample sizes.
In brief, we find that scalable measurements are always possible for TPDs and APDs but not always for SLDs.

\subsection{Robustness guarantees against experimental imperfections}
\label{sec:robustness_results}

To analyze how robust the two-copy Bell sampling protocol for measuring QWEs (recall Fig.~\ref{fig:two_copy_bell_measurement}) is against experimental imperfections, we continue to develop our theory from Sec.~\ref{sec:bell_sampling}.
The first step is to revisit the two-copy observables.

\bigskip
\refstepcounter{mycounter}
\noindent
\textbf{Lemma~\arabic{mycounter}\label{lem:operator_norms}} (Operator norms of {\add two-copy} QWE observables)
\emph{The operator norms of the observables from Eqs.~\eqref{eq:sld_observable}, \eqref{eq:apd_observable}, and \eqref{eq:tpd_observable} are given by}
$\|\twocopyobservable_i\|_\infty = 3^i\binom{n}{i}/2^n$,  
$\|\twocopyobservable'_i\|_\infty =1$, and $\|\tilde \twocopyobservable_i\|_\infty = 1$. \smallskip

\emph{Proof:}
The upper bounds follow from an application of the triangle inequality to the respective definitions and from the fact that Pauli and $\Swap$ operators have unit  operator norm.
Conversely, the matching lower bounds follow from
$\bra{\boldsymbol \Psi} \twocopyobservable_i \ket{\boldsymbol \Psi}  = 3^i \binom{n}{i}/2^n$ and 
$ \bra{\mathbf 0} \twocopyobservable'_i \ket{\mathbf{0}} =\bra{\mathbf 0} \tilde{\twocopyobservable}_i \ket{\mathbf{0}} = 1$,
where  $\ket{\boldsymbol \Psi} = \ket{\Psi^-}^{\otimes n}$ and $\ket{\mathbf{0}}= \ket{0}^{\otimes 2n}$. \hfill $\square$

\bigskip

In other words, luckily, the TPD  and  APD observables  $\tilde{\twocopyobservable}_i$ and $\twocopyobservable'_i$, respectively, have operator norms that are constant and small.
The operator norm of the SLD observable $\twocopyobservable_i$, on the other hand, blows up exponentially in general but not if $i$ is constant.
In light of the next result, one can thus robustly confirm that a promised pure input state is $m$-uniform if $m$ is a sufficiently small constant.
Indeed, operator norms relate to the sensitivity to experimental imperfections as follows.

\bigskip
\refstepcounter{mycounter}
\noindent
\textbf{Theorem~\arabic{mycounter}\label{thm:robustness}} (Robustness guarantees for 
{\add learning enumerators})
\emph{Let $\rho $ be an $n$-qubit state, $\mathfrak{a}_i[\rho] \in \{a_i[\rho], a'_i[\rho], \tilde{a}_i[\rho]\}$ one of its QWEs, and $\twocopyobservablegeneric_i \in \{\twocopyobservable_i, \twocopyobservable'_i, \tilde \twocopyobservable_i \}$ the corresponding 
{\add two-copy}
observable from Eq.~\eqref{eq:sld_observable}, \eqref{eq:apd_observable}, or \eqref{eq:tpd_observable}.
From Bell sampling experiments, we can estimate  $\mathfrak{a}_i[\rho] $ up to a systematic error of $\varepsilon>0$ if experimental imperfections are below $\varepsilon/\| \twocopyobservablegeneric_i\|_\infty$.
More precisely,  $\varepsilon =  \| \twocopyobservablegeneric_i\|_\infty \| \sigma - \rho\otimes \rho \|_{1}$, where $\sigma$ denotes the effective experimental $2n$-qubit state  that incorporates all preparation and readout errors (including those from $\Cnot$ gates).
Therefore, TPDs and APDs can be {\add learned} robustly  while SLDs---in the worst case---cannot.
}
\smallskip

\emph{Proof:}
Since we have effectively implemented $\sigma$ instead of $\rho\otimes \rho$,
our estimator $\hat{\mathfrak{a}}_i = \Tr[\twocopyobservablegeneric_i \sigma]$ for $\mathfrak{a}_i[\rho] = \Tr[\twocopyobservablegeneric_i (\rho \otimes \rho)]$ is biased. 
Indeed, the systematic error is given by
\begin{align}
    | \hat{\mathfrak{a}}_i- \mathfrak{a}_i[\rho]| = \|\twocopyobservablegeneric_i (\sigma - \rho\otimes \rho)  \|_1
\end{align}
and Hölder's inequality yields
\begin{align}
    | \hat{\mathfrak{a}}_i- \mathfrak{a}_i[\rho]| \le   \| \twocopyobservablegeneric_i\|_\infty \| \sigma - \rho\otimes \rho \|_{1}  =\varepsilon \,,
\end{align}
which completes the proof. \hfill $\square$

\bigskip
  
How does the error $\| \sigma - \rho\otimes \rho \|_{1} $ relate to experiments?
As $\sigma$ is the imperfect experimental realization of $\rho\otimes \rho$,
there must be some quantum channel $\mathcal{E}$ such that $\sigma = \mathcal{E} [\rho \otimes \rho]  $.
We find the bound
\begin{align} \label{eq:trace_distance_and_diamond_norm}
    \norm{ \sigma - \rho\otimes \rho}_1 \leq \norm {\mathcal{E} - \mathcal{I}}_{1 \rightarrow 1}\,,
\end{align}
as an immediate consequence of the definition
\begin{align} \label{eq:diamond_norm_definition}
\norm{\mathcal{E} - \mathcal{I}}_{1 \rightarrow 1}
= \max_{ \|\omega\|_1 \le 1  }
\norm{ \mathcal{E}  (\omega)-
 \mathcal{I} (\omega)
}_1
\end{align}
of the one-to-one norm, which lower-bounds the well-known diamond norm. 
In the latter norm, the distance between two Pauli error channels equals twice the total variation distance of the corresponding Pauli error probability vectors~\cite{magesan_characterizing_quantum_2012}.
Thus,  
\begin{align}
    \norm{\mathcal{E}-\mathcal{I}}_{{1 \rightarrow 1}} \le   2(1-p_0)\,,
\end{align}
if $\mathcal{E}$ is a Pauli channel with {\add error-absence} probability $p_0$.
For example, for the local depolarizing noise channel
$\mathcal{E} = \mathcal{E}_p^{\otimes n}$,
we have $p_0=(1-3p/4)^n$.
Thus, for every constant value $p>0$  of noise per qubits, $p_0$ is exponentially suppressed in $n$, i.e., the QWEs of $\rho$ and $\mathcal{E}_p^{\otimes n}[\rho]$ {\add may}, 
unsurprisingly,
{\add  exhibit significant} differences.
Nevertheless,  as the measured state $\sigma = \mathcal{E}_ p^{\otimes n}[\rho] \otimes \mathcal{E}_ p^{\otimes n}[\rho]$ is still a two-copy state,
the Bell sampling protocol 
{\add remains to give}
rise to an unbiased measurement of QWEs,
however, for the noisy state $ \mathcal{E}_ p^{\otimes n}[\rho]$.
For examples of many-body quantum states with a noteworthy amount of noise yet interesting QWEs, see App.~\ref{app:manybody_examples}.

A more serious issue is if the two copies become entangled with each other.
Let us model such a scenario via a collection of $m$ coherent overrotations
\begin{align}
    \mathcal{E} = \mathop{\circ}\limits_{i=1}^m \mathcal{U}_i \,,
\end{align}
where each quantum channel $\mathcal{U}_i $ implements the unitary
\begin{align} \label{eq:unitary_parametrization}
    U_i = \exp\left( -\text{i} t_i H_i \right)
\end{align}
for some small times $t_i > 0$ and error Hamiltonians $H_i$ acting on an arbitrary number of qubits. 
Using standard arguments \footnote{For example, for $m=1$ and abbreviating $\omega = \rho\otimes \rho$, we find $U \omega U ^\dagger - \omega  =   U \omega U^\dagger ( \mathbbm 1- U) - (\mathbbm{1}-U)\omega$, which implies
$\|U \omega U ^\dagger - \omega\|_1 \le \|  U \omega U^\dagger ( \mathbbm 1- U) \|_1+ \|(\mathbbm 1 - U) \omega\|_1$ by  the triangle inequality.
Next, applying Hölder's inequality yields 
$\|  U \omega U^\dagger ( \mathbbm 1- U) \|_1  \le 
\|  U\|_\infty  \| \omega \|_ 1 \|U^\dagger \|_\infty  \|( \mathbbm 1- U) \|_\infty$ 
and $\|(\mathbbm 1 - U) \omega\|_1 \le \|\mathbbm 1 -U \|_\infty \|\omega\|_1$.
Since $U$ is a unitary and $\omega$ is a state, we have $\| U \|_\infty =\| U^\dagger \|_\infty = \|\omega\|_1 = 1$.
Finally,  $\|U \omega U ^\dagger - \omega\|_1 \le  2\min_\phi  \| {U} - \mathbbm{1}\text{e}^{\text{i}\phi} \|_\infty $ follows from  minimizing the norm over the global phase, which completes the proof of Eq.~\eqref{eq:imperfect_copies_derivation_2}
for $m=1$. The general case of $m\ge 1$ follows similarly using telescope sums.},
we find the bound
\begin{align} \label{eq:imperfect_copies_derivation_2}
    \norm{ \sigma - \rho\otimes \rho}_1 
    &\le 2\sum_{i=1} ^m  \min_{\phi\in[0,2\pi)} \norm{ {U}_i - \mathbbm{1}\text{e}^{\text{i}\phi}} _\infty \, . 
\end{align} 
For each 
term in Eq.~\eqref{eq:imperfect_copies_derivation_2}, we can insert  Eq.~\eqref{eq:unitary_parametrization}, which yields 
\begin{align}
    \min_{\phi} \norm{ {U}_i - \mathbbm{1}\text{e}^{\text{i}\phi}} _\infty&=\min_{\phi}\norm{ \text{e}^{-\text{i}(t_i H_i+\phi)} - \mathbbm1}_\infty \\
    \nonumber
    &= 2\min_{\phi} \norm{ \, \sin \left( \frac{t_i H_i+\phi}{2}\right)}_\infty \\
    \nonumber
    &\le t_i \min_{\phi\in \mathbb{R}}\norm{H_i+\phi} _\infty\\
    &=t_i \frac{\lambda_{\max} [H_i]-\lambda_{\min} [H_i]}{2}\,,
    \nonumber
\end{align}
where have used the unitary equivalence of the operator norm as well as the bound $|\sin(x)|\leq|x|$. Here, the difference $\Delta [H_i] = \lambda_{\max}[H_i]-\lambda_{\min}[H_i]$ is the \emph{spectral width} of the error Hamiltonian $H_i$.
We have thus established the bound
\begin{align}
    \norm{ \sigma - \rho\otimes \rho}_1\leq \sum_{i=1}^m  t_i \, \Delta [H_i] \, ,
\end{align}
which makes the {\add physical} intuition \emph{``the error on the prepared state remains small if  errors in the preparation circuits are neither too frequent nor too strong''} mathematically precise.

\subsection{Sample complexity bounds}
\label{sec:sample_complexity}
Sample complexity bounds are essential for predicting the runtime of an experiment prior to its execution.
Let us denote by $N$ the number of Bell samples $ \boldalpha_1,\ldots, \boldalpha_N \in \{\Phi^\pm, \Psi^\pm\}^n$ taken from two copies of an $n$-qubit state $\rho$ by repeatedly executing the 
{\add learning protocol}
described in Fig.~\ref{fig:two_copy_bell_measurement} above.
The next result settles the question how large one has to select $N$ in order to ensure accurate estimation of the various QWEs.

\bigskip
\refstepcounter{mycounter}
\noindent
\textbf{Theorem~\arabic{mycounter}\label{thm:sample_complexity_bounds}}  
(Sample complexities for {\add learning enumerators})
\emph{Let $\rho $ be an $n$-qubit state, $\mathfrak{a}_i[\rho] \in \{a_i[\rho], a'_i[\rho], \tilde{a}_i[\rho]\}$ one of its QWEs, and $\twocopyobservablegeneric_i \in \{\twocopyobservable_i, \twocopyobservable'_i, \tilde \twocopyobservable_i \}$ the corresponding
{\add two-copy}
observable from Eq.~\eqref{eq:sld_observable}, \eqref{eq:apd_observable}, or \eqref{eq:tpd_observable}.
Let $\varepsilon, \delta>0$ be fixed target accuracy parameters.
For every $i \in \{0,\ldots, n\}$, it is sufficient to gather
\begin{align}
    N = \frac{\Delta [\twocopyobservablegeneric_i]^2}{2\varepsilon^2} \ln\left(\tfrac{2}{\delta}\right)
\end{align}
many Bell samples from $\rho\otimes \rho$ to ensure a statistical error of  $\vert \hat{\mathfrak{a}}_i -\mathfrak a_i[\rho] \vert< \varepsilon$ with probability at least $1-\delta$,
where $\Delta [\twocopyobservablegeneric_i]\leq 2\|\twocopyobservablegeneric_i\|_\infty$ is the spectral width of $\twocopyobservablegeneric_i$. 
Moreover, if we want that $\vert \hat{\mathfrak{a}}_i -\mathfrak a_i[\rho] \vert< \varepsilon$ holds for all $i\in \{0,\dots,n\}$ simultaneously, then
\begin{align}
    N = \frac{ {\add \max_i\Delta [\twocopyobservablegeneric_i] ^2 }}  {2\varepsilon^2} \ln\left(\tfrac{2(n+1)}{\delta}\right)
\end{align}
Bell samples from $\rho\otimes \rho$ are sufficient.
}
\smallskip

\emph{Proof:}
For every sample ${\boldalpha} \in \{\Phi^\pm, \Psi^\pm \}^n$,
the single-shot estimator $\bra{\boldalpha} \twocopyobservablegeneric_i \ket{\boldalpha}$ takes values in the spectrum of $\twocopyobservablegeneric_i$.
Therefore, Höffding’s inequality establishes the first claim.
The second claim follows similarly by applying a union bound argument 
{\add and the triangle inequality}. 
\hfill $\square$
\bigskip

Theorem~\ref{thm:sample_complexity_bounds} has a few immediate consequences. 
For one, we have $\Delta [\tilde \twocopyobservable_i] = 1$ since $\tilde \twocopyobservable_i$ is a projector.
Hence, we only need at most $N= \varepsilon^{-2} \ln(2/\delta)/2$ samples to estimate any triplet probability $\tilde a _i [\rho]$ of interest.
This should not come as a surprise as two-copy Bell sampling allows us to directly sample from the TPD $\tilde{\mathbf{a}}[\rho]$.
Similarly, for every subsystem size $i$ of interest, estimating the averaged purity $a'_i[\rho]$ requires only up to $N=2 \varepsilon^{-2} \ln(2/ \delta)$ samples as $\Delta[\twocopyobservable'_i] = 2$, by Lem.~\ref{lem:operator_norms}.
For SLDs, on the other hand, the situation is more delicate because $\| \twocopyobservable_i\|_\infty= 3^i\binom{n}{i}/2^n$ can be exponentially large.
Therefore, in general, exponentially many Bell samples are required to estimate the $i$-body sector length $a_i[\rho]$.
For any constant value of $i$, however, $\|\twocopyobservable_i\|_\infty$ is exponentially \emph{small} in $n$.
As such, the total amount of squared $i$-body Pauli correlations, i.e., $A_i[\rho] = 2^n a_i[\rho]$ can be estimated using just $N=(3^i \binom{n}{i})^2 \varepsilon^{-2} \ln(2/\delta)/2 = O(n^{2i}\varepsilon^{-2}\ln(1/\delta))$ many Bell samples. 
This shows that checking 
{\add constant}-$m$ uniformity of a {\add promised-to-be-pure} input state is not only robust (Thm.~\ref{thm:robustness}) but also sample efficient (Thm.~\ref{thm:sample_complexity_bounds}).
A related application involves determining a lower bound for the distance of a QECC.
We should point out, however, that---while this does constitute a valid strategy with the additional single-setting benefit (App.~\ref{app:single_setting_benefits})---directly measuring all $\sum_{i=1}^m 3^i\binom{n}{i}$ low-weight Pauli operators may require fewer samples.

Let us stress the fact that Thm.~\ref{thm:sample_complexity_bounds} only concerns the worst-case sample complexity for measuring QWEs.
While this suffices to guarantee sample-efficient 
{\add learnability} 
of TPDs and APDs,
we need to take a closer look in the case of SLDs.
To this end, we point out that the statistics arise from a multinomial distribution.
As such, we can explicitly compute the variances
\begin{align}
    \sigma_i^2=\frac{1}{N}\sum_{j=0}^n |\tilde T ^{-1}_{i,j}|^2 \tilde a_j[\rho]- \frac{1}{N}\left(\sum_{j=0}^n \tilde T ^{-1}_{i,j} \tilde a_j[\rho]\right)^2
\end{align}
of the individual estimators $\hat{a}_i$ for $a_i[\rho]$,
from which the total variance of the SLD follows as
\begin{align} \label{eq:sld_variance}
    \sigma^2=\mathbb E\left[\sum_{i=0}^n (a_i[\rho]-\hat a_i)^2\right]=\sum_{i=0}^n  \sigma_i^2\,,
\end{align}
where the expectation value is over different realizations of an $N$-sample experiment.
Since $\tilde T_{i,j}^{-1}$ takes exponentially large values for small triplet counts $j\lesssim n/2$~\footnote{See \url{https://mc-zen.github.io/qsalto/?n=500} for an illustration of this fact in an interactive online version of Fig.~\ref{fig:macwilliams} for $n=500$ qubits.
}, 
the variance $\sigma^2$ of the SLD highly depends on where the TPD concentrates.
While Cor.~\ref{cor:triplet_mean_bound} guarantees that the mean of the TPD is never smaller than $3n/4$, it does not prevent the existence of heavy tails of the TPD around $j \approx n/2$ triplets.
Indeed, such issues occur, e.g.,  for \emph{Dicke states} with a constant filling fraction and for GHZ states, see App.~\ref{app:manybody_examples} for a visual impression.

\begin{figure}
    \centering
    \includegraphics{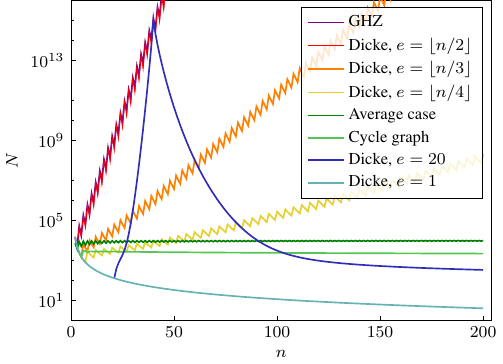}
    \caption{Number $N$ of Bell samples that is required for reaching a variance $\sigma^2=10^{-4}$  of the Shor--Laflamme quantum weight enumerator (aka sector length) distribution (SLD) for various states as a function of the number $n$ of qubits.
    For some states, e.g., half-filled Dicke states [Eq.~\eqref{eq:dicke_state_definition}], estimating SLDs up to  constant additive error has exponential sample complexity.
    In the average case of a $2$-design, however, a constant number of Bell samples suffices.
    }
    \label{fig:sample_requirements}
\end{figure}

Fixing a target variance of $\sigma^2 = 10^{-4}$ for the SLD, we solve Eq.~\eqref{eq:sld_variance} for the number $N$ of required Bell samples and plot the result in Fig.~\ref{fig:sample_requirements} as a function of the number $n$ of qubits.
The sample requirements highly depend on the specific state, for which we insert the corresponding QWEs into Eq.~\eqref{eq:sld_variance}.
We investigate various Dicke state families (whose QWEs are treated in Sec.~\ref{sec:dicke_entanglement_stability} below), 
GHZ and cycle graph states (whose QWEs are taken from Ref.~\cite{miller_shor_laflamme_2023}),
and average-case states.
For the latter, we compute $N$ in Fig.~\ref{fig:sample_requirements} by inserting~\cite{mele_introduction_to_2024}
\begin{align} \label{eq:sld_2design}
    \mathbb E_{\ket{\psi}\sim \mu_2}  a_i \left[  \ket{\psi}\!\bra{\psi}  \right] 
    = 
    \begin{cases}
        \tfrac{1}{2^n} & \text { if } i=0 ,\\ 
        \tfrac{3^i \binom{n}{i}}{2^n(2^n+1)} & \text{ if } i>0,
    \end{cases}
\end{align}
into Eq.~\eqref{eq:sld_variance}, where $\mu_2$ is an arbitrary \emph{2-design}, e.g., that of uniformly-random stabilizer or Haar-random states~\cite{kliesch_theory_of_2021}.
Also note that the output distribution of brickwork circuits is an approximate 2-design after linear depth~\cite{harrow_approximate_unitary_2023}.

We can see in Fig.~\ref{fig:sample_requirements} that the curves for GHZ (purple) and half-filled Dicke states (red), which lie almost on top of each other, 
are quickly diverging:
even for a moderate number of $n=50 $ qubits, one would need more than $10^{16}$ Bell samples to achieve an SLD variance of $\sigma^2 = 10^{-4}$.
Also for third-filled (orange) and quarter-filled (yellow) Dicke states, the sample requirements scale exponentially in $n$.
For a constant number $e$ of Dicke excitations, e.g., $e=20$ (dark blue) and $e=1$ (bright blue), 
on the other hand, the sample requirements asymptotically~\footnote{For the $e=20$ curve, 
however, we observe an initial increase of $N$ with $n$. This is due to the competing effects, e.g., for $n=21$ ($n=40$), the state is local-unitary equivalent to $e=1$ ($e=n/2$). These two extreme cases have opposite behaviors as described in the main text.}
decrease with $n$.
Indeed, for $n>100$ qubits, less than a few thousands (tens) of Bell samples are required for $e=20$ ($e=1$).
Finally, the sample requirements of average-case (dark green) and cycle graph states (bright green) show almost no dependence on $n$, namely $N\approx 10^4$ and $N \lesssim 3 \times 10^3$, respectively.

The exponential behavior of the sample requirements for GHZ and Dicke states with a constant filling factor are easily explained by the aforementioned heavy tails of the respective TPDs around $j\approx n/2$ triplets.
When the number $e$ of Dicke excitations is constant, however, $N$ decreases in the regime $n>2e$ because the   QWEs become increasingly similar to those of a pure product state (this was formally proven for $e=1$ in App.~D of Ref.~\cite{miller_shor_laflamme_2023}).
For the latter, a single Bell sample is sufficient because the TPD of $\ket{0}^{\otimes n}$ has a vanishing variance.
Furthermore, we attribute the observation that average-case and cycle graph states feature the same qualitative behavior to an earlier observation~\cite{miller_shor_laflamme_2023} that cycle graph and other cluster states have SLDs that are closely matched by random states.
For them, the TPD only features a single peak at $j = 3n/4$ triplets that is sufficiently concentrated to prevent an uncontrolled spread of statistical errors via Eq.~\eqref{eq:sld_variance}.
For a depiction of the QWEs of Dicke, cycle graph, and GHZ states on $n=200$ qubits, see App.~\ref{app:manybody_examples}.

Our discussion of Fig.~\ref{fig:sample_requirements} shows that the situation is not as concerning as it might seem based on an initial look at Thm.~\ref{thm:sample_complexity_bounds}.
While some states do require exponentially many Bell samples to estimate the SLD up to constant additive error, for most states, the sample requirements are manageable. Moreover, such states include cluster states, which are highly entangled but can be easily prepared using a constant-depth  circuit.

\section{Theoretically analyzing the persistence of entanglement against local noise}
\label{sec:dicke_entanglement_stability}

In this final section,
{\add we} showcase how the machinery of QWEs can be leveraged
to theoretically predict how much noise certain entangled states are able to withstand without becoming fully separable.
Specifically, we analyze Dicke states in the context of local depolarizing noise.
We expand on  earlier strategies~\cite{miller_shor_laflamme_2023} in two ways:
first, we exploit that not only the $n$-body sector length and the purity criterion can be tested via SLDs but also the concurrence and the fidelity criterion---a connection that we have established in the present work.
Second, we demonstrate for concrete examples that the strategy of Ref.~\cite{miller_shor_laflamme_2023} is not just limited to stabilizer states.

Denote by $n$ and $e$  the number of qubits and excitations, respectively,
of a \emph{Dicke state} vector  
\begin{align} \label{eq:dicke_state_definition}
    \ket{D^n_e} = \tfrac{1}{\sqrt{\binom{n}{e}}} \sum_{ \substack{\mathbf{x} \in \FF_2^n : \, \vert \mathbf{x}\vert=e} } \ket{\mathbf{x}} \,,
\end{align}
where $\vert \mathbf{x} \vert$ is the Hamming weight of $\mathbf{x}\in \FF_2^n$.
Leveraging the Schmidt decomposition of $\ket{D^n_e}$~\cite{bergmann_entanglement_criteria_2013}, we find that the purity of every subsystem $S$ containing $\vert S \vert = i$ qubits (and therefore also their average) is given by
\begin{align} \label{eq:apd_dicke_states}
    a'_i\big[  \Psi_e  \big] = \sum_{j=0}^e \left(\tfrac{\binom{i}{j}\binom{n-i}{e-j}}{\binom{n}{e}}\right)^2 \,,
\end{align}
where we have introduced the notation $\Psi_e = \ket{D^n_e} \! \bra{D^n_e}$.
Applying the matrix from Eq.~\eqref{eq:trafo_apd_to_sld}, we find $\mathbf{a}[\Psi_e] = T'^{-1} \mathbf{a'}[\Psi_e] $.
Observe that in the SLD basis, local depolarizing noise acts diagonally [Eq.~\eqref{eq:sector_length_decay}], i.e., $\mathbf{a}\left[\mathcal{E}^{\otimes n}_p[\Psi_e]\right] = E(p) \mathbf{a}[\Psi_e]$,
where  
\begin{align}
    E(p)  = \diag \left( 1, (1-p)^2 ,\ldots, (1-p)^{2n} \right) 
\,    .
\end{align}
By transforming  back to Rains' unitary QWEs ({\add aka} APD) or forth to Rains' shadow QWEs ({\add aka} TPD), we can compute 
\begin{align}
    \label{eq:dicke_sld_decay}
    \mathbf{a}[\mathcal{E}^{\otimes n}_p[\Psi_e]] &=  E(p) T'^{-1} \mathbf{a'}[\Psi_e] \, , \\
    \label{eq:dicke_apd_decay}
    \mathbf{a'}[\mathcal{E}^{\otimes n}_p[\Psi_e]] &= T' E(p) T'^{-1} \mathbf{a'}[\Psi_e]\, , \\ 
    \label{eq:dicke_spd_decay}
    \text{and } \hspace{2mm}
    \mathbf{\tilde a}[\mathcal{E}^{\otimes n}_p[\Psi_e]] &= \tilde T E(p) T'^{-1} \mathbf{a'}[\Psi_e] 
\end{align}
from Eq.~\eqref{eq:apd_dicke_states}.
This, in turn, allows us to numerically find the largest noise value $p$ for which certain criteria certify the presence of entanglement.
\begin{figure}
    \centering
    \includegraphics{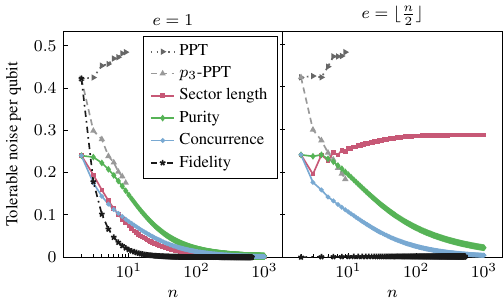}
    \caption{Noise robustness of the presence of entanglement in $W$ states $(e=1)$ and half-filled Dicke states $(e=\lfloor \tfrac{n}{2}\rfloor$) for various entanglement criteria described in the main text as a function of the number $n$ of qubits.
    If the noise level $p$ is below any of the curves, the locally-depolarized state $\mathcal{E}_p^{\otimes n}[\Psi_e]$ is necessarily entangled according to the corresponding criterion.
    }
    \label{fig:dicke_robustness}
\end{figure}
In Fig.~\ref{fig:dicke_robustness}, we show the results for $W$ states ($e=1$) and half-filled Dicke states ($e=\lfloor \tfrac{n}{2}\rfloor$).
The $n$-body sector length criterion  (pink) and the fidelity criterion $\bra{D^n_e} \rho \ket{D^n_e} > f^n_e$ (black) can be tested on the level of SLDs 
via Eq.~\eqref{eq:nbody_criterion} and Eq.~\eqref{eq:overlap_decay}, respectively, where the fidelity bounds $f^n_1$ and $f^n_{n/2}$ are taken from Ref.~\cite{haffner_scalable_multiparticle_2005} and Ref.~\cite{bergmann_entanglement_criteria_2013}, respectively. 
For $f^n_{n/2}$, we assume $n$ to be even.
The purity criterion, $a'_n[\rho]=\Tr[\rho^2] > \tfrac{1}{n}\sum_{i} \Tr[ \Tr_i[\rho]^2] = a'_{n-1}[\rho]$ (green), is testable via Eq.~\eqref{eq:dicke_apd_decay}.
By Thm.~\ref{thm:main},
$\tilde a_j[\mathcal{E}^{\otimes n}_p[\Psi_e]] $ can be interpreted as the probability to observe $j$ triplets in a Bell sampling experiment with two noisy copies of $\Psi_e$.
Thus, we can also find the largest value of $p$ for which the lower bound on the concurrence $C[ \mathcal{E}^{\otimes n}_p[\Psi_e] ]$ from Ineq.~\eqref{eq:lower_bound_concurrence} is still positive (blue).

Using brute force density matrix calculations, we also compute the largest value of $p$ for which the \emph{positive partial transposition} 
(PPT) criterion (dark gray) 
and its relaxation to the $p_3$-PPT criterion (bright gray) certify entanglement~\cite{peres_separability_criterion_1996, horodecki_separability_1996, zhou_single_copies_2020, elben_mixed_state_2020}.
We point out that the choice of bipartition is of crucial importance here.
For the purity criterion, we find it to be best practice to probe entanglement across a bipartition of size $n-1$ versus $1$ while bipartitions of size $ \lceil\tfrac{n}{2}\rceil$ versus $\lfloor \tfrac{n}{2}\rfloor $ lead to better noise thresholds for PPT and $p_3$-PPT.
We observe in Fig.~\ref{fig:dicke_robustness} that the PPT criterion, which we can only evaluate for a small number qubits, leads to the best lower bound on the noise threshold.
For very large qubit numbers, we can still apply the QWE machinery and we find that the purity criterion leads to better results than the concurrence criterion [Eq.~\eqref{eq:concurrence_criterion}] for both $W$ states and half-filled Dicke states.
The $n$-body sector length criterion yields poor results for $W$ states, however, we find that half-filled Dicke states can withstand an astonishing amount ($p \ge 0.28$ for all $n\ge 52$) of depolarizing noise per qubit before 
$a_n\big[\mathcal{E}^{\otimes n}_p[ \Psi_{\lfloor{n}/{2}\rfloor}]\big]$ falls below $2^{-n}$.

Having said this,
we stress that this noise threshold only concerns the presence of entanglement according to that criterion.
Detecting entanglement in the laboratory, however, remains challenging as we just saw in Fig.~\ref{fig:sample_requirements} in Sec.~\ref{sec:complexity_and_robustness}.

In conclusion, entanglement criteria based on QWEs have several {\add theoretical} advantages over other criteria such as PPT and $p_3$-PPT:
(i) they can handle local noise as easily as global noise;
(ii) they can be easily applied to very large system sizes; and
(iii) they can test multiple criteria at the same time.

\section{Discussion}

In this work, we have established a concrete physical interpretation of the longstanding mathematical concept known as \emph{quantum shadow enumerators}:
they can be understood as the \emph{triplet probability distribution} that arises in a two-copy Bell sampling experiment (Thm.~\ref{thm:main}).
This insight led us to a fruitful connection between \emph{quantum error correction} and \emph{entanglement},
where shadow enumerators and Bell sampling occupy pivotal roles, respectively.

Prioritizing practicality, we made strides to strengthen this new connection.
For one, we have showcased---both experimentally and theoretically---how the weight enumerator machinery from quantum error correction can be repurposed as a powerful tool for entanglement characterization and, more generally, for benchmarking the performance of quantum devices.
Conversely, we have explained and experimentally demonstrated how various notions of quantum weight enumerators (e.g., the numbers of fixed-Pauli-weight stabilizer and logical operators of a probed quantum error-correcting code) can be directly measured via two-copy Bell sampling.
In this context, we have also clarified how error correction relates back to Bell sampling:
for {\add self-dual CSS} codes, Bell sampling can be regarded as a \emph{logical} protocol, which grants compatibility with the error correction capabilities of the code whose weight enumerators are being measured.
In vivid terms, the experiment corrects itself.
We demonstrated this possibility on our trapped-ion quantum processor using the example of the seven-qubit color code.
Furthermore, to support quantum state analyses beyond the paradigm of quantum error correction, we have developed and implemented natural error mitigation strategies, which make heavy use of the quantum weight enumerator machinery themselves.

{\add 
Besides the rich information one gains about entanglement and QEC-related properties,
our protocol has yet another useful application:
when the QWEs of a target state are known in advance, 
directly measuring QWEs constitutes an extremely sample-efficient one-sided test---with the additional practical benefit of being single-setting---for bugfixing quantum circuits and spotting implementation errors early on.
In this context, it is important to note that QWEs are local-unitary invariants~\cite{rains_quantum_weight_1998}, hence, incorrect QWEs cannot stem from incorrect   single-qubit unitaries in the final layer of a preparation circuit.
This informs where to look for mistakes.
}

Acknowledging the impossibility of completely eliminating errors,
we have proven rigorous robustness guarantees against experimental imperfections (Thm.~\ref{thm:robustness}).
In a similar spirit, we have derived concrete sample complexity bounds (Thm.~\ref{thm:sample_complexity_bounds}),
thereby revealing a surprising distinction in the hardness of estimating various instances of quantum weight enumerators via Bell sampling:
while triplet probabilities (aka shadow enumerators) and averaged purities (aka unitary enumerators)
can always be measured in a scalable manner,
estimating sector lengths (aka Shor--Laflamme enumerators) has an exponential worst-case sample complexity.
By investigating the origins of this issue, we were able to show that, fortunately, the average-case sample complexity remains unaffected.
In conjunction with our reduction of the computational postprocessing complexity from exponential to polynomial, these advances pave the way for blue-sky research into the fundamental {\add entanglement} structure of large-scale quantum states.
This is complemented by the fact that Bell sampling is a single-setting protocol that allows for extremely fast data acquisition in practice.

Moving past the direct applicability of our developments, we envision several future research directions.
With two-copy Bell measurements, one can efficiently probe certain quadratic properties of a quantum state $\rho$, however, there also exist
many higher-moment entanglement criteria~\cite{zhou_single_copies_2020, elben_mixed_state_2020, liu_detecting_entanglement_2022, rico_entanglement_detection_2024, vermersch_many_body_2024}.
In this context, one could explore existing generalizations~\cite{rains_polynomial_invariants_2000, huber_positive_maps_2021,  munne_sdp_bounds_2024} of the weight enumerator machinery in the $\rho^{\otimes k}$ scenario for $k\ge 3$.
In a different direction, one may consider properties that are not quadratic within a single state, but rather involve multiple states. 
A more general notion of quantum weight enumerators pertains to $\Tr[\rho\sigma]$ instead of $\Tr[\rho^2]$, allowing for a straightforward translation of the purity aspects of this work into the trace overlap between different states~\cite{shor_quantum_analog_1997, rains_quantum_weight_1998}.
This plausibly opens up new methods  and applications in state comparison and fidelity estimation.
In this work, we have placed heavy emphasis on the connection between the ``big three'' quantum weight enumerators, but this does not mean that they constitute everything there is to be discovered about enumerators. 
There are conceivably other, yet undiscovered, enumerator variants where the triplet probability distribution framework may reveal interesting connections.
For example, recent proposals for `circuit' and `signed' enumerators could potentially benefit from the TPD connection~\cite{kukliansky_quantum_circuit_2025, rall_signed_quantum_2017}.

Finally, we highlight a well-known connection between Bell sampling and the Hong-Ou-Mandel effect from quantum optics \cite{hong_measurement_of_1987, moura_alves_multipartite_entanglement_2004, daley_measuring_entanglement_2012, garcia_escartin_swap_test_2013, islam_measuring_entanglement_2015}.
An important application of this effect is to assess the indistinguishability of two photons, which can be regarded as a quantum-optical analog to estimating state overlaps on digital quantum computers.
The latter can be achieved by performing transversal Bell measurements on two different input states.
In this work, we have focused on the digital version of this protocol. 
It would be interesting to explore the implications of our developments for quantum optics and, more broadly, for analog quantum simulations.

\smallskip

\emph{Note added.}
After discovering the physical interpretation of shadow enumerators as triplet probabilities in two-copy Bell measurements, we became aware of a complementary result that interprets them as probabilities in \textsc{Swap} tests~\cite{shi_exploring_quantum_2024}. 
We regard these two viewpoints as independent perspectives on an interesting problem.

\bigskip

\begin{acknowledgments}  
The authors express their gratitude toward Marine Demarty, Sofienne Jerbi, Ferdinand Schmidt-Kaler, David C.\  Spierings, and Nikolai Wyderka for their insightful discussions.

This work has been supported by the Quantum Flagship (Millenion, for this is the result of a joint-node collaboration, and PasQuans2), the 
{\add BMFTR} (RealistiQ, QSolid, MUNIQC-Atoms, {\add DAQC, QuSol, Hybrid++, PasQuops}), the Munich Quantum Valley (K-8), 
the BMWK (EniQmA), the QuantERA (HQCC), the Cluster of Excellence MATH+, the DFG (CRC 183 {\add and SPP 2514:\,563402549}), the Einstein Foundation (Einstein Research Unit on Quantum Devices), Berlin Quantum, and the ERC (DebuQC).
This research was sponsored by IARPA and the Army Research Office, under the Entangled Logical Qubits program, and was accomplished under Cooperative Agreement Number W911NF-23-2-0212. The views and conclusions contained in this document are those of the authors and should not be interpreted as representing the official policies, either expressed or implied, of IARPA, the Army Research Office, or the U.S. Government. The U.S. Government is authorized to reproduce and distribute reprints for Government purposes notwithstanding any copyright notation herein. 

The Innsbruck team acknowledges funding by the European Union under the European Research Council (ERC, QUDITS, 101039522), by the European Union’s Horizon Europe research and innovation programme under grant agreement No 101114305 (``MILLENION-SGA1'' EU Project), and by the Office of the Director of National Intelligence (ODNI), Intelligence Advanced Research Projects Activity (IARPA), via US Army Research Office (ARO) grant No.\ W911NF-21-1-0007. Views and opinions expressed are however those of the authors only and do not necessarily reflect those of the European Union or the European Research Council Executive Agency. Neither the European Union nor the granting authority can be held responsible for them. We also acknowledge support by the Austrian Science Fund (FWF Grant-DOI 10.55776/F71) (SFB BeyondC), the Austrian Research Promotion Agency under Contracts Number 897481 (HPQC), and the Institut für Quanteninformation GmbH.

\end{acknowledgments}

\bibliography{references}

@book{nielsen_quantum_computation_2000,
  author = {Nielsen, M. A. and Chuang, I. L.},
  publisher = {Cambridge University Press},
  title = {{Quantum Computation and Quantum Information}},
  opturl = {https://doi.org/10.1017/CBO9780511976667},
  doi = {10.1017/CBO9780511976667},
  year = {2000}
}

@book{lidar_quantum_error_2013,
  title={{Quantum error correction}},
  author={Lidar, D. A. and Brun, T. A.},
  year={2013},
  publisher={Cambridge University Press}
}

@article{terhal_quantum_error_2015,
  title = {Quantum error correction for quantum memories},
  author = {Terhal, B. M.},
  journal = {Rev. Mod. Phys.},
  volume = {87},
  issue = {2},
  pages = {307--346},
  numpages = {40},
  year = {2015},
  month = {Apr},
  publisher = {American Physical Society},
  doi = {10.1103/RevModPhys.87.307}
}

@article{campbell_roads_towards_2017, 
    title ={Roads towards fault-tolerant universal quantum computation},
    author={E. T. Campbell and B. M. Terhal and C. Vuillot},
    doi={10.1038/nature23460},
    journal={Nature},
    volume={549},
    pages={172-179}, 
    year={2017}
}

@article{emerson_scalable_noise_2005,
doi = {10.1088/1464-4266/7/10/021},
opturl = {https://dx.doi.org/10.1088/1464-4266/7/10/021},
year = {2005},
month = {sep},
publisher = {},
volume = {7},
number = {10},
pages = {S347},
author = {J. Emerson and R. Alicki and K. Zyczkowski},
title = {Scalable noise estimation with random unitary operators},
journal = {J. Opt. B},
}

@article{eisert_quantum_certification_2020,
   title={Quantum certification and benchmarking},
   volume={2},
   ISSN={2522-5820},
   opturl={http://dx.doi.org/10.1038/s42254-020-0186-4},
   DOI={10.1038/s42254-020-0186-4},
   number={7},
   journal={Nature Rev. Phys.},
   publisher={Springer Science and Business Media LLC},
   author={Eisert, J. and Hangleiter, D. and Walk, N. and Roth, I. and Markham, D. and Parekh, R. and Chabaud, U. and Kashefi, E.},
   year={2020},
   month=jun, pages={382} 
}

@article{kliesch_theory_of_2021,
  title = {Theory of Quantum System Certification},
  author = {Kliesch, M. and Roth, I.},
  journal = {PRX Quantum},
  volume = {2},
  issue = {1},
  pages = {010201},
  numpages = {53},
  year = {2021},
  month = {Jan},
  publisher = {American Physical Society},
  doi = {10.1103/PRXQuantum.2.010201},
  opturl = {https://link.aps.org/doi/10.1103/PRXQuantum.2.010201}
}

@article{helsen_general_framework_2022,
  title = {General Framework for Randomized Benchmarking},
  author = {Helsen, J. and Roth, I. and Onorati, E. and Werner, A.H. and Eisert, J.},
  journal = {PRX Quantum},
  volume = {3},
  issue = {2},
  pages = {020357},
  numpages = {54},
  year = {2022},
  month = {Jun},
  publisher = {American Physical Society},
  doi = {10.1103/PRXQuantum.3.020357},
  opturl = {https://link.aps.org/doi/10.1103/PRXQuantum.3.020357}
}

@article{eisert_mind_the_2025,
      title={{\add Mind the gaps: The fraught road to quantum advantage}}, 
      author={{\add J. Eisert and J. Preskill}},
      year={2025},
      eprint={2510.19928},
      archivePrefix={arXiv},
      url={https://arxiv.org/abs/2510.19928}, 
}

@article{flammia_direct_fidelity_2011,
  title = {{Direct fidelity estimation from few Pauli measurements}},
  author = {Flammia, S. T. and Liu, Y.-K.},
  journal = {Phys. Rev. Lett.},
  volume = {106},
  issue = {23},
  pages = {230501},
  numpages = {4},
  year = {2011},
  month = {Jun},
  publisher = {American Physical Society},
  doi = {10.1103/PhysRevLett.106.230501},
  opturl = {https://link.aps.org/doi/10.1103/PhysRevLett.106.230501}
}

@article{mooney_generation_and_2021,
doi = {10.1088/2399-6528/ac1df7},
opturl = {https://dx.doi.org/10.1088/2399-6528/ac1df7},
year = {2021},
month = {sep},
publisher = {IOP Publishing},
volume = {5},
number = {9},
pages = {095004},
author = {G. J. Mooney and G. A.~L. White and C. D. Hill and L. C.~L. Hollenberg},
title = {{Generation and verification of 27-qubit Greenberger-Horne-Zeilinger states in a superconducting quantum computer}},
journal = {J. Phys. Commun.},
}

@article{moses_a_race_2023,
  title = {A Race-Track Trapped-Ion Quantum Processor},
  author = {Moses, S. A. and Baldwin, C. H. and Allman, M. S. and Ancona, R. and Ascarrunz, L. and others},
  journal = {Phys. Rev. X},
  volume = {13},
  issue = {4},
  pages = {041052},
  numpages = {25},
  year = {2023},
  month = {Dec},
  publisher = {American Physical Society},
  doi = {10.1103/PhysRevX.13.041052},
  opturl = {https://link.aps.org/doi/10.1103/PhysRevX.13.041052}
}

@article{kam_characterization_of_2024,
  title = {Characterization of entanglement on superconducting quantum computers of up to 414 qubits},
  author = {Kam, J. F. and Kang, H. and Hill, C. D. and Mooney, G. J. and Hollenberg, L. C. L.},
  journal = {Phys. Rev. Res.},
  volume = {6},
  issue = {3},
  pages = {033155},
  numpages = {23},
  year = {2024},
  month = {Aug},
  publisher = {American Physical Society},
  doi = {10.1103/PhysRevResearch.6.033155},
  opturl = {https://link.aps.org/doi/10.1103/PhysRevResearch.6.033155}
}

@article{javadiabhari_big_cats_2025,
      title={{\add Big cats: entanglement in 120 qubits and beyond}}, 
      author={{\add A. Javadi-Abhari and S. Martiel and A. Seif and M. Takita and K. X. Wei}},
      year={2025},
      eprint={2510.09520},
      archivePrefix={arXiv},
      optprimaryClass={quant-ph},
      url={https://arxiv.org/abs/2510.09520}, 
}

@article{bermudez_assessing_the_2017,
  title = {Assessing the Progress of Trapped-Ion Processors Towards Fault-Tolerant Quantum Computation},
  author = {Bermudez, A. and Xu, X. and Nigmatullin, R. and O'Gorman, J. and Negnevitsky, V. and Schindler, P. and Monz, T. and Poschinger, U. G. and Hempel, C. and Home, J. and Schmidt-Kaler, F. and Biercuk, M. and Blatt, R. and Benjamin, S. and M\"uller, M.},
  journal = {Phys. Rev. X},
  volume = {7},
  issue = {4},
  pages = {041061},
  numpages = {41},
  year = {2017},
  month = {Dec},
  publisher = {American Physical Society},
  doi = {10.1103/PhysRevX.7.041061},
  opturl = {https://link.aps.org/doi/10.1103/PhysRevX.7.041061}
}

@article{egan_fault_tolerant_2021,
  author = {Egan, L. and Debroy, D. M. and Noel, C. and Risinger, A. and Zhu, D. and Biswas, D. and Newman, M. and Li, M. and Brown, K. R. and Cetina, M. and Monroe, C.},
  title = {Fault-tolerant control of an error-corrected qubit},
  journal = {Nature},
  year = {2021},
  volume = {598},
  number = {7880},
  pages = {281--286},
  month = {Oct},
  doi = {10.1038/s41586-021-03928-y},
  opturl = {https://doi.org/10.1038/s41586-021-03928-y},
  issn = {1476-4687},
  publisher = {Nature Publishing Group},
}

@article{ryan_anderson_realization_of_2021,
  title = {{Realization of real-time fault-tolerant quantum error correction}},
  author = {R.-Anderson, C. and Bohnet, J. G. and Lee, K. and Gresh, D. and Hankin, A. and others},
  journal = {Phys. Rev. X},
  volume = {11},
  issue = {4},
  pages = {041058},
  numpages = {29},
  year = {2021},
  month = {Dec},
  publisher = {American Physical Society},
  doi = {10.1103/PhysRevX.11.041058},
  opturl = {https://link.aps.org/doi/10.1103/PhysRevX.11.041058}
}

@article{krinner_realizing_repeated_2022,
    title={Realizing repeated quantum error correction in a distance-three surface code},
    doi={10.1038/s41586-022-04566-8},
    author={S. Krinner and N. Lacroix and A. Remm and A. Di Paolo and E. Genois and others},
    journal={Nature}, 
    volume=605,
    pages={669-674},
    year=2022
}

@article{google_suppressing_quantum_2023,
    title={Suppressing quantum errors by scaling a surface code logical qubit},
    author={{Google Quantum AI}},
    doi={10.1038/s41586-022-05434-1},
    journal={Nature},
    volume={614},
    pages={676-681},
    year={2023}
}

@article{gupta_encoding_a_2024,
  author = {Gupta, R. S. and Sundaresan, N. and Alexander, T. and Wood, C. J. and Merkel, S. T. and others},
  title = {Encoding a magic state with beyond break-even fidelity},
  journal = {Nature},
  volume = {625},
  number = {7994},
  pages = {259--263},
  year = {2024},
  month = {jan},
  doi = {10.1038/s41586-023-06846-3},
  opturl = {https://doi.org/10.1038/s41586-023-06846-3},
  issn = {1476-4687},
  id = {Gupta2024}
}

@article{bluvstein_logical_quantum_2024,
    author = {Bluvstein, D. and Evered, S. J. and Geim, A. A. and Li, S. H. and Zhou, H. and others},
    title = {Logical quantum processor based on reconfigurable atom arrays},
    journal = {Nature},
    volume = {626},
    number = {7997},
    pages = {58--65},
    year = {2024},
    doi = {10.1038/s41586-023-06927-3},
    opturl = {https://doi.org/10.1038/s41586-023-06927-3},
    issn = {1476-4687},
    date = {2024-02-01},
}

@article{chiu_continuous_operation_2025,
  author    = {{\add N.-C. Chiu and E. C. Trapp and J. Guo and M. H. Abobeih and L. M. Stewart and S. Hollerith and P. L. Stroganov and M. Kalinowski and A. A. Geim and S. J. Evered and S. H. Li and X. Lyu and L. M. Peters and D. Bluvstein and T. T. Wang and M. Greiner and V. Vuleti{\'c} and M. D. Lukin}},
  title     = {{\add Continuous operation of a coherent 3,000-qubit system}},
  journal   = {Nature},
  year      = {2025},
  volume=646,
  doi       = {10.1038/s41586-025-09596-6},
  pages={1075–1080}
}

@article{acharya_quantum_error_2025,
  author  = {Acharya, R. and others},
  title   = {Quantum error correction below the surface code threshold},
  journal = {Nature},
  year    = {2025},
  volume  = {638},
  number  = {8052},
  pages   = {920--926},
  doi     = {10.1038/s41586-024-08449-y},
  url     = {https://doi.org/10.1038/s41586-024-08449-y},
  issn    = {1476-4687}
}

@article{ohliger_efficient_and_2013,
    doi = {10.1088/1367-2630/15/1/015024},
    opturl = {https://dx.doi.org/10.1088/1367-2630/15/1/015024},
    year = {2013},
    month = {jan},
    publisher = {IOP Publishing},
    volume = {15},
    number = {1},
    pages = {015024},
    author = {M. Ohliger and V. Nesme and J. Eisert},
    title = {Efficient and feasible state tomography of quantum many-body systems},
    journal = {New J. Phys.},
}

@inproceedings{aaronson_shadow_tomography_2018,
    author = {Aaronson, S.},
    title = {Shadow tomography of quantum states},
    year = {2018},
    isbn = {9781450355599},
    publisher = {Association for Computing Machinery},
    address = {New York, NY, USA},
    opturl = {https://doi.org/10.1145/3188745.3188802},
    doi = {10.1145/3188745.3188802},
    booktitle = {Proceedings of the 50th Annual ACM SIGACT Symposium on Theory of Computing},
    pages = {325},
    numpages = {14},
    location = {Los Angeles, CA, USA},
    series = {STOC 2018}
}

@article{huang_predicting_many_2020,
    author = {Huang, H.-Y. and  Kueng, R. and Preskill, J.},
    year = {2020},
    title = {{Predicting many properties of a quantum system from very few measurements}},
    journal = {Nature Phys.},
    pages = {1050},
    volume = {16},
    issue = {10},
    doi ={10.1038/s41567-020-0932-7}
}

@article{elben_the_randomized_2023,
  author    = {A. Elben and S. T. Flammia and H.-Y. Huang and R. Kueng and J. Preskill and B. Vermersch and P. Zoller},
  title     = {The randomized measurement toolbox},
  journal={Nature Rev. Phys.}, 
  year      = {2023},
  volume    = {5},
  number    = {1},
  pages     = {9--24},
  doi       = {10.1038/s42254-022-00535-2},
  opturl       = {https://doi.org/10.1038/s42254-022-00535-2}
}

@INPROCEEDINGS{dalvi_one_time_2024,
  author={{\add A. S. Dalvi   and J. Whitlow and  M. D'Onofrio and L. Riesebos and  T. Chen  and  S. Phiri and K. R. Brown and J. M. Baker}},
  booktitle={IEEE Trans. Quantum Eng.}, 
  title={{\add One-time compilation of device-level instructions for quantum subroutines}}, 
  year={2024},
  volume={01},
  number={},
  pages={873-884},
  doi={10.1109/QCE60285.2024.00107}}

@misc{montanaro_learning_stabilizer_2017,
      title={{Learning stabilizer states by Bell sampling}}, 
      author={A. Montanaro},
      year={2017},
      eprint={1707.04012},
      archivePrefix={arXiv}
}

@misc{gottesman_talk_2008,
  author       = {{\add S. Aaronson and D. Gottesman}},
  title        = {{\add Identifying stabilizer states}},
  year         = {2008},
  howpublished = {\url{http://pirsa.org/08080052/}},
  note         = {{\add Perimeter Institute Recorded Seminar Archive}}
}

@INPROCEEDINGS{chen_exponential_separations_2022,
  author={Chen, S. and Cotler, J. and Huang, H.-Y. and Li, J.},
  booktitle={2021 IEEE 62nd Annual Symposium on Foundations of Computer Science (FOCS)}, 
  title={Exponential Separations Between Learning With and Without Quantum Memory}, 
  year={2022},
  volume={},
  number={},
  pages={574-585},
  doi={10.1109/FOCS52979.2021.00063}
}

@article{huang_quantum_advantage_2022,
author = {H.-Y. Huang  and M. Broughton  and J. Cotler  and S. Chen  and J. Li  and M. Mohseni  and H. Neven  and R. Babbush  and R. Kueng  and J. Preskill  and J. R. McClean },
title = {Quantum advantage in learning from experiments},
journal = {Science},
volume = {376},
number = {6598},
pages = {1182-1186},
year = {2022},
doi = {10.1126/science.abn7293},
opturl = {https://www.science.org/doi/abs/10.1126/science.abn7293},
}

@article{aharonov_quantum_algorithmic_2022,
  author = {Aharonov, D. and Cotler, J. and Qi, X.-L.},
  title = {Quantum algorithmic measurement},
  journal = {Nature Comm.},
  year = {2022},
  volume = {13},
  number = {1},
  pages = {887},
  doi = {10.1038/s41467-021-27922-0},
  opturl = {https://doi.org/10.1038/s41467-021-27922-0},
  issn = {2041-1723}
}

@misc{chen_a_hierarchy_2021,
      title={A Hierarchy for Replica Quantum Advantage}, 
      author={S. Chen and J. Cotler and H.-Y. Huang and J. Li},
      year={2021},
      eprint={2111.05874},
      archivePrefix={arXiv}
}

@article{king_triply_efficient_2025,
  title = {Triply Efficient Shadow Tomography},
  author = {King, Robbie and Gosset, David and Kothari, Robin and Babbush, Ryan},
  journal = {PRX Quantum},
  volume = {6},
  issue = {1},
  pages = {010336},
  numpages = {25},
  year = {2025},
  month = {Feb},
  publisher = {American Physical Society},
  doi = {10.1103/PRXQuantum.6.010336},
  url = {https://link.aps.org/doi/10.1103/PRXQuantum.6.010336}
}

@article{huang_information_theoretic_2021,
  title = {{Information-theoretic bounds on quantum advantage in machine learning}},
  author = {Huang, H.-Y. and Kueng, R. and Preskill, J.},
  journal = {Phys. Rev. Lett.},
  volume = {126},
  issue = {19},
  pages = {190505},
  numpages = {7},
  year = {2021},
  month = {May},
  publisher = {American Physical Society},
  doi = {10.1103/PhysRevLett.126.190505},
  opturl = {https://link.aps.org/doi/10.1103/PhysRevLett.126.190505}
}

@article{scali_the_topology_2024,
   title={The topology of data hides in quantum thermal states},
   volume={1},
   pages={036106},
   DOI={10.1063/5.0209201},
   number={3},
   journal={APL Quantum},
   publisher={AIP Publishing},
   author={Scali, S. and Umeano, C. and Kyriienko, O.},
   year={2024},
   month=jul }

@article{hangleiter_bell_sampling_2024,
  title = {{Bell sampling from quantum circuits}},
  author = {Hangleiter, D. and Gullans, M. J.},
  journal = {Phys. Rev. Lett.},
  volume = {133},
  issue = {2},
  pages = {020601},
  numpages = {7},
  year = {2024},
  month = {Jul},
  publisher = {American Physical Society},
  doi = {10.1103/PhysRevLett.133.020601},
  opturl = {https://link.aps.org/doi/10.1103/PhysRevLett.133.020601}
}

@article{gross_schur_weyl_2021,
  author = {D. Gross and S. Nezami and M. Walter},
  title = {{Schur-Weyl duality for the Clifford group with applications: Property testing, a robust Hudson theorem, and de Finetti representations}},
  journal = {Commun. Math. Phys.},
  year = {2021},
  volume = {385},
  number = {3},
  pages = {1325--1393},
  month = {Aug},
  doi = {10.1007/s00220-021-04118-7},
  opturl = {https://doi.org/10.1007/s00220-021-04118-7},
  issn = {1432-0916},
  publisher = {Springer},
}

@article{haug_scalable_measures_2023,
  title = {Scalable measures of magic resource for quantum computers},
  author = {Haug, T. and Kim, M. S.},
  journal = {PRX Quantum},
  volume = {4},
  issue = {1},
  pages = {010301},
  numpages = {23},
  year = {2023},
  month = {Jan},
  publisher = {American Physical Society},
  doi = {10.1103/PRXQuantum.4.010301},
  opturl = {https://link.aps.org/doi/10.1103/PRXQuantum.4.010301}
}

@article{leone_learning_t_2024,
  doi = {10.22331/q-2024-05-27-1361},
  opturl = {https://doi.org/10.22331/q-2024-05-27-1361},
  title = {{Learning $T$-doped stabilizer states}},
  author = {Leone, L. and Oliviero, S. F. E. and Hamma, A.},
  journal = {{Quantum}},
  issn = {2521-327X},
  publisher = {{Verein zur F{\"{o}}rderung des Open Access Publizierens in den Quantenwissenschaften}},
  volume = {8},
  pages = {1361},
  month = {May},
  year = {2024}
}

@article{haug_efficient_quantum_2024,
  title = {Efficient quantum algorithms for stabilizer entropies},
  author = {Haug, T. and Lee, S. and Kim, M. S.},
  journal = {Phys. Rev. Lett.},
  volume = {132},
  issue = {24},
  pages = {240602},
  numpages = {7},
  year = {2024},
  month = {Jun},
  publisher = {American Physical Society},
  doi = {10.1103/PhysRevLett.132.240602},
  opturl = {https://link.aps.org/doi/10.1103/PhysRevLett.132.240602}
}

@article{bittel_a_complete_2025,
      title={{\add A complete theory of the Clifford commutant}}, 
      author={{\add L. Bittel and J. Eisert and L. Leone and A. A. Mele and S. F. E. Oliviero}},
      year={2025},
      eprint={2504.12263},
      archivePrefix={arXiv},
      optprimaryClass={quant-ph},
      url={https://arxiv.org/abs/2504.12263}, 
}

@article{bittel_operational_interpretation_2025,
      title={{\add Operational interpretation of the stabilizer entropy}}, 
      author={{\add L. Bittel and L. Leone}},
      year={2025},
      eprint={2507.22883},
      archivePrefix={arXiv},
      optprimaryClass={quant-ph},
      url={https://arxiv.org/abs/2507.22883}, 
}

@article{cotler_quantum_virtual_2019,
  title = {Quantum virtual cooling},
  author = {Cotler, J. and Choi, S. and Lukin, A. and Gharibyan, H. and Grover, T. and Tai, M. . and Rispoli, M. and Schittko, R. and Preiss, P. M. and Kaufman, A. M. and Greiner, M. and Pichler, H. and Hayden, P.},
  journal = {Phys. Rev. X},
  volume = {9},
  issue = {3},
  pages = {031013},
  numpages = {11},
  year = {2019},
  month = {Jul},
  publisher = {American Physical Society},
  doi = {10.1103/PhysRevX.9.031013},
  opturl = {https://link.aps.org/doi/10.1103/PhysRevX.9.031013}
}

@article{koczor_exponential_error_2021,
  title = {Exponential error suppression for near-term quantum devices},
  author = {Koczor, B.},
  journal = {Phys. Rev. X},
  volume = {11},
  issue = {3},
  pages = {031057},
  numpages = {30},
  year = {2021},
  month = {Sep},
  publisher = {American Physical Society},
  doi = {10.1103/PhysRevX.11.031057},
  opturl = {https://link.aps.org/doi/10.1103/PhysRevX.11.031057}
}

@article{huggins_virtual_distillation_2021,
  title = {Virtual Distillation for Quantum Error Mitigation},
  author = {Huggins, W. J. and McArdle, S. and O'Brien, T. E. and Lee, J. and Rubin, N. C. and Boixo, S. and Whaley, K. B. and Babbush, R. and McClean, J. R.},
  journal = {Phys. Rev. X},
  volume = {11},
  issue = {4},
  pages = {041036},
  numpages = {25},
  year = {2021},
  publisher = {American Physical Society},
  doi = {10.1103/PhysRevX.11.041036},
  opturl = {https://link.aps.org/doi/10.1103/PhysRevX.11.041036}
}

@article{hakoshima_localized_virtual_2024,
  title = {Localized virtual purification},
  author = {Hakoshima, H. and Endo, S. and Yamamoto, K. and Matsuzaki, Y. and Yoshioka, N.},
  journal = {Phys. Rev. Lett.},
  volume = {133},
  issue = {8},
  pages = {080601},
  numpages = {7},
  year = {2024},
  month = {Aug},
  publisher = {American Physical Society},
  doi = {10.1103/PhysRevLett.133.080601},
  opturl = {https://link.aps.org/doi/10.1103/PhysRevLett.133.080601}
}

@article{shor_scheme_for_1995,
  title = {Scheme for reducing decoherence in quantum computer memory},
  author = {Shor, P. W.},
  journal = {Phys. Rev. A},
  volume = {52},
  issue = {4},
  pages = {R2493--R2496},
  numpages = {0},
  year = {1995},
  month = {Oct},
  publisher = {American Physical Society},
  doi = {10.1103/PhysRevA.52.R2493},
  opturl = {https://link.aps.org/doi/10.1103/PhysRevA.52.R2493}
}

@article{laflamme_perfect_quantum_1996,
  title = {{Perfect quantum error correcting code}},
  author = {Laflamme, R. and Miquel, C. and Paz, J. P. and Zurek, W. H.},
  journal = {Phys. Rev. Lett.},
  volume = {77},
  issue = {1},
  pages = {198--201},
  numpages = {0},
  year = {1996},
  month = {Jul},
  publisher = {American Physical Society},
  doi = {10.1103/PhysRevLett.77.198},
  opturl = {https://link.aps.org/doi/10.1103/PhysRevLett.77.198}
}

@article{steane_multiple_particle_1996, 
author={Steane, A.}, 
year=1996,
title={Multiple-particle interference and quantum error correction}, 
journal={Proc. R. Soc. Lond. A},
volume=452,
pages={2551},
doi={10.1098/rspa.1996.0136}
}

@article{steane_error_correcting_1996,
  title = {{Error correcting codes in quantum theory}},
  author = {Steane, A. M.},
  journal = {Phys. Rev. Lett.},
  volume = {77},
  issue = {5},
  pages = {793--797},
  numpages = {0},
  year = {1996},
  month = {Jul},
  publisher = {American Physical Society},
  doi = {10.1103/PhysRevLett.77.793},
  opturl = {https://link.aps.org/doi/10.1103/PhysRevLett.77.793}
}

@article{knill_theory_of_1997,
  title = {Theory of quantum error-correcting codes},
  author = {Knill, E. and Laflamme, R.},
  journal = {Phys. Rev. A},
  volume = {55},
  issue = {2},
  pages = {900--911},
  numpages = {0},
  year = {1997},
  month = {Feb},
  publisher = {American Physical Society},
  doi = {10.1103/PhysRevA.55.900} 
}

@article{shor_quantum_analog_1997,
  title = {{Quantum analog of the MacWilliams identities for classical coding theory}},
  author = {Shor, P. and Laflamme, R.},
  journal = {Phys. Rev. Lett.},
  volume = {78},
  issue = {8},
  pages = {1600--1602},
  numpages = {0},
  year = {1997},
  publisher = {American Physical Society},
  doi = {10.1103/PhysRevLett.78.1600},
  opturl = {https://link.aps.org/doi/10.1103/PhysRevLett.78.1600}
}

@ARTICLE{calderbank_quantum_error_1998,
  author={Calderbank, A. R. and Rains, E. M. and Shor, P. M. and Sloane, N. J. A.},
  journal={IEEE Trans. Inf. Th.}, 
  title={{Quantum error correction via codes over GF(4)}}, 
  year={1998},
  volume={44},
  number={4},
  pages={1369-1387},
  doi={10.1109/18.681315}
}

@ARTICLE{rains_quantum_weight_1998,
  author={Rains, E. M.},
  journal={IEEE Trans. Inf. Th.}, 
  title={Quantum weight enumerators}, 
  year={1998},
  volume={44},
  number={4},
  pages={1388-1394},
  doi={10.1109/18.681316}}

@ARTICLE{rains_quantum_shadow_1999,
  author={Rains, E. M.},
  journal={IEEE Trans. Inf. Th.}, 
  title={Quantum shadow enumerators}, 
  year={1999},
  volume={45},
  number={7},
  pages={2361-2366},
  doi={10.1109/18.796376}
}

@article{macwilliams_a_theorem_1962,
  author    = {F. J. MacWilliams},
  title     = {{A theorem on the distribution of weights in a systematic code}},
  journal   = {The Bell System Tech. J.},
  volume    = {42},
  pages     = {79--94},
  year      = {1963},
  publisher = {Bell Telephone Laboratories, Inc.},
  doi       = {10.1002/j.1538-7305.1963.tb04003.x}
}

@book{macwilliams_the_theory_1978,
  author = {MacWilliams, F. J. and Sloane, N. J. A.},
  bibopturl = {https://www.bibsonomy.org/bibtex/2a53f5d82eee126a9b3fbb3e8e5932ffb/heartsoar},
  edition = {2nd},
  publisher = {North-Holland Publishing Company},
  title = {{The theory of error-correcting codes}},
  year = {1978}
}

@book{pless_handbook_of_1998,
  title={Handbook of coding theory},
  editor={Pless, V. S. and Huffman, W. C.},
  year={1998},
  publisher={Elsevier},
  address={Amsterdam, Netherlands},
  isbn={978-0-444-50423-3},
  volumes={2}
}

@book{nebe_self_dual_2006,
  title     = {Self-dual codes and invariant theory},
  author    = {G. Nebe and E. M. Rains and N. J. A. Sloane},
  series    = {Alg. Comp. Math.},
  volume    = {17},
  publisher = {Springer},
  year      = {2006},
  isbn      = {978-3-540-28451-5},
  doi       = {10.1007/3-540-30731-1},
  opturl       = {https://doi.org/10.1007/3-540-30731-1}
}

@article{bravyi_high_threshold_2024,
  title = {High-threshold and low-overhead fault-tolerant quantum memory},
  author = {Bravyi, S. and Cross, A. W. and Gambetta, J. M. and Maslov, D. and Rall, P. and Yoder, T. J.},
  journal = {Nature},
  volume = {627},
  number = {8005},
  pages = {778--782},
  year = {2024},
  doi = {10.1038/s41586-024-07107-7},
  opturl = {https://doi.org/10.1038/s41586-024-07107-7}
}

@article{horodecki_method_for_2002,
  title = {{Method for direct detection of quantum entanglement}},
  author = {Horodecki, P. and Ekert, A.},
  journal = {Phys. Rev. Lett.},
  volume = {89},
  issue = {12},
  pages = {127902},
  numpages = {4},
  year = {2002},
  month = {Aug},
  publisher = {American Physical Society},
  doi = {10.1103/PhysRevLett.89.127902},
  opturl = {https://link.aps.org/doi/10.1103/PhysRevLett.89.127902}
}

@article{horodecki_quantum_entanglement_2009,
  title = {Quantum entanglement},
  author = {Horodecki, R. and Horodecki, P. and Horodecki, M. and Horodecki, K.},
  journal = {Rev. Mod. Phys.},
  volume = {81},
  issue = {2},
  pages = {865--942},
  numpages = {0},
  year = {2009},
  month = {Jun},
  publisher = {American Physical Society},
  doi = {10.1103/RevModPhys.81.865},
  opturl = {https://link.aps.org/doi/10.1103/RevModPhys.81.865}
}

@article{wong_potential_multipartite_2001,
  title = {Potential multiparticle entanglement measure},
  author = {Wong, A. and Christensen, N.},
  journal = {Phys. Rev. A},
  volume = {63},
  issue = {4},
  pages = {044301},
  numpages = {4},
  year = {2001},
  month = {Mar},
  publisher = {American Physical Society},
  doi = {10.1103/PhysRevA.63.044301},
  opturl = {https://link.aps.org/doi/10.1103/PhysRevA.63.044301}
}

@article{scott_multipartite_entanglement_2004,
  title = {Multipartite entanglement, quantum-error-correcting codes, and entangling power of quantum evolutions},
  author = {Scott, A. J.},
  journal = {Phys. Rev. A},
  volume = {69},
  issue = {5},
  pages = {052330},
  numpages = {10},
  year = {2004},
  month = {May},
  publisher = {American Physical Society},
  doi = {10.1103/PhysRevA.69.052330},
  opturl = {https://link.aps.org/doi/10.1103/PhysRevA.69.052330}
}

@article{carvalho_decoherence_and_2004,
  title = {{Decoherence and multipartite entanglement}},
  author = {Carvalho, A. R. R. and Mintert, F. and Buchleitner, A.},
  journal = {Phys. Rev. Lett.},
  volume = {93},
  issue = {23},
  pages = {230501},
  numpages = {4},
  year = {2004},
  publisher = {American Physical Society},
  doi = {10.1103/PhysRevLett.93.230501},
  opturl = {https://link.aps.org/doi/10.1103/PhysRevLett.93.230501}
}

@article{mintert_concurrence_of_2005,
    title = {{Concurrence of mixed multipartite quantum states}},
    author = {Mintert, F. and Ku\ifmmode \acute{s}\else \'{s}\fi{}, M. and Buchleitner, A.},
    journal = {Phys. Rev. Lett.},
    volume = {95},
    issue = {26},
    pages = {260502},
    numpages = {4},
    year = {2005},
    publisher = {American Physical Society},
    doi = {10.1103/PhysRevLett.95.260502},
    opturl = {https://link.aps.org/doi/10.1103/PhysRevLett.95.260502}
}

@article{mintert_observable_entanglement_2007,
  title = {{Observable entanglement measure for mixed quantum states}},
  author = {Mintert, F. and Buchleitner, A.},
  journal = {Phys. Rev. Lett.},
  volume = {98},
  issue = {14},
  pages = {140505},
  numpages = {3},
  year = {2007},
  publisher = {American Physical Society},
  doi = {10.1103/PhysRevLett.98.140505},
  opturl = {https://link.aps.org/doi/10.1103/PhysRevLett.98.140505}
}

@article{aolita_scalable_method_2008,
  title = {{Scalable method to estimate experimentally the entanglement of multipartite systems}},
  author = {Aolita, L. and Buchleitner, A. and Mintert, F.},
  journal = {Phys. Rev. A},
  volume = {78},
  issue = {2},
  pages = {022308},
  numpages = {4},
  year = {2008},
  publisher = {American Physical Society},
  doi = {10.1103/PhysRevA.78.022308},
  opturl = {https://link.aps.org/doi/10.1103/PhysRevA.78.022308}
}

@article{beckey_computable_and_2021,
  title = {{Computable and operationally meaningful multipartite entanglement measures}},
  author = {Beckey, J. L. and Gigena, N. and Coles, P. J. and Cerezo, M.},
  journal = {Phys. Rev. Lett.},
  volume = {127},
  issue = {14},
  pages = {140501},
  numpages = {7},
  year = {2021},
  publisher = {American Physical Society},
  doi = {10.1103/PhysRevLett.127.140501},
  opturl = {https://link.aps.org/doi/10.1103/PhysRevLett.127.140501}
}

@article{cullen_calculating_concentratable_2022,
  title = {Calculating concentratable entanglement in graph states},
  author = {Cullen, A. R. and Kok, P.},
  journal = {Phys. Rev. A},
  volume = {106},
  issue = {4},
  pages = {042411},
  numpages = {9},
  year = {2022},
  month = {Oct},
  publisher = {American Physical Society},
  doi = {10.1103/PhysRevA.106.042411},
  opturl = {https://link.aps.org/doi/10.1103/PhysRevA.106.042411}
}

@article{beckey_multipartite_entanglement_2023,
  title = {{Multipartite entanglement measures via Bell-basis measurements}},
  author = {J. L. Beckey and G. Pelegr\'{\i} and S. Foulds and N. J. Pearson},
  journal = {Phys. Rev. A},
  volume = {107},
  issue = {6},
  pages = {062425},
  numpages = {15},
  year = {2023},
  publisher = {American Physical Society},
  doi = {10.1103/PhysRevA.107.062425},
  opturl = {https://link.aps.org/doi/10.1103/PhysRevA.107.062425}
}

@article{schatzki_hierarchy_of_2024,
  title = {Hierarchy of multipartite correlations based on concentratable entanglement},
  author = {Schatzki, L. and Liu, G. and Cerezo, M. and Chitambar, E.},
  journal = {Phys. Rev. Res.},
  volume = {6},
  issue = {2},
  pages = {023019},
  numpages = {7},
  year = {2024},
  month = {Apr},
  publisher = {American Physical Society},
  doi = {10.1103/PhysRevResearch.6.023019},
  opturl = {https://link.aps.org/doi/10.1103/PhysRevResearch.6.023019}
}

@article{schmid_experimental_direct_2008,
  title = {{Experimental direct observation of mixed state entanglement}},
  author = {C. {Schmid et al.}},
  journal = {Phys. Rev. Lett.},
  volume = {101},
  issue = {26},
  pages = {260505},
  numpages = {4},
  year = {2008},
  publisher = {American Physical Society},
  doi = {10.1103/PhysRevLett.101.260505},
  opturl = {https://link.aps.org/doi/10.1103/PhysRevLett.101.260505}
}

@article{islam_measuring_entanglement_2015,
    author = {R. {Islam et al.}},
    year = {2015},
    title = {{Measuring entanglement entropy in a quantum many-body system}},
    journal = {Nature},
    pages = {77},
    volume = {528},
    issue = {7580},
    doi = {10.1038/nature15750}
}

@article{bluvstein_a_quantum_2022,
  author = {Bluvstein, D. and Levine, H. and Semeghini, G. and Wang, T. T. and Ebadi, S. and others},
  title = {A quantum processor based on coherent transport of entangled atom arrays},
  journal = {Nature},
  year = {2022},
  month = {April},
  volume = {604},
  number = {7906},
  pages = {451-456},
  doi = {10.1038/s41586-022-04592-6},
  opturl = {https://doi.org/10.1038/s41586-022-04592-6},
  issn = {1476-4687},
}

@article{ryan_anderson_implementing_fault_2022,
      title={{Implementing fault-tolerant entangling gates on the five-qubit code and the color code}}, 
      author = {C. Ryan-Anderson and N. C. Brown and M. S. Allman and B. Arkin and G. Asa-Attuah and others},
      year={2022},
      eprint={2208.01863},
      archivePrefix={arXiv},
      opturl={https://arxiv.org/abs/2208.01863}, 
}

@article{heussen_strategies_for_2023,
  title = {Strategies for a practical advantage of fault-tolerant circuit design in noisy trapped-ion quantum computers},
  author = {Heu\ss{}en, S. and Postler, L. and Rispler, M. and Pogorelov, I. and Marciniak, C. D. and Monz, T. and Schindler, P. and M\"uller, M.},
  journal = {Phys. Rev. A},
  volume = {107},
  issue = {4},
  pages = {042422},
  numpages = {33},
  year = {2023},
  month = {Apr},
  publisher = {American Physical Society},
  doi = {10.1103/PhysRevA.107.042422},
  opturl = {https://link.aps.org/doi/10.1103/PhysRevA.107.042422}
}

@article{postler_demonstration_of_2022,
  author = {L. Postler and S. Heußen and I. Pogorelov and M. Rispler and T. Feldker and M. Meth and C. D. Marciniak and R. 
  Stricker and M. Ringbauer and R. Blatt and P. Schindler and M. M{\"u}ller and T. Monz},
  title = {Demonstration of fault-tolerant universal quantum gate operations},
  journal = {Nature},
  year = {2022},
  month = {May},
  volume = {605},
  number = {7911},
  pages = {675--680},
  doi = {10.1038/s41586-022-04721-1},
  opturl = {https://doi.org/10.1038/s41586-022-04721-1}
}

@article{postler_demonstration_of_2024,
  title = {{Demonstration of fault-tolerant Steane quantum error correction}},
  author = {Postler, L. and Butt, F. and Pogorelov, I. and Marciniak, C. D. and Heu\ss{}en, S. and Blatt, R. and Schindler, P. and Rispler, M. and M\"uller, M. and Monz, T.},
  journal = {PRX Quantum},
  volume = {5},
  issue = {3},
  pages = {030326},
  numpages = {19},
  year = {2024},
  month = {Aug},
  publisher = {American Physical Society},
  doi = {10.1103/PRXQuantum.5.030326},
  opturl = {https://link.aps.org/doi/10.1103/PRXQuantum.5.030326}
}

@article{zen_quantum_circuit_2024,
  title = {Quantum circuit discovery for fault-tolerant logical state preparation with reinforcement learning},
  author = {R.~Zen and J.~Olle and L.~Colmenarez and M.~Puviani and M.~M\"uller  and F.~Marquardt},
  journal = {Phys. Rev. X},
  volume = {15},
  issue = {4},
  pages = {041012},
  numpages = {42},
  year = {2025},
  month = {Oct},
  publisher = {American Physical Society},
  doi = {10.1103/gqpr-dgz7},
  url = {https://link.aps.org/doi/10.1103/gqpr-dgz7}
}

@article{peham_automated_synthesis_2025,
  title = {Automated Synthesis of Fault-Tolerant State Preparation Circuits for Quantum Error-Correction Codes},
  author = {T.~Peham  and  L.~Schmid  and L.~Berent and M.~M\"uller and R.~Wille},
  journal = {PRX Quantum},
  volume = {6},
  issue = {2},
  pages = {020330},
  numpages = {32},
  year = {2025},
  month = {May},
  publisher = {American Physical Society},
  doi = {10.1103/PRXQuantum.6.020330},
  url = {https://link.aps.org/doi/10.1103/PRXQuantum.6.020330}
}

@misc{da_silva_demonstration_of_2024,
      title={Demonstration of logical qubits and repeated error correction with better-than-physical error rates}, 
      author = {M.~P.~da Silva and C.~R.~Anderson and J.~M.~Bello-Rivas and A.~Chernoguzov and J.~M.~Dreiling and others},
      year={2024},
      eprint={2404.02280},
      archivePrefix={arXiv},
      opturl={https://arxiv.org/abs/2404.02280}, 
}

@misc{mayer_benchmarking_logical_2024,
      title={{Benchmarking logical three-qubit quantum Fourier transform encoded in the Steane code on a trapped-ion quantum computer}}, 
      author = {K. Mayer and C. R.-Anderson and N. Brown and E. Durso-Sabina and C. H. Baldwin and others},
      year={2024},
      eprint={2404.08616},
      archivePrefix={arXiv},
      opturl={https://arxiv.org/abs/2404.08616}, 
}

@article{ryan_anderson_high_fidelity_2024,
author = {C.~Ryan-Anderson  and N.~C.~Brown  and C.~H.~Baldwin  and J.~M.~Dreiling  and C.~Foltz  and J.~P.~Gaebler  and others },
title = {High-fidelity teleportation of a logical qubit using transversal gates and lattice surgery},
journal = {Science},
volume = {385},
number = {6715},
pages = {1327-1331},
year = {2024},
doi = {10.1126/science.adp6016},
url = {https://www.science.org/doi/pdf/10.1126/science.adp6016},
}

@article{pogorelov_experimental_fault_2025,
  author    = {I. Pogorelov and F. Butt and L. Postler and Ch. D. Marciniak and P. Schindler and M. Müller and T. Monz},
  title     = {Experimental fault-tolerant code switching},
  journal   = {Nature Physics},
  volume    = {21},
  number    = {2},
  pages     = {298--303},
  year      = {2025},
  month     = feb,
  issn      = {1745-2481},
  doi       = {10.1038/s41567-024-02727-2},
  url       = {https://doi.org/10.1038/s41567-024-02727-2}
}

@misc{valentini_demonstration_of_2024,
      title={Demonstration of two-dimensional connectivity for a scalable error-corrected ion-trap quantum processor architecture}, 
      author = {M. Valentini and M. W. van Mourik and F. Butt and J. Wahl and M. Dietl and others},
      year={2024},
      eprint={2406.02406},
      archivePrefix={arXiv} 
}

@article{zhou_low_overhead_2025,
  author={H. Zhou and C. Zhao and M. Cain and D. Bluvstein and N.~Maskara and C. Duckering and H.-Y. Hu and S.-T. Wang and A. Kubica and M. D. Lukin},
  title     = {Low-overhead transversal fault tolerance for universal quantum computation},
  journal   = {Nature},
  year      = {2025},
  volume    = {646},
  number    = {8084},
  pages     = {303--308},
  doi       = {10.1038/s41586-025-09543-5},
  url       = {https://doi.org/10.1038/s41586-025-09543-5},
  issn      = {1476-4687}
}

@article{butt_fault_tolerant_2024,
  title = {Fault-Tolerant Code-Switching Protocols for Near-Term Quantum Processors},
  author = {Butt, F. and Heu\ss{}en, S. and Rispler, M. and M\"uller, M.},
  journal = {PRX Quantum},
  volume = {5},
  issue = {2},
  pages = {020345},
  numpages = {26},
  year = {2024},
  month = {May},
  publisher = {American Physical Society},
  doi = {10.1103/PRXQuantum.5.020345},
  opturl = {https://link.aps.org/doi/10.1103/PRXQuantum.5.020345}
}

@article{bravyi_homological_product_2014,
title={Homological product codes},
author={S. Bravyi and M. B. Hastings},
journal={Proc. of the 46th ACM Symp. Th.  Comp. (STOC 2014)}, 
pages={273-282},year=2014,
doi={10.1145/2591796.2591870}
}

@article{breuckmann_quantum_low_2021,
  title = {Quantum low-density parity-check codes},
  author = {Breuckmann, N. P. and Eberhardt, J. N.},
  journal = {PRX Quantum},
  volume = {2},
  issue = {4},
  pages = {040101},
  numpages = {19},
  year = {2021},
  month = {Oct},
  publisher = {American Physical Society},
  doi = {10.1103/PRXQuantum.2.040101},
  opturl = {https://link.aps.org/doi/10.1103/PRXQuantum.2.040101}
}

@inproceedings{panteleev_asymptotically_good_2022,
author = {Panteleev, P. and Kalachev, G.},
title = {{Asymptotically good quantum and locally testable classical LDPC codes}},
year = {2022},
publisher = {Association for Computing Machinery},
address = {New York, NY, USA},
doi = {10.1145/3519935.3520017},
booktitle = {Proceedings of the 54th Annual ACM SIGACT Symposium on Theory of Computing},
pages = {375},
numpages = {14},
location = {Rome, Italy},
series = {STOC 2022}
}

@ARTICLE{leverrier_decoding_quantum_2023,
  author={Leverrier, A. and Z{\'e}mor, G.},
  journal={IEEE Trans. Inf. Th.}, 
  title={{Decoding quantum Tanner codes}}, 
  year={2023},
  volume={69},
  number={8},
  pages={5100-5115},
  doi={10.1109/TIT.2023.3267945}}

@article{bombin_topological_quantum_2006a,
  title = {{Topological quantum distillation}},
  author = {Bombin, H. and Martin-Delgado, M. A.},
  journal = {Phys. Rev. Lett.},
  volume = {97},
  issue = {18},
  pages = {180501},
  numpages = {4},
  year = {2006},
  month = {Oct},
  publisher = {American Physical Society},
  doi = {10.1103/PhysRevLett.97.180501},
  opturl = {https://link.aps.org/doi/10.1103/PhysRevLett.97.180501}
}

@article{bombin_topological_quantum_2006b,
  title = {Topological quantum error correction with optimal encoding rate},
  author = {Bombin, H. and Martin-Delgado, M. A.},
  journal = {Phys. Rev. A},
  volume = {73},
  issue = {6},
  pages = {062303},
  numpages = {5},
  year = {2006},
  month = {Jun},
  publisher = {American Physical Society},
  doi = {10.1103/PhysRevA.73.062303},
  opturl = {https://link.aps.org/doi/10.1103/PhysRevA.73.062303}
}

@article{cabello_optimal_preparation_2011,
  title = {Optimal preparation of graph states},
  author = {Cabello, A. and Danielsen, L. E. and L\'opez-Tarrida, A. J. and Portillo, J. R.},
  journal = {Phys. Rev. A},
  volume = {83},
  issue = {4},
  pages = {042314},
  numpages = {7},
  year = {2011},
  month = {Apr},
  publisher = {American Physical Society},
  doi = {10.1103/PhysRevA.83.042314},
  opturl = {https://link.aps.org/doi/10.1103/PhysRevA.83.042314}
}

@article{kubischta_family_of_2023,
  title = {{Family of quantum codes with exotic transversal gates}},
  author = {Kubischta, E. and Teixeira, I.},
  journal = {Phys. Rev. Lett.},
  volume = {131},
  pages = {240601},
  numpages = {6},
  year = {2023},
  month = {Dec},
  publisher = {American Physical Society},
  doi = {10.1103/PhysRevLett.131.240601},
  opturl = {https://link.aps.org/doi/10.1103/PhysRevLett.131.240601}
}

@misc{kubischta_quantum_weight_2023,
      title={Quantum Weight Enumerators for Real Codes with $X$ and $Z$ Exactly Transversal}, 
      author={E. Kubischta and I. Teixeira and J. M. Silvester},
      year={2024},
      eprint={2306.12526},
      archivePrefix={arXiv},
      opturl={https://arxiv.org/abs/2306.12526}, 
}

@article{bravyi_how_much_2025,
  title = {{How much entanglement is needed for quantum error correction?}},
  author = {Bravyi, S. and Lee, D. and Li, Z. and Yoshida, B.},
  journal = {Phys. Rev. Lett.},
  volume = {134},
  issue = {21},
  pages = {210602},
  numpages = {6},
  year = {2025},
  month = {May},
  publisher = {American Physical Society},
  doi = {10.1103/PhysRevLett.134.210602},
  url = {https://link.aps.org/doi/10.1103/PhysRevLett.134.210602}
}

@inproceedings{gleason_weight_polynomials_1970,
  author    = {A. M. Gleason},
  title     = {{Weight polynomials of self-dual codes and the MacWilliams identities}},
  booktitle = {Actes du Congr\`es Internat.~Math.},
  volume    = {3},
  pages     = {211--215},
  year      = {1970},
  publisher = {Gauthier-Villars},
  address   = {Paris}
}

@article{huber_absolutely_maximally_2017,
  title = {{Absolutely maximally entangled states of seven qubits do not exist}},
  author = {Huber, F. and G\"uhne, O. and Siewert, J.},
  journal = {Phys. Rev. Lett.},
  volume = {118},
  issue = {20},
  pages = {200502},
  numpages = {5},
  year = {2017},
  month = {May},
  publisher = {American Physical Society},
  doi = {10.1103/PhysRevLett.118.200502},
  opturl = {https://link.aps.org/doi/10.1103/PhysRevLett.118.200502}
}

@article{aschauer_local_invariants_2004,
    title = {Local invariants for multi-partite entangled states allowing for a simple entanglement criterion},
    author = {H. Aschauer and J. Calsamiglia and M. Hein and H. J. Briegel},
    journal = {Quant. Inf. Comp.},
	volume = {4},
	pages = {383},
    opturl = {https://doi.org/10.48550/arXiv.quant-ph/0306048},
    doi = {10.48550/arXiv.quant-ph/0306048},
    year= {2004}
}

@article{de_vincente_multipartite_entanglement_2011,
  title = {Multipartite entanglement detection from correlation tensors},
  author = {de Vicente, J. I. and Huber, M.},
  journal = {Phys. Rev. A},
  volume = {84},
  issue = {6},
  pages = {062306},
  numpages = {10},
  year = {2011},
  month = {Dec},
  publisher = {American Physical Society},
  doi = {10.1103/PhysRevA.84.062306},
  opturl = {https://link.aps.org/doi/10.1103/PhysRevA.84.062306}
}

@article{laskowski_correlation_tensor_2011,
  title = {Correlation-tensor criteria for genuine multiqubit entanglement},
  author = {Laskowski, W. and Markiewicz, M. and Paterek, T. and \ifmmode \dot{Z}\else \.{Z}\fi{}ukowski, M.},
  journal = {Phys. Rev. A},
  volume = {84},
  issue = {6},
  pages = {062305},
  numpages = {8},
  year = {2011},
  month = {Dec},
  publisher = {American Physical Society},
  doi = {10.1103/PhysRevA.84.062305},
  opturl = {https://link.aps.org/doi/10.1103/PhysRevA.84.062305}
}

@article{kloeckl_characterizing_multipartite_2015,
  title = {Characterizing multipartite entanglement without shared reference frames},
  author = {Kl\"ockl, C. and Huber, M.},
  journal = {Phys. Rev. A},
  volume = {91},
  issue = {4},
  pages = {042339},
  numpages = {8},
  year = {2015},
  month = {Apr},
  publisher = {American Physical Society},
  doi = {10.1103/PhysRevA.91.042339},
  opturl = {https://link.aps.org/doi/10.1103/PhysRevA.91.042339}
}

@article{morelli_correlation_constraints_2024,
  title = {{Correlation constraints and the Bloch geometry of two qubits}},
  author = {Morelli, S. and Eltschka, C. and Huber, M. and Siewert, J.},
  journal = {Phys. Rev. A},
  volume = {109},
  issue = {1},
  pages = {012423},
  numpages = {8},
  year = {2024},
  month = {Jan},
  publisher = {American Physical Society},
  doi = {10.1103/PhysRevA.109.012423},
  opturl = {https://link.aps.org/doi/10.1103/PhysRevA.109.012423}
}

@article{wyderka_characterizing_quantum_2020,
doi = {10.1088/1751-8121/ab7f0a},
opturl = {https://dx.doi.org/10.1088/1751-8121/ab7f0a},
year = {2020},
month = {jul},
publisher = {IOP Publishing},
volume = {53},
number = {34},
pages = {345302},
author = {N. Wyderka and O. G{\"u}hne},
title = {Characterizing quantum states via sector lengths},
journal = {J. Phys. A},
}

@article{tran_correlations_between_2016,
  title = {Correlations between outcomes of random measurements},
  author = {Tran, M. C. and Daki\ifmmode \acute{c}\else \'{c}\fi{}, B. and Laskowski, W. and Paterek, T.},
  journal = {Phys. Rev. A},
  volume = {94},
  issue = {4},
  pages = {042302},
  numpages = {9},
  year = {2016},
  month = {Oct},
  publisher = {American Physical Society},
  doi = {10.1103/PhysRevA.94.042302},
  opturl = {https://link.aps.org/doi/10.1103/PhysRevA.94.042302}
}

@article{kaszlikowski_quantum_correlation_2008,
  title = {{Quantum correlation without classical correlations}},
  author = {Kaszlikowski, D. and Sen(De), A. and Sen, U. and Vedral, V. and Winter, A.},
  journal = {Phys. Rev. Lett.},
  volume = {101},
  issue = {7},
  pages = {070502},
  numpages = {4},
  year = {2008},
  month = {Aug},
  publisher = {American Physical Society},
  doi = {10.1103/PhysRevLett.101.070502},
  opturl = {https://link.aps.org/doi/10.1103/PhysRevLett.101.070502}
}

@article{eltschka_maximum_nbody_2020,
  doi = {10.22331/q-2020-02-10-229},
  opturl = {https://doi.org/10.22331/q-2020-02-10-229},
  title = {{Maximum {$N$}-body correlations do not in general imply genuine multipartite entanglement}},
  author = {Eltschka, C. and Siewert, J.},
  journal = {{Quantum}},
  issn = {2521-327X},
  publisher = {{Verein zur F{\"{o}}rderung des Open Access Publizierens in den Quantenwissenschaften}},
  volume = {4},
  pages = {229},
  year = {2020}
}

@article{hein_multiparticle_entanglement_2004,
  title = {Multiparty entanglement in graph states},
  author = {Hein, M. and Eisert, J. and Briegel, H. J.},
  journal = {Phys. Rev. A},
  volume = {69},
  issue = {6},
  pages = {062311},
  numpages = {20},
  year = {2004},
  month = {Jun},
  publisher = {American Physical Society},
  doi = {10.1103/PhysRevA.69.062311},
  opturl = {https://link.aps.org/doi/10.1103/PhysRevA.69.062311}
}

@INPROCEEDINGS{miller_graphstatevis_interactive_2021,
  author={Miller, M. and Miller, D.},
  booktitle={IEEE Trans. Quantum Eng.}, 
  title={{GraphStateVis: Interactive visual analysis of qubit graph states and their stabilizer groups}}, 
  year={2021},
  volume={},
  number={},
  pages={378-384},
  doi={10.1109/QCE52317.2021.00057}}

@article{miller_shor_laflamme_2023,
author = {D. Miller and D. Loss and I. Tavernelli and H. Kampermann and D. Bru{\ss} and N. Wyderka},
doi = {10.1088/1751-8121/ace8d4},
opturl = {https://dx.doi.org/10.1088/1751-8121/ace8d4},
year = {2023},
publisher = {IOP Publishing},
volume = {56},
number = {33},
pages = {335303},
title = {{Shor-Laflamme distributions of graph states and noise robustness of entanglement}},
journal = {J. Phys. A},
}

@article{hastings_measuring_renyi_2010,
  title = {{Measuring Renyi entanglement entropy in quantum Monte Carlo simulations}},
  author = {Hastings, M. B. and Gonz\'alez, I. and Kallin, A. B. and Melko, R. G.},
  journal = {Phys. Rev. Lett.},
  volume = {104},
  issue = {15},
  pages = {157201},
  numpages = {4},
  year = {2010},
  month = {Apr},
  publisher = {American Physical Society},
  doi = {10.1103/PhysRevLett.104.157201},
  opturl = {https://link.aps.org/doi/10.1103/PhysRevLett.104.157201}
}

@article{bridgeman_hand_waving_2017,
doi = {10.1088/1751-8121/aa6dc3},
opturl = {https://dx.doi.org/10.1088/1751-8121/aa6dc3},
year = {2017},
month = {may},
publisher = {IOP Publishing},
volume = {50},
number = {22},
pages = {223001},
author = {J. C. Bridgeman and C. T. Chubb},
title = {Hand-waving and interpretive dance: an introductory course on tensor networks},
journal = {J. Phys. A},
}

@article{mele_introduction_to_2024,
  doi = {10.22331/q-2024-05-08-1340},
  opturl = {https://doi.org/10.22331/q-2024-05-08-1340},
  title = {Introduction to {H}aar {m}easure {t}ools in {q}uantum {i}nformation: {A} {b}eginner's {t}utorial},
  author = {Mele, A. A.},
  journal = {{Quantum}},
  issn = {2521-327X},
  publisher = {{Verein zur F{\"{o}}rderung des Open Access Publizierens in den Quantenwissenschaften}},
  volume = {8},
  pages = {1340},
  month = may,
  year = {2024}
}

@article{subasi_entanglement_spectroscopy_2019,
doi = {10.1088/1751-8121/aaf54d},
opturl = {https://dx.doi.org/10.1088/1751-8121/aaf54d},
year = {2019},
month = {jan},
publisher = {IOP Publishing},
volume = {52},
number = {4},
pages = {044001},
author = {Y. Subasi and L. Cincio and P. J. Coles},
title = {Entanglement spectroscopy with a depth-two quantum circuit},
journal = {J. Phys. A},
}

@article{quek_exponentially_tighter_2022,
    author = {Y. Quek and D. Stilck Franca and S. Khatri and J. Jakob Meyer and J. Eisert},
    title = {{Exponentially tighter bounds on limitations of quantum error mitigation}},
    year = {2024},
    doi={10.1038/s41567-024-02536-7},
    pages=1648,
    volume=20,
    journal = {Nature Phys.},
    opturl = {https://doi.org/10.48550/arXiv.2210.11505}
}

@ARTICLE{conway_a_new_1990,
  author={Conway, J. H. and Sloane, N. J. A.},
  journal={IEEE Trans. Inf. Th.}, 
  title={A new upper bound on the minimal distance of self-dual codes}, 
  year={1990},
  volume={36},
  number={6},
  pages={1319-1333},
  doi={10.1109/18.59931}}

@article{wootters_entanglement_of_1998,
  title = {{Entanglement of formation of an arbitrary state of two qubits}},
  author = {Wootters, W. K.},
  journal = {Phys. Rev. Lett.},
  volume = {80},
  issue = {10},
  pages = {2245--2248},
  numpages = {0},
  year = {1998},
  month = {Mar},
  publisher = {American Physical Society},
  doi = {10.1103/PhysRevLett.80.2245},
  opturl = {https://link.aps.org/doi/10.1103/PhysRevLett.80.2245}
}

@article{rungta_universal_state_2001,
  title = {Universal state inversion and concurrence in arbitrary dimensions},
  author = {Rungta, P. and Bu\ifmmode \check{z}\else \v{z}\fi{}ek, V. and Caves, C. M. and Hillery, M. and Milburn, G. J.},
  journal = {Phys. Rev. A},
  volume = {64},
  issue = {4},
  pages = {042315},
  numpages = {13},
  year = {2001},
  month = {Sep},
  publisher = {American Physical Society},
  doi = {10.1103/PhysRevA.64.042315},
  opturl = {https://link.aps.org/doi/10.1103/PhysRevA.64.042315}
}

@article{hall_multipartite_reduction_2005,
  title = {Multipartite reduction criteria for separability},
  author = {Hall, W.},
  journal = {Phys. Rev. A},
  volume = {72},
  issue = {2},
  pages = {022311},
  numpages = {5},
  year = {2005},
  month = {Aug},
  publisher = {American Physical Society},
  doi = {10.1103/PhysRevA.72.022311},
  opturl = {https://link.aps.org/doi/10.1103/PhysRevA.72.022311}
}

@ARTICLE{rains_polynomial_invariants_2000,
  author={Rains, E. M.},
  journal={IEEE Trans. Inf. Th.}, 
  title={Polynomial invariants of quantum codes}, 
  year={2000},
  volume={46},
  number={1},
  pages={54-59},
  doi={10.1109/18.817508}}

@ARTICLE{ashikhmin_upper_bounds_1999,
  author={Ashikhmin, A. and Litsyu, S.},
  journal={IEEE Trans. Inf. Th.}, 
  title={Upper bounds on the size of quantum codes}, 
  year={1999},
  volume={45},
  number={4},
  pages={1206-1215},
  doi={10.1109/18.761270}}

@ARTICLE{ashikhmin_quantum_error_2000,
  author={Ashikhmin, A. E. and Barg, A. M. and Knill, E. and Litsyn, S. N.},
  journal={IEEE Trans. Inf. Th.}, 
  title={{Quantum error detection. I. Statement of the problem}}, 
  year={2000},
  volume={46},
  number={3},
  pages={778-788},
  doi={10.1109/18.841162}}

@article{cao_quantum_lego_2024,
  title = {{Quantum Lego expansion pack: Enumerators from tensor networks}},
  author = {Cao, C.-J. and Gullans, M. J. and Lackey, B. and Wang, Z.},
  journal = {PRX Quantum},
  volume = {5},
  issue = {3},
  pages = {030313},
  numpages = {43},
  year = {2024},
  month = {Jul},
  publisher = {American Physical Society},
  doi = {10.1103/PRXQuantum.5.030313},
  opturl = {https://link.aps.org/doi/10.1103/PRXQuantum.5.030313}
}

@software{pato_planqtn_a_2025,
 author = {Pato, B. and Vanlerberghe, J. and Cao, C. and Lackey, B. and Brown, K.},
 title = {{PlanqTN, a Python library and interactive web app implementing the quantum LEGO framework}},
 year = 2025,
 publisher = {Zenodo},
 doi = {10.5281/zenodo.16761072},
 url = {https://doi.org/10.5281/zenodo.16761072},
}

@article{schuster_operator_growth_2023,
  title = {{Operator growth in open quantum systems}},
  author = {Schuster, T. and Yao, N. Y.},
  journal = {Phys. Rev. Lett.},
  volume = {131},
  issue = {16},
  pages = {160402},
  numpages = {6},
  year = {2023},
  month = {Oct},
  publisher = {American Physical Society},
  doi = {10.1103/PhysRevLett.131.160402},
  opturl = {https://link.aps.org/doi/10.1103/PhysRevLett.131.160402}
}

@ARTICLE{cao_quantum_weight_2024,
  author={Cao, C.-J. and Lackey, B.},
  journal={IEEE Trans. Inf. Th.}, 
  title={Quantum weight enumerators and tensor networks}, 
  year={2024},
  volume={70},
  number={5},
  pages={3512-3528},
  doi={10.1109/TIT.2023.3340503}}

@article{braccia_computing_exact_2024,
  author    = {P. Braccia and P. Bermejo and L. Cincio and M. Cerezo},
  title     = {Computing exact moments of local random quantum circuits via tensor networks},
  journal   = {Quantum Machine Intelligence},
  volume    = {6},
  number    = {2},
  pages     = {54},
  year      = {2024},
  month     = sep,
  issn      = {2524-4914},
  doi       = {10.1007/s42484-024-00187-8},
  url       = {https://doi.org/10.1007/s42484-024-00187-8}
}

@inproceedings{vardy_algorithmic_complexity_1997,
  title={Algorithmic complexity in coding theory and the minimum distance problem},
  author={Vardy, A.},
  booktitle={Proceedings of the twenty-ninth annual ACM symposium on Theory of computing},
  pages={92--109},
  year={1997}
}

@ARTICLE{shi_exploring_quantum_2024,
  author={Shi, Fei and Guo, Kaiyi and Zhang, Xiande and Zhao, Qi},
  journal={IEEE Trans. Inf. Theory}, 
  title={Exploring Quantum Weight Enumerators From the n-Qubit Parallelized SWAP Test}, 
  year={2026},
  volume={72},
  number={2},
  pages={1220-1231},
  doi={10.1109/TIT.2025.3634135}}

@article{pogorelov_compact_ion_2021,
  title = {{Compact ion-trap quantum computing demonstrator}},
  author = {Pogorelov, I. and Feldker, T. and Marciniak, C. D. and Postler, L. and Jacob, G. and others},
  journal = {PRX Quantum},
  volume = {2},
  issue = {2},
  pages = {020343},
  numpages = {23},
  year = {2021},
  publisher = {American Physical Society},
  doi = {10.1103/PRXQuantum.2.020343},
  opturl = {https://link.aps.org/doi/10.1103/PRXQuantum.2.020343}
}

@article{terhal_bell_inequalities_2000,
    author = {B. M. Terhal},
    title = {Bell inequalities and the separability criterion},
    journal = {Phys. Lett. A},
    volume = {271},
    number = {5},
    pages = {319},
    year = {2000},
    issn = {0375-9601},
    opturl = {https://doi.org/10.1016/S0375-9601(00)00401-1},
    doi = {10.1016/S0375-9601(00)00401-1},
}

@article{guehne_detection_of_2002,
  title = {Detection of entanglement with few local measurements},
  author = {G\"uhne, O. and Hyllus, P. and Bru\ss{}, D. and Ekert, A. and Lewenstein, M. and Macchiavello, C. and Sanpera, A.},
  journal = {Phys. Rev. A},
  volume = {66},
  issue = {6},
  pages = {062305},
  numpages = {5},
  year = {2002},
  publisher = {American Physical Society},
  doi = {10.1103/PhysRevA.66.062305},
  opturl = {https://doi.org/10.1103/PhysRevA.66.062305}
}

@article{bourennane_experimental_detection_2004,
  title = {{Experimental detection of multipartite entanglement using witness operators}},
  author = {Bourennane, M. and Eibl, M. and Kurtsiefer, C. and Gaertner, S. and Weinfurter, H. and G\"uhne, O. and Hyllus, P. 
  and Bru\ss{}, D. and Lewenstein, M. and Sanpera, A.},
  journal = {Phys. Rev. Lett.},
  volume = {92},
  issue = {8},
  pages = {087902},
  numpages = {4},
  year = {2004},
  publisher = {American Physical Society},
  doi = {10.1103/PhysRevLett.92.087902},
  opturl = {https://doi.org/10.1103/PhysRevLett.92.087902}
}

@article{guehne_entanglement_detection_2009,
    author = {O. G{\"u}hne and G. T{\'o}th},
    title = {Entanglement detection},
    journal = {Phys. Rep.},
    volume = {474},
    number = {1},
    pages = {1},
    numpages = {75},
    year = {2009},
    issn = {0370-1573},
    opturl = {https://doi.org/10.1016/j.physrep.2009.02.004},
    doi = {10.1016/j.physrep.2009.02.004}
}

@article{haffner_scalable_multiparticle_2005,
  author    = {H{\"{a}}ffner, H. and H{\"{a}}nsel, W. and Roos, C. F. and Benhelm, J. and Chek-al-kar, D. and others},
  title     = {Scalable multiparticle entanglement of trapped ions},
  journal   = {Nature},
  year      = {2005},
  month     = {dec},
  volume    = {438},
  number    = {7068},
  pages     = {643--646},
  doi       = {10.1038/nature04279},
  opturl       = {https://doi.org/10.1038/nature04279},
  issn      = {1476-4687},
}

@article{bergmann_entanglement_criteria_2013,
doi = {10.1088/1751-8113/46/38/385304},
opturl = {https://dx.doi.org/10.1088/1751-8113/46/38/385304},
year = {2013},
publisher = {IOP Publishing},
volume = {46},
number = {38},
pages = {385304},
author = {M. Bergmann and O. G{\"u}hne},
title = {{Entanglement criteria for Dicke states}},
journal = {J. Phys. A},
}

@article{peres_separability_criterion_1996,
  title = {{Separability criterion for density matrices}},
  author = {Peres, A.},
  journal = {Phys. Rev. Lett.},
  volume = {77},
  issue = {8},
  pages = {1413--1415},
  numpages = {0},
  year = {1996},
  publisher = {American Physical Society},
  doi = {10.1103/PhysRevLett.77.1413},
  opturl = {https://link.aps.org/doi/10.1103/PhysRevLett.77.1413}
}

@article{horodecki_separability_1996,
title = {{Separability of mixed states: necessary and sufficient conditions}},
journal = {Phys. Lett. A},
volume = {223},
number = {1},
pages = {1-8},
year = {1996},
issn = {0375-9601},
doi = {https://doi.org/10.1016/S0375-9601(96)00706-2},
opturl = {https://www.sciencedirect.com/science/article/pii/S0375960196007062},
author = {M. Horodecki and P. Horodecki and R. Horodecki},
}

@article{zhou_single_copies_2020,
  title = {{Single-copies estimation of entanglement negativity}},
  author = {Zhou, Y. and Zeng, P. and Liu, Z.},
  journal = {Phys. Rev. Lett.},
  volume = {125},
  issue = {20},
  pages = {200502},
  numpages = {6},
  year = {2020},
  publisher = {American Physical Society},
  doi = {10.1103/PhysRevLett.125.200502},
  opturl = {https://link.aps.org/doi/10.1103/PhysRevLett.125.200502}
}

@article{elben_mixed_state_2020,
  title = {{Mixed-state entanglement from local randomized measurements}},
  author = {Elben, A. and Kueng, R. and Huang, H.-Y. and van Bijnen, R. and Kokail, C. and Dalmonte, M. and Calabrese, P. and Kraus, B. and Preskill, J. and Zoller, P. and Vermersch, B.},
  journal = {Phys. Rev. Lett.},
  volume = {125},
  issue = {20},
  pages = {200501},
  numpages = {6},
  year = {2020},
  publisher = {American Physical Society},
  doi = {10.1103/PhysRevLett.125.200501},
  opturl = {https://link.aps.org/doi/10.1103/PhysRevLett.125.200501}
}

@article{fischer_ancilla_free_2022,
  title = {Ancilla-free implementation of generalized measurements for qubits embedded in a qudit space},
  author = {Fischer, L. E. and Miller, D. and Tacchino, F. and Barkoutsos, P. K. and Egger, D. J. and Tavernelli, I.},
  journal = {Phys. Rev. Res.},
  volume = {4},
  issue = {3},
  pages = {033027},
  numpages = {17},
  year = {2022},
  month = {Jul},
  publisher = {American Physical Society},
  doi = {10.1103/PhysRevResearch.4.033027},
  opturl = {https://link.aps.org/doi/10.1103/PhysRevResearch.4.033027}
}

@article{stricker_experimental_single_2022,
  title = {Experimental Single-Setting Quantum State Tomography},
  author = {Stricker, R. and Meth, M. and Postler, L. and Edmunds, C. and Ferrie, C. and Blatt, R. and Schindler, P. and Monz, T. and Kueng, R. and Ringbauer, M.},
  journal = {PRX Quantum},
  volume = {3},
  issue = {4},
  pages = {040310},
  numpages = {34},
  year = {2022},
  month = {Oct},
  publisher = {American Physical Society},
  doi = {10.1103/PRXQuantum.3.040310},
  opturl = {https://link.aps.org/doi/10.1103/PRXQuantum.3.040310}
}

@article{magesan_characterizing_quantum_2012,
  title = {{Characterizing quantum gates via randomized benchmarking}},
  author = {Magesan, E. and Gambetta, J. M. and Emerson, J.},
  year = 2012,
  journal = {Phys. Rev. A},
  publisher = {American Physical Society},
  volume = 85,
  pages = {042311},
  doi = {10.1103/PhysRevA.85.042311},
  opturl = {https://link.aps.org/doi/10.1103/PhysRevA.85.042311}
}

@article{werner_quantum_states_1989,
  title = {{Quantum states with Einstein-Podolsky-Rosen correlations admitting a hidden-variable model}},
  author = {Werner, R. F.},
  journal = {Phys. Rev. A},
  volume = {40},
  issue = {8},
  pages = {4277--4281},
  numpages = {0},
  year = {1989},
  month = {Oct},
  publisher = {American Physical Society},
  doi = {10.1103/PhysRevA.40.4277},
  opturl = {https://link.aps.org/doi/10.1103/PhysRevA.40.4277}
}

@article{mckay_efficient_z_2017,
  title = {Efficient $Z$ gates for quantum computing},
  author = {McKay, D. C. and Wood, C. J. and Sheldon, S. and Chow, J. M. and Gambetta, J. M.},
  journal = {Phys. Rev. A},
  volume = {96},
  issue = {2},
  pages = {022330},
  numpages = {8},
  year = {2017},
  month = {Aug},
  publisher = {American Physical Society},
  doi = {10.1103/PhysRevA.96.022330},
  opturl = {https://link.aps.org/doi/10.1103/PhysRevA.96.022330}
}

@article{maslov_basic_circuit_2017,
    doi = {10.1088/1367-2630/aa5e47},
    opturl = {https://dx.doi.org/10.1088/1367-2630/aa5e47},
    year = {2017},
    month = {feb},
    publisher = {IOP Publishing},
    volume = {19},
    number = {2},
    pages = {023035},
    author = {D. Maslov},
    title = {Basic circuit compilation techniques for an ion-trap quantum machine},
    journal = {New J. Phys.},
}

@article{sorensen_quantum_computation_1999,
  title = {Quantum computation with ions in thermal motion},
  author = {S\o{}rensen, A. and M\o{}lmer, K.},
  journal = {Phys. Rev. Lett.},
  volume = {82},
  issue = {9},
  pages = {1971--1974},
  numpages = {0},
  year = {1999},
  month = {Mar},
  publisher = {American Physical Society},
  doi = {10.1103/PhysRevLett.82.1971},
  opturl = {https://link.aps.org/doi/10.1103/PhysRevLett.82.1971}
}

@article{skornia_nonclassical_interference_2001,
  title = {Nonclassical interference effects in the radiation from coherently driven uncorrelated atoms},
  author = {Skornia, C. and Zanthier, J. von and Agarwal, G. S. and Werner, E. and Walther, H.},
  journal = {Phys. Rev. A},
  volume = {64},
  issue = {6},
  pages = {063801},
  numpages = {5},
  year = {2001},
  month = {Nov},
  publisher = {American Physical Society},
  doi = {10.1103/PhysRevA.64.063801},
  opturl = {https://link.aps.org/doi/10.1103/PhysRevA.64.063801}
}

@article{sorensen_probabilistic_generation_2003,
  title = {Probabilistic generation of entanglement in optical cavities},
  author = {S\o{}rensen, A. S. and M\o{}lmer, K.},
  journal = {Phys. Rev. Lett.},
  volume = {90},
  issue = {12},
  pages = {127903},
  numpages = {4},
  year = {2003},
  month = {Mar},
  publisher = {American Physical Society},
  doi = {10.1103/PhysRevLett.90.127903},
  opturl = {https://link.aps.org/doi/10.1103/PhysRevLett.90.127903}
}

@article{thiel_generation_of_2007,
  title = {{Generation of Symmetric Dicke States of Remote Qubits with Linear Optics}},
  author = {Thiel, C. and von Zanthier, J. and Bastin, T. and Solano, E. and Agarwal, G. S.},
  journal = {Phys. Rev. Lett.},
  volume = {99},
  issue = {19},
  pages = {193602},
  numpages = {4},
  year = {2007},
  month = {Nov},
  publisher = {American Physical Society},
  doi = {10.1103/PhysRevLett.99.193602},
  opturl = {https://link.aps.org/doi/10.1103/PhysRevLett.99.193602}
}

@article{chen_carving_complex_2015,
  title = {{Carving complex many-atom entangled states by single-photon detection}},
  author = {Chen, W. and Hu, J. and Duan, Y. and Braverman, B. and Zhang, H. and Vuleti\ifmmode \acute{c}\else \'{c}\fi{}, V.},
  journal = {Phys. Rev. Lett.},
  volume = {115},
  issue = {25},
  pages = {250502},
  numpages = {5},
  year = {2015},
  month = {Dec},
  publisher = {American Physical Society},
  doi = {10.1103/PhysRevLett.115.250502},
  opturl = {https://link.aps.org/doi/10.1103/PhysRevLett.115.250502}
}

@article{welte_cavity_carving_2017,
  title = {{Cavity carving of atomic Bell states}},
  author = {Welte, S. and Hacker, B. and Daiss, S. and Ritter, S. and Rempe, G.},
  journal = {Phys. Rev. Lett.},
  volume = {118},
  issue = {21},
  pages = {210503},
  numpages = {6},
  year = {2017},
  month = {May},
  publisher = {American Physical Society},
  doi = {10.1103/PhysRevLett.118.210503},
  opturl = {https://link.aps.org/doi/10.1103/PhysRevLett.118.210503}
}

@article{welte_photon_mediated_2018,
  title = {{Photon-Mediated Quantum Gate between Two Neutral Atoms in an Optical Cavity}},
  author = {Welte, S. and Hacker, B. and Daiss, S. and Ritter, S. and Rempe, G.},
  journal = {Phys. Rev. X},
  volume = {8},
  issue = {1},
  pages = {011018},
  numpages = {11},
  year = {2018},
  month = {Feb},
  publisher = {American Physical Society},
  doi = {10.1103/PhysRevX.8.011018},
  opturl = {https://link.aps.org/doi/10.1103/PhysRevX.8.011018}
}

@article{davis_painting_nonclassical_2018,
  title = {{Painting Nonclassical States of Spin or Motion with Shaped Single Photons}},
  author = {Davis, E. J. and Wang, Z. and Safavi-Naeini, A. H. and Schleier-Smith, M. H.},
  journal = {Phys. Rev. Lett.},
  volume = {121},
  issue = {12},
  pages = {123602},
  numpages = {6},
  year = {2018},
  month = {Sep},
  publisher = {American Physical Society},
  doi = {10.1103/PhysRevLett.121.123602},
  opturl = {https://link.aps.org/doi/10.1103/PhysRevLett.121.123602}
}

@article{dordevic_entanglement_transport_2021,
author = {T. \DJ{}or\dj{}evi\'c  and P. Samutpraphoot  and P. L. Ocola  and H. Bernien  and B. Grinkemeyer  and I. Dimitrova  and V. Vuleti{\'c}  and M. D. Lukin },
title = {{Entanglement transport and a nanophotonic interface for atoms in optical tweezers}},
journal = {Science},
volume = {373},
number = {6562},
pages = {1511-1514},
year = {2021},
doi = {10.1126/science.abi9917},
opturl = {https://www.science.org/doi/abs/10.1126/science.abi9917},
}

@article{richter_collective_photon_2023,
  title = {Collective photon emission of two correlated atoms in free space},
  author = {Richter, S. and Wolf, S. and von Zanthier, J. and Schmidt-Kaler, F.},
  journal = {Phys. Rev. Res.},
  volume = {5},
  issue = {1},
  pages = {013163},
  numpages = {5},
  year = {2023},
  month = {Mar},
  publisher = {American Physical Society},
  doi = {10.1103/PhysRevResearch.5.013163},
  opturl = {https://link.aps.org/doi/10.1103/PhysRevResearch.5.013163}
}

@article{ramette_carving_entangled_2025,
  title = {Carving entangled multiparticle states with exponentially improved fidelity},
  author = {J.~Ramette  and J.~Sinclair and Z.~Li and V.~Vuleti{\'c}},
  journal = {Phys. Rev. A},
  volume = {111},
  issue = {5},
  pages = {052426},
  numpages = {7},
  year = {2025},
  month = {May},
  publisher = {American Physical Society},
  doi = {10.1103/PhysRevA.111.052426},
  url = {https://link.aps.org/doi/10.1103/PhysRevA.111.052426}
}

@misc{rall_signed_quantum_2017,
      title={Signed quantum weight enumerators characterize qubit magic state distillation}, 
      author={P. Rall},
      year={2017},
      eprint={1702.06990},
      archivePrefix={arXiv} 
}

@article{miller_hardware_tailored_2024,
  author    = {D. Miller and L. E. Fischer and K. Levi and E. J. Kuehnke and I. O. Sokolov and P. Kl. Barkoutsos and J. Eisert and I. Tavernelli},
  title     = {Hardware-tailored diagonalization circuits},
  journal   = {npj Quant. Inf.},
  volume    = {10},
  number    = {1},
  pages     = {122},
  year      = {2024},
  month     = {nov},
  issn      = {2056-6387},
  doi       = {10.1038/s41534-024-00901-1},
  url       = {https://doi.org/10.1038/s41534-024-00901-1},
}

@article{bertoni_shallow_shadows_2024,
  title = {{Shallow shadows: Expectation estimation using low-depth random Clifford circuits}},
  author = {Bertoni, C. and Haferkamp, J. and Hinsche, M. and Ioannou, M. and Eisert, J. and Pashayan, H.},
  journal = {Phys. Rev. Lett.},
  volume = {133},
  issue = {2},
  pages = {020602},
  numpages = {7},
  year = {2024},
  month = {Jul},
  publisher = {American Physical Society},
  doi = {10.1103/PhysRevLett.133.020602},
  opturl = {https://link.aps.org/doi/10.1103/PhysRevLett.133.020602}
}

@ARTICLE{kukliansky_quantum_circuit_2025,
  author={Kukliansky, A. and Lackey, B.},
  journal={IEEE Trans. Inf. Th.}, 
  title={{Quantum circuit tensors and enumerators with applications to quantum fault tolerance}}, 
  year={2025},
  volume={71},
  number={6},
  pages={4406-4427},
  doi={10.1109/TIT.2025.3555189}}

@article{liu_detecting_entanglement_2022,
  title = {{Detecting entanglement in quantum many-body systems via permutation moments}},
  author = {Liu, Z. and Tang, Y. and Dai, H. and Liu, P. and Chen, S. and Ma, X.},
  journal = {Phys. Rev. Lett.},
  volume = {129},
  issue = {26},
  pages = {260501},
  numpages = {6},
  year = {2022},
  month = {Dec},
  publisher = {American Physical Society},
  doi = {10.1103/PhysRevLett.129.260501} 
}

@article{vermersch_many_body_2024,
  title = {{Many-body entropies and 
  entanglement from polynomially many 
  local measurements}},
  author = {Vermersch, B. and Ljubotina, M. and Cirac, J. I. and Zoller, P. and Serbyn, M. and Piroli, L.},
  journal = {Phys. Rev. X},
  volume = {14},
  issue = {3},
  pages = {031035},
  numpages = {30},
  year = {2024},
  month = {Aug},
  publisher = {American Physical Society},
  doi = {10.1103/PhysRevX.14.031035},
  opturl = {https://link.aps.org/doi/10.1103/PhysRevX.14.031035}
}

@article{hong_measurement_of_1987,
  title = {Measurement of subpicosecond time intervals between two photons by interference},
  author = {Hong, C. K. and Ou, Z. Y. and Mandel, L.},
  journal = {Phys. Rev. Lett.},
  volume = {59},
  issue = {18},
  pages = {2044--2046},
  numpages = {0},
  year = {1987},
  month = {Nov},
  publisher = {American Physical Society},
  doi = {10.1103/PhysRevLett.59.2044},
  opturl = {https://link.aps.org/doi/10.1103/PhysRevLett.59.2044}
}

@article{garcia_escartin_swap_test_2013,
  title = {{Swap test and Hong-Ou-Mandel effect are equivalent}},
  author = {Garcia-Escartin, J. C. and Chamorro-Posada, P.},
  journal = {Phys. Rev. A},
  volume = {87},
  issue = {5},
  pages = {052330},
  numpages = {10},
  year = {2013},
  month = {May},
  publisher = {American Physical Society},
  doi = {10.1103/PhysRevA.87.052330},
  opturl = {https://link.aps.org/doi/10.1103/PhysRevA.87.052330}
}

@article{daley_measuring_entanglement_2012,
  title = {{Measuring entanglement growth in quench dynamics of bosons in an optical lattice}},
  author = {Daley, A. J. and Pichler, H. and Schachenmayer, J. and Zoller, P.},
  journal = {Phys. Rev. Lett.},
  volume = {109},
  issue = {2},
  pages = {020505},
  numpages = {5},
  year = {2012},
  month = {Jul},
  publisher = {American Physical Society},
  doi = {10.1103/PhysRevLett.109.020505},
  opturl = {https://link.aps.org/doi/10.1103/PhysRevLett.109.020505}
}

@article{moura_alves_multipartite_entanglement_2004,
  title = {{Multipartite Entanglement Detection in Bosons}},
  author = {Moura Alves, C. and Jaksch, D.},
  journal = {Phys. Rev. Lett.},
  volume = {93},
  issue = {11},
  pages = {110501},
  numpages = {4},
  year = {2004},
  month = {Sep},
  publisher = {American Physical Society},
  doi = {10.1103/PhysRevLett.93.110501},
  opturl = {https://link.aps.org/doi/10.1103/PhysRevLett.93.110501}
}

@article{fischer_dual_frame_2024,
  title = {Dual-frame optimization for informationally complete quantum measurements},
  author = {Fischer, L. E. and Dao, T. and Tavernelli, I. and Tacchino, F.},
  journal = {Phys. Rev. A},
  volume = {109},
  issue = {6},
  pages = {062415},
  numpages = {12},
  year = {2024},
  month = {Jun},
  publisher = {American Physical Society},
  doi = {10.1103/PhysRevA.109.062415},
  opturl = {https://link.aps.org/doi/10.1103/PhysRevA.109.062415}
}

@article{harrow_approximate_unitary_2023,
  author  = {A. W. Harrow and S. Mehraban},
  title   = {{Approximate unitary $t$-designs by short random quantum circuits Using nearest-neighbor and long-range gates}},
  journal = {Commun. Math. Phys.},
  volume  = {401},
  number  = {2},
  pages   = {1531--1626},
  year    = {2023},
  doi     = {10.1007/s00220-023-04675-z},
  opturl     = {https://doi.org/10.1007/s00220-023-04675-z},
  issn    = {1432-0916}
}

@misc{munne_sdp_bounds_2024,
      title={{SDP bounds on quantum codes}}, 
      author={G. Angl{\`e}s  Munn{\'e} and A. Nemec and F. Huber},
      year={2024},
      eprint={2408.10323},
      archivePrefix={arXiv},
      optoptprimaryClass={quant-ph} 
}

@article{huber_positive_maps_2021,
    author = {Huber, F.},
    title = "{Positive maps and trace polynomials from the symmetric group}",
    journal = {J. Math. Phys.},
    volume = {62},
    number = {2},
    pages = {022203},
    year = {2021},
    month = {02},
    issn = {0022-2488},
    doi = {10.1063/5.0028856}
}

@article{rico_entanglement_detection_2024,
  title = {Entanglement detection with trace polynomials},
  author = {Rico, A. and Huber, F.},
  journal = {Phys. Rev. Lett.},
  volume = {132},
  issue = {7},
  pages = {070202},
  numpages = {6},
  year = {2024},
  month = {Feb},
  publisher = {American Physical Society},
  doi = {10.1103/PhysRevLett.132.070202} 
}

@article{schatzki_tensor_rank_2024,
  title = {{\add Tensor rank and other multipartite entanglement measures of graph states}},
  author = {{\add L. Schatzki   and  L. Ma  and  E. Solomonik and  E. Chitambar}},
  journal = {Phys. Rev. A},
  volume = {110},
  issue = {3},
  pages = {032409},
  numpages = {12},
  year = {2024},
  month = {Sep},
  publisher = {American Physical Society},
  doi = {10.1103/PhysRevA.110.032409},
  url = {https://link.aps.org/doi/10.1103/PhysRevA.110.032409}
}

@article{sang_approximate_quantum_2024,
  title = {{\add Approximate Quantum Error Correcting Codes from Conformal Field Theory}},
  author = {{\add S.~Sang and T.~H.~Hsieh and Y.~Zou}},
  journal = {Phys. Rev. Lett.},
  volume = {133},
  issue = {21},
  pages = {210601},
  numpages = {7},
  year = {2024},
  month = {Nov},
  publisher = {American Physical Society},
  doi = {10.1103/PhysRevLett.133.210601},
  url = {https://link.aps.org/doi/10.1103/PhysRevLett.133.210601}
}

@article{zhang_probing_mixed_2025,
      title={{\add Probing mixed-state phases on a quantum computer via Renyi correlators and variational decoding}}, 
      author={{\add Y.~Zhang and T.~H.~Hsieh and Y.~B.~Kim and Y.~Zou}},
      year={2025},
      eprint={2505.02900},
      archivePrefix={arXiv},
      optprimaryClass={quant-ph},
      url={https://arxiv.org/abs/2505.02900}, 
}

@article{leone_stabilizer_renyi_2022,
  title = {{\add Stabilizer R\'enyi entropy}},
  author = {{\add L.~Leone and S.~F.~E.~Oliviero, and A.~Hamma}},
  journal = {Phys. Rev. Lett.},
  volume = {128},
  issue = {5},
  pages = {050402},
  numpages = {5},
  year = {2022},
  month = {Feb},
  publisher = {American Physical Society},
  doi = {10.1103/PhysRevLett.128.050402},
  url = {https://link.aps.org/doi/10.1103/PhysRevLett.128.050402}
}

@mastersthesis{miller_small_quantum_2019,
  title        = {Small Quantum Networks in the Qudit Stabilizer Formalism},
  author       = {Miller, Daniel},
  year         = {2019},
  school       = {Heinrinch-Heine Universität Düsseldorf}, 
  note         = {arXiv:1910.09551 [quant-ph]},
  url          = {https://arxiv.org/abs/1910.09551}
}

\appendix

{\add \section{Elementary examples of QWEs}
\label{app:example}
}

In this appendix, we illuminate the general theory of \emph{quantum weight enumerators}  (QWEs) using two 
simple examples,
to improve the accessibility of the various interpretations and applications reviewed in Sec.~\ref{sec:qwe_macwilliams} of the main text.
Consider the two state vectors $\ket{0,0}$ and $\ket{\Phi^+}= \tfrac{1}{\sqrt{2}}(\ket{0,0} + \ket{1,1})$.
An important difference between them is that $\ket{0,0}$ is only correlated in the $Z$-basis while
$\ket{\Phi^+}$ is perfectly (anti)correlated in three different bases, namely $\bra{0,0} (Z\otimes Z) \ket{0,0} = 1$, $\bra{\Phi^+} (X\otimes X) \ket{\Phi^+} = \bra{\Phi^+} (Z\otimes Z) \ket{\Phi^+} = 1$, and $ \bra{\Phi^+} (Y\otimes Y) \ket{\Phi^+} = -1$.
The term ``\emph{weight}'' in QWE refers to the weight of multi-qubit Pauli operators. 
The expressions above are examples of weight-2 Pauli correlations.
It turns out that the only other non-vanishing correlations are given by $\bra{0,0} (I\otimes I) \ket{0,0}  = \bra{\Phi^+} (I\otimes I)\ket{\Phi^+} = 1$ (weight~0) and $\bra{0,0} (Z\otimes I )\ket{0,0}  =\bra{0,0} (I\otimes Z) \ket{0,0} = 1$ (weight~1).
The Shor--Laflamme QWE distribution (SLD) allows us to distinguish the two states by summarizing the information above into normalized vectors of the form $\mathbf{a}=(a_0,a_1,a_2)$ 
that are defined in full generality in Eq.~\eqref{def:sld} of the main text.
In our example,
we have
\begin{align}
    \mathbf{a} \left[\ket{0,0}\bra{0,0}\right]       &= (0.25, \ 0.5,\  0.25) \hspace{2mm} \text{ and} \\ 
    \mathbf{a} \left[\ket{\Phi^+}\bra{\Phi^+}\right] &= (0.25, \ 0.0,\  0.75).
\end{align}
Note that $a_2[\ket{\Phi^+}\bra{\Phi^+}] = 0.75 $ is three times as large as $a_2[\ket{0,0}\bra{0,0}] =0.25$ reflecting the fact that $\ket{\Phi^+}$ is perfectly correlated in three times as many bases as $\ket{0,0}$.
This suffices to prove that the state $\ket{\Phi^+}$ is entangled: 
indeed, the well-known entanglement criterion for $n$-qubit states $\rho$, 
\begin{align}
    \Tr[\rho (X^{\otimes n})  ]^2 + \Tr[\rho (Z^{\otimes n})]^2 > 1 \Longrightarrow \rho  {\text{    entangled}} \, ,
\end{align}
applies here.
More generally, QWEs provide a framework for a powerful generalization of this entanglement criterion, see Eq.~\eqref{eq:nbody_criterion} of the main text.

Another way to demonstrate that $\ket{\Phi^+}$ is entangled is by examining the purities of its subsystems.
The set $\{1,2\}$, which labels the qubits has four subsets: $\diameter, \{1\}, \{2\}$, and $\{1,2\}$.
Thus, any two-qubit state $\rho$ has four marginals: the (trivial) zero-qubit state $\rho_{\diameter} = 1$, the one-qubit states $\rho_{\{1\}}$ and $\rho_{\{2\}}$, and the global state $\rho_{\{1,2\}}=\rho$.
For $\ket{0,0}$ and $\ket{\Phi^+}$, the single-qubit marginals are $\ket{0}\!\bra{0}$ (pure) and $\mathbbm 1/2$ (maximally mixed), respectively.
Thus, the single-qubit subsystem purities are given by $\text{Tr}[\ket{0}\!\bra{0}^2]=1$ and $\Tr[(\mathbbm 1/2)^2]= 0.5$, respectively.
This information is captured by Rains' unitary QWEs, which are vectors of the form $\mathbf{a}'=(a'_0, a'_1, a'_2)$
that are defined in full generality in Eq.~\eqref{def:apd} of the main text.
The entries of $\mathbf{a}'$ are (averaged) subsystem purities, where the subscripts refer to subsystem sizes.
In our example, we have 
\begin{align}
    \mathbf{a}' \left[\ket{0,0}\bra{0,0}\right]       &= (1.0, \ 1.0,\  1.0) \hspace{2mm} \text{ and} \\ 
    \mathbf{a}' \left[\ket{\Phi^+}\bra{\Phi^+}\right] &= (1.0, \ 0.5,\  1.0).
\end{align}
Since the entries of $\mathbf{a}' \left[\ket{\Phi^+}\bra{\Phi^+}\right] $ obey $a'_2 > a'_1$, we can conclude that $\ket{\Phi ^+ }$ is entangled, see Eq.~\eqref{eq:purity_criterion} 
{\add in the main text}.
These two features, Pauli correlations and subsystem purities, are in fact two sides of the same medal:
one can always convert the SLD into the vector of Rains' unitary QWEs, and vice versa~\cite{rains_quantum_weight_1998}.
This relationship is thoroughly discussed in Sec.~\ref{sec:qwe_macwilliams} of the main text.

\vspace{8mm}

\section{Overlap decay in terms of SLDs}
\label{app:overlap_decay}

{\add In this appendix, we prove Proposition~\ref{lem:overlap_decay} from the main text.}
Let $\rho$ be an arbitrary $n$-qubit state.
Expanding it in the Pauli basis yields 
\begin{equation}
\rho =  \sum_{P\in\{I,X,Y,Z\}^{\otimes n}}  \tfrac{\Tr[\rho P]}{2^n} P .
\end{equation}
By linearity of the locally depolarizing channel $\mathcal{E}_p^{\otimes n}[\cdot]$, we find 
$\mathcal{E}_p^{\otimes n}[\rho] = \sum_{P}  \tfrac{\Tr[\rho P]}{2^n}\mathcal{E}_p^{\otimes n}[P]$.
Since also $\Tr[\cdot]$ is linear,
the overlap decay between $\rho$ and $\mathcal{E}_p^{\otimes n}[\rho] $  follows as
\begin{align}\label{eq:overlap_decay_proof}
    \Tr\left[  (\mathcal{E}_p^{\otimes n} [ \rho]) \rho \right ]
  =   \sum_{P,Q }  (1-p)^{\wt(P)} \tfrac{\Tr[\rho P]}{2^n} \tfrac{\Tr[\rho Q]}{2^n}  \Tr[ P Q] \, ,
\end{align}
{\add where we have used $ \mathcal{E}_p^{\otimes n}[P] = (1-p)^{\wt(P)}P$.}
Further exploiting the fact  $\Tr[PQ]= \delta_{P,Q}2^n$, 
we  find
\begin{align}
    \Tr\left[  (\mathcal{E}_p^{\otimes n} [ \rho]) \rho \right ]
&= \sum_{P  \in \{I,X,Y,Z\}^{\otimes n}} (1-p)^{\wt(P)}  \tfrac{\Tr[\rho P]^2}{2^n} \, .
\end{align} 
{\add Therefore,} we can write
\begin{align}
\Tr\left[  (\mathcal{E}_p^{\otimes n} [ \rho]) \rho \right ]
    &= \sum_{j=0}^n  (1-p)^{j}  \sum_{\substack{P \in \{I,X,Y,Z\}^{\otimes n} \\ {\wt(P) = j}  }}  \tfrac{\Tr[\rho P]^2}{2^n} 
    \nonumber
    \\
    &= \sum_{j=0}^n  (1-p)^{j} a_j [\rho] \, ,
\end{align}
which completes the proof.

{ \add
\section{Derivation of dual and shadow enumerators}
\label{app:derivation_pwd_dual_and_shadow}

In this appendix, we provide a unified derivation of the well-known fact that---for a QECC with stabilizer group $\mathcal{S}$---the abstract definitions of dual Shor--Laflamme QWEs and Rains' shadow QWEs from Eqs.~\eqref{def:sld_dual} and~\eqref{def:rains_qwe} of the main text
simplify to the Pauli weight distributions over the logical Pauli group $\mathcal{S}^\perp$
and its coset $\tilde{\mathcal{S}}$, respectively.

In both cases, we will use the so-called $\Swap$ trick~\cite{hastings_measuring_renyi_2010},
which asserts for arbitrary $n$-qubit operators $A$ and $B$, 
\begin{align}
    \Tr[AB] = \Tr[(A\otimes B)\Omega]\, ,
\end{align}
where $\Omega = \prod_{i=1}^n \Swap_{i,n+i}$ is the operator that exchanges two $n$-qubit registers.
Inserting $A=P\rho$ and $B=P\sigma$ for   $P\in\{I,X,Y,Z\}^{\otimes n}$ and two $n$-qubit states $\rho $ and $\sigma$,
yields
\begin{align} \label{eq:swap_trick}
    \Tr[\rho P \sigma P] = \Tr [(\rho \otimes \sigma) (P\otimes P) \Omega] \, .
\end{align}
The Pauli decomposition $\Omega =  \tfrac{1}{2^n}\sum_{Q\in \{I,X,Y,Z\}^{\otimes n}} Q\otimes Q$ 
allows us to write (after a little algebra),
\begin{align} \label{eq:swap_operator_shifted}
    (P\otimes P) \Omega = \frac{1}{2^n}\sum_{Q\in \{I,X,Y,Z\}^{\otimes n}} (-1 )^{\delta _{PQ,QP}} Q\otimes Q \,.
\end{align}
Combining Eqs.~\eqref{eq:swap_trick} and~\eqref{eq:swap_operator_shifted}, finally yields
\begin{align} \label{eq:swap_trick_sum}
\Tr[\rho P \sigma P] =\sum_{Q\in \{I,X,Y,Z\}^{\otimes n}}  \hspace{-5mm}
 \frac{(-1)^{\delta_{PQ,QP}}\Tr[\rho Q]\Tr[ \sigma Q]}{2^n} \, .
\end{align}

Let us begin with deriving the dual SLD.
Inserting Eq.~\eqref{eq:swap_trick_sum} with $\rho=\sigma$ into Def.~\eqref{def:sld_dual} of the main text, we find 
\begin{align} \label{eq:sld_dual_step1}
    b_i[\rho] = \frac{1}{4^n}  \sum_{\substack{P, Q\in \{I,X,Y,Z\}^{\otimes n} \\\wt(P)=i}} \Tr[\rho Q]^2 (-1)^{\delta_{PQ,QP}} \, .
\end{align}
If, more specifically, $\rho=\rho^\text{stab}_\text{QECC}$  is the code state from Eq.~\eqref{eq:rho_qecc_stab} of the main text,
then the map $ \{I,X,Y,Z\}^{\otimes n}
    \rightarrow \RR$,
\begin{equation} \label{eq:indicator_function_dual}
\begin{aligned}
  &  Q \longmapsto \Tr[\rho_\text{QECC}^\text{stab}Q]^2 = \begin{cases}
        1 &\text{ if } Q \in \mathcal{S}, \\
        0 &\text{ if } Q \not \in \mathcal{S},
    \end{cases}
\end{aligned}
\end{equation}
is the indicator function on the stabilizer group $\mathcal{S}$ of the QECC.
Here, we sightly abuse notation and treat stabilizer groups 
as subsets of $\{I,X,Y,Z\}^{\otimes n}$, effectively ignoring phases, which are irrelevant in this context.
Using Eq.~\eqref{eq:indicator_function_dual}, we can simplify Eq.~\eqref{eq:sld_dual_step1} into 
\begin{align}\label{eq:sld_dual_step2}
 b_i[ \rho_\text{QECC}^\text{stab} ] 
 =   \frac{1}{4^n}  \sum_{\substack{P \in \{I,X,Y,Z\}^{\otimes n} \\\wt(P)=i}}  \sum_{S\in\mathcal{S}} (-1)^{\delta_{PS,SP}} \, .
\end{align}
Clearly, the inner sum in Eq.~\eqref{eq:sld_dual_step2} equates to $|\mathcal{S}| = 2^{n-k}$ 
if $P$ is contained 
in the dual $\mathcal{S}^\perp$, which is defined in Eq.~\eqref{def:normalizer} of the main text.
Conversely,
if $P\not \in \mathcal{S}^\perp$, the inner sum in Eq.~\eqref{eq:sld_dual_step2} vanishes.
Therefore, the map 
 $  \{I,X,Y,Z\}^{\otimes n}
    \rightarrow \RR,  $ 
\begin{align}
   P \longmapsto \frac{1}{|\mathcal{S}|} \sum_{S \in \mathcal{S}} (-1)^{\delta_{PS,SP}}
\end{align}
is the indicator function on $\mathcal{S}^{\perp}$. 
In conclusion, $b_i[\rho^\text{stab}_\text{QECC}]$ is the probability that a uniformly random logical operator $P\in \mathcal{S}^\perp$ has $\wt(P)=i$, 
which proves Eq.~\eqref{eq:sld_dual_step3} in the main text.

Finally, let us derive an analogous result for Rains' shadow QWEs.
This time inserting $\sigma =\tilde\rho$ into 
Eq.~\eqref{eq:swap_trick_sum},
we find
\begin{align} \label{eq:pauli_twirl_2}
    \Tr[\rho P \tilde \rho P ] & = 
      \frac{1}{2^n}  \sum_{\substack{ Q \in\{I,X,Y,Z\}^{\otimes n} }} \Tr[\rho Q] \Tr[\tilde \rho  Q ] (-1)^{\delta_{PQ,QP}} \, .
\end{align}
Further using 
$\Tr[\rhotranspose Q] = (-1)^{\delta_{Q,\qtranspose}} \Tr[\rho Q]$ and
accounting for a potential minus sign (if $Q$ and $\mathbf{Y}=Y^{\otimes n}$ anticommute),
we can write
\begin{align} \label{eq:shadow_step1}
    \Tr[\rho P \tilde \rho P ] & = 
      \frac{1}{2^n}  \sum_{\substack{ Q }}  \Tr[\rho Q]^2 (-1)^{\delta_{Q,\qtranspose }  \hspace{0.25mm} +  \hspace{0.25mm} \delta_{Q \mathbf{Y} , \mathbf{Y}  Q \vphantom{\qtranspose}}  \hspace{0.25mm}+ \hspace{0.25mm} \delta_{PQ,QP \vphantom{\qtranspose}}} \, .
\end{align} 
Since $Q=\qtranspose$ iff  the number of $Y$-operators in $Q$  is even and  $Q\mathbf{Y} = \mathbf{Y} Q$ iff the combined number of $X$- and $Z$-operators in $Q$ is even, we obtain
\begin{align}  \label{eq:shadow_step2}
    (-1)^{\delta_{Q,\qtranspose}  \hspace{.75mm} +  \hspace{.75mm} \delta_{Q\mathbf{Y} ,\mathbf{Y} Q \vphantom{\qtranspose}}} = (-1)^{\wt{(Q)}} \, .
\end{align}
As above, we now specialize to the case $\rho=\rho^\text{stab}_\text{QECC}$.
Inserting Eqs.~\eqref{eq:shadow_step1} and~\eqref{eq:shadow_step2} into Def.~\eqref{def:rains_qwe} of the main text,
yields
 \begin{align}  \label{eq:shadow_step3}
     \tilde a_i[\rho_\text{QECC}^\text{stab}]  
 =   \frac{1}{4^n}  \sum_{\substack{P \in \{I,X,Y,Z\}^{\otimes n} \\\wt(P)=i}}  \sum_{S\in\mathcal{S}} (-1)^{\wt(S) + \delta_{PS,SP}}
 \end{align}
in direct analogy to Eq.~\eqref{eq:sld_dual_step2}.
This time, the inner sum in Eq.~\eqref{eq:shadow_step3} equates to $|\mathcal{S}| = 2^{n-k}$ if $P\in\{I,X,Y,Z\}^{\otimes n}$ is contained in the shadow $\tilde{\mathcal{S}}$, which is defined in Eq.~\eqref{def:shadow_coset} of the main text.
Conversely, if $P$ is not contained in $\tilde{\mathcal{S}}$, then the inner sum in Eq.~\eqref{eq:shadow_step3} vanishes.
Therefore, the map 
 $  \{I,X,Y,Z\}^{\otimes n} \rightarrow \RR,   $ 
\begin{align}
   P \longmapsto \frac{1}{|\mathcal{S}|} \sum_{S \in \mathcal{S}} (-1)^{\wt(S)+\delta_{PS,SP}}
\end{align}
is the indicator function on Rains' shadow $\tilde{\mathcal{S}}$. 
This shows that $\tilde a_i[\rho^\text{stab}_\text{QECC}]$ is the probability that a uniformly random $P\in \tilde{\mathcal{S}}$ has $\wt(P)=i$, 
which proves Eq.~\eqref{eq:shadow_step4} in the main text.


}

{\add 

\section{The shadow is a coset of the dual}
\label{app:proof_shadow_characterization}

In this appendix, we prove Prop.~\ref{lem:shadow_characterization} from the main text.}
First, let $Q\in \tilde {\mathcal{S}}$ be an arbitrary element in the shadow and consider the operator $L=PQ$.
For all stabilizers $S\in \mathcal{S}$ we thus have $PS =SP (-1) ^{\wt(S)}$, $QS =SQ (-1) ^{\wt(S)}$, and therefore $LS = SL$, 
i.e., $L\in \mathcal{S}^\perp$ is a logical Pauli operator.
We have thus shown the first inclusion $\tilde{\mathcal{S}} \subset P \mathcal{S}^\perp$.
For the other inclusion, consider $Q\in P\mathcal{S}^\perp$,
i.e., $Q=PL$ for some $L \in \mathcal{S}$. 
Then, we have $SQ = SPL =PSL(-1)^{\wt(S)}   =PLS(-1)^{\wt(S)}  = QS(-1)^{\wt(S)}$, which proves $\tilde{\mathcal{S}} \supseteq  P\mathcal{S}^\perp$. 

{\add 
Concerning the addendum, consider the case  $\tilde{\mathcal{S}} = \mathcal{S}^\perp$ first.
Then, 
comparing Eqs.~\eqref{def:normalizer}
and \eqref{def:shadow_coset} of the main text
shows}
that $\mathcal{S}$ cannot contain elements whose weight is odd. 
Conversely, if $\mathcal{S}$ is generated by   even-weight stabilizers, then all elements in the Abelian group $\mathcal{S}$ have an even weight 
{\add (Lem.~8 in Ref.~\cite{schatzki_tensor_rank_2024}),}
and the anticommutativity condition in Eq.~\eqref{def:shadow_coset} is vacuous. 
This then shows $\tilde{\mathcal{S}}= \mathcal{S}^\perp$ and completes the proof. 

\section{Notational overview}
\label{app:notation}

For convenience, this appendix explicitly repeats the matrix entries for all nine transforms that are displayed in Fig.~\ref{fig:macwilliams} of the main text.
As always, let $n$ be the number of qubits. 
The MacWilliams transforms are given by 
\begin{align} 
\tag{\ref{eq:macwilliams_matrix}}
    M_{i,j} &= \frac{1}{2^n} \sum_{l=0}^i
    \binom{n-j}{i-l} \binom{j}{l} (-1)^l 3^{i-l}\,,
\\
\tag{\ref{eq:macwilliams_antidiagonal}}
     M'_{i,j} &= \delta_{i,n-j} \,,
 \\
\text{and} \hspace{2mm}     
\tag{\ref{eq:macwilliams_diagonal}}
    \tilde{M}_{i,j} &=   (-1)^{n+i} \delta_{i,j} 
\end{align}
{\add in the bases of \emph{Shor--Laflamme} (SLDs), \emph{averaged subsystem purity} (APDs), and \emph{triplet probability distributions} (TPDs),} respectively.
To transform SLDs into APDs and back again,
one has to use the lower-triangular matrices with entries 
\begin{align} 
\tag{\ref{eq:trafo_sld_to_apd}}
    T'_{i,j} &= \frac{2^{n-i}}{\binom{n}{i}} \binom{n-j}{n-i}
    \\ 
\text{and} \hspace{2mm}     
\tag{\ref{eq:trafo_apd_to_sld}}
    T'^{-1}_{i,j} &= \frac{\binom{n}{j}}{2^{n-j}}
     \binom{n-j}{n-i} (-1)^{i+j}\,.
\end{align}
To transform SLDs into TPDs (Rains' shadow QWEs) and back again,
one must apply the Krawtchouk 
matrices whose entries are given by
\begin{align}
\tag{\ref{eq:trafo_sld_to_tpd_via_signed_mwt}}
    \tilde T_{i,j} &= \frac{1}{2^n} \sum_{l=0}^i
    \binom{n-j}{i-l} \binom{j}{l} (-1)^{j-l}  3^{i-l}
\\
\text{and} \hspace{2mm}     
\tag{\ref{eq:trafo_tpd_to_sld_via_signed_mwt}}
     \tilde T ^{-1}_{i,j} &= \frac{1}{2^n}\sum_{l=0}^i
     \binom{n-j}{i-l} \binom{j}{l} (-3)^{i-l}\,.
\end{align}
Finally, one can also directly convert APDs into TPDs, and vice versa,
via
\begin{align} \tag{\ref{eq:trafo_apd_to_tpd}}
     \tilde T ' _{i,j}  &= \frac{1}{2^n}
     \binom{n}{j} \sum_{l=0}^i
     \binom{n-j}{i-l}\binom{j}{l} (-1)^{j-l} 
\\
\text{and} \hspace{2mm}     
    \tag{\ref{eq:trafo_tpd_to_apd}}
    \tilde T '^{-1}_{i,j} & = 
\frac{1} {\binom{n}{i}}
\sum_{l=0}^i
\binom{n-j}{i-l}
\binom{j}{l}
{\left(-1\right)^{i-l}} \,.
\end{align} 
As the colors   in Fig.~\ref{fig:macwilliams} of the main text illustrate,
the only matrices from the above that never unfavorably amplify numerical imprecision (theory) and statistical errors (experiment) of the QWEs are $M'$, $\tilde M$, and $\tilde T '^{-1}$.

\section{Operator norms of QWE transformations}
\label{app:operator_norms}
\begin{figure}[t!]
    \centering
    \includegraphics{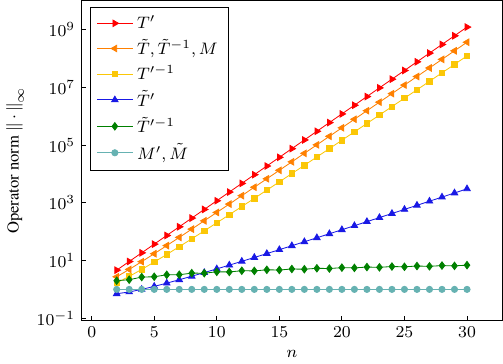}
    \caption{Operator norms of the linear maps acting on $\RR^{n+1}$ from Fig.~\ref{fig:macwilliams} as a function of the number of qubits. Except for $\tilde T'^{-1}$, $M'$, and $\tilde M$, all norms are growing exponentially.
    Hence, in general, one has to work with an increased numerical precision when transforming between QWEs.  }
    \label{fig:trafo_operator_norms}
\end{figure}

In Lemma~\ref{lem:operator_norms} of the main text,
we have analytically computed the operator norms of the 
{\add two-copy} 
observables $ \twocopyobservable_i, \twocopyobservable'_i,$ and $\tilde \twocopyobservable_i$, which are all highly degenerate.
Indeed, in Sec.~\ref{sec:bell_sampling}
{\add of the main text}
we have shown that these operators are diagonal in the basis of Bell pairs $\ket{\boldalpha}$ [Eq.~\eqref{eq:bell_pairs}] and that their eigenvalues only depend on the number of singlets in $\ket{\boldalpha}$ but not on their position within $\ket{\boldalpha}$.
The core of the efficiency of the quantum weight enumerator machinery is to lift this degeneracy by replacing $\twocopyobservable_0,\ldots, \twocopyobservable_n$ and $\twocopyobservable'_0, \ldots, \twocopyobservable'_n$ with $\tilde{T}^{-1}$ and $\tilde{T}'^{-1}$, respectively.
In the process, the operator norms do not remain the same.
In Fig.~\ref{fig:trafo_operator_norms}, we show the operator norms of 
$\tilde{T}^{-1}$ and $\tilde{T}'^{-1}$ alongside those of the other transformations from 
{\add App.~\ref{app:notation}.}
We see that $\|{\tilde T ^{-1}}\|_\infty $ (red) grows exponentially while $\|{\tilde T '^{-1}}\|_\infty $ (green) increases 
{\add sublinearly}.
This is easily explained as the behavior inherited by their corresponding $2n$-qubit observables.
Note that we have strong numerical evidence for $\|{\tilde T '^{-1}}\|_\infty \lesssim 1.25 \sqrt{n}$.
Concerning the other transformations, we see that the exponentially-growing operator norms of $\tilde T, \tilde T^{-1}$, and $M$ all coincide (orange)
{\add since these matrices are obtained from one another merely by modifying the signs of certain columns or rows.}
The worst operator norm is that of $T'$ (red), which transforms SLDs into APDs.
Also that of $\tilde T'$ (blue) grows exponentially albeit less quickly.
Finally, we see $ \| M'\|_\infty = \|\tilde M\|_\infty = 1$ 
({\add cyan}) 
as these are the bases
{\add in which the self-inverse MacWilliams transform  is a symmetric matrix and, therefore, orthogonal}.

\section{\add Cost reduction for MacWillams' transform}
\label{app:trafo_recursion}

From a naive perspective, {\add precomputing} the transformation matrices $M$, $\tilde T $, $\tilde T ^{-1}$,  $\tilde T '$, and $\tilde T '^{-1}$ from their definitions in App.~\ref{app:notation} has a   cost of ${O}(n^4)$.
Here we show how this complexity can be reduced to ${O}(n^2)$ by identifying useful recurrence relations between matrix entries.
{\add With these simplifications,
precomputing on a standard laptop   all $6(n+1)$ single-shot estimators  from Sec.~\ref{sec:measure_sld} of the main text requires, e.g., less than three seconds for $n=1000$ qubits per copy.}

In order to compute $\tilde T '^{-1}$ for example, we start by omitting the factor $1/\binom{n}{i}$ which can be reapplied later. According to Eq.~\eqref{eq:trafo_tpd_to_apd}, the first row then consists of all ones while the first column contains the binomial coefficients with alternating signs, i.e.,  $(-1)^i\binom{n}{i}$ for $i\in\{0,\dots,n\}$.
The crucial insight is that in each $2\times 2$ block of this matrix, the lower-right entry is just the sum of the other three entries.
Following this prescription, the entire matrix can be build up from top-to-bottom and left-to-right in ${O}(n^2)$ steps. 
Finally, each row is divided by the absolute value of its first entry, thus completing the calculation of $\tilde T '^{-1}$. 

Similar instructions are found for the remaining matrices. Specifically, $M$ has almost the same boundary conditions as $\tilde T'^{-1}$ (up to normalization), only that $(-1)^i\binom{n}{i}$ appears in the last column instead of the first one. The matrix can then be completed according to $M_{i, j} = 3M_{i-1,j+1}+M_{i-1,j}+M_{i,j+1}$.
Moreover, we again stress the fact that the two matrices $\tilde T $ and $\tilde T ^{-1}$ arise from $M$ by changing signs of odd rows and columns, respectively. It is therefore not necessary to
to recompute  $\tilde T $ and $\tilde T ^{-1}$ from their definitions when  $M$ is already known.

This approach is limited by numerical precision when adding numbers of largely differing orders of magnitude, however, this issue can be avoided by using integers until the application of normalization factors. In general, storing most of the transformation matrices with exponential operator norm in standard IEEE 754 64-bit floating point format is only possible until $n=1029$. 
Until then, we find that the recursive method yields an enormous speed-up compared to the naive approach.

\section{Benefits of single-setting protocols}
\label{app:single_setting_benefits}

{\add In this appendix, we discuss different strategies for learning the \emph{sector length distribution} (SLD) of an $n$-qubit state $\rho$.
In particular, we discuss for which experimental setups the single-setting property of the two-copy Bell sampling protocol,
which is developed in the main text of this paper,
constitutes a game-changing advantage over randomized protocols.
}

Recall from Eq.~\eqref{def:sld} that the $i$-body sector length $a_i[\rho]$ of an $n$-qubit state $\rho$  is defined as the normalized sum of $\Tr[\rho P ]^2 $ over all $3^i\binom{n}{i}$ Pauli operators $P$ with $\wt(P)=i$.
Thus, a brute-force approach for 
{\add learning}
the entire SLD $\mathbf{a}[\rho]$ is to first measure all Pauli expectation values (or just their squares) and then to compute the corresponding sums.
Clearly, such a 
{\add learning protocol}
suffers from an exponential postprocessing cost.
But even if we had infinite classical computational power, 
how can we obtain all $\Tr[\rho P]^2$ in the first place?
A naive strategy uses \emph{classical shadows}~\cite{ohliger_efficient_and_2013,aaronson_shadow_tomography_2018, huang_predicting_many_2020, elben_the_randomized_2023, fischer_dual_frame_2024} 
(unrelated to Rains' shadow),
where individual copies of $\rho$ are measured after applying a random unitary $U$ drawn from a suitable ensemble of readout circuits.
While each such measurement enables for all $\mathbf{r}\in \FF_2^n$ access to $\Tr[\rho U^\dagger Z^{\mathbf{r}}U]$,
which is the expectation value of a Pauli operator if $U$ is a Clifford circuit,
the classical shadows approach has several drawbacks:
(i) measuring all $\Tr[\rho P]^2$ to constant additive precision provably requires exponentially many samples in $n$~\cite{chen_a_hierarchy_2021};
(ii) the readout circuits are often---but not always~\cite{miller_hardware_tailored_2024, bertoni_shallow_shadows_2024}---challenging to implement reliably; and 
(iii) for many experimental setups, the randomized nature of classical shadows is highly undesirable in practice as updating the readout circuit on a shot-by-shot basis can slow down experimental data acquisition by several orders of magnitude. 

{\add 
While issue (iii) can be solved in principle via sophisticated techniques such as partial compilation~\cite{dalvi_one_time_2024},
doing so requires expensive hardware and sophisticated software implementations.
Moreover, such solutions can only reduce the slowdown without ever fully eliminating it.
}

In some regards---it circumvents issue (ii) and (iii)---a better alternative is based on single-setting informationally complete measurements.
These can be realized by embedding each qubit into a ququart at the readout stage~\cite{fischer_ancilla_free_2022, stricker_experimental_single_2022}
and  enable access to $\Tr[\rho P] $ for all $P\in \{I,X,Y,Z\}^{\otimes n}$ simultaneously
without the need of ever altering the readout circuit.
However, just as classical shadows, also single-setting informationally complete measurements fall into the paradigm of \emph{single-copy access protocols}, which all suffer from problem~(i).
Two-copy Bell sampling, on the other hand, allows us to measure $\Tr[\rho P]^2$ for all $P$ up to constant error with a number of shots that is constant in $n$~\cite{huang_information_theoretic_2021}.
However, it was hitherto unclear if and how one can also avoid the classical postprocessing cost and  under what circumstances errors on the individual $\Tr[\rho P]^2$ may or may not conspire when they are added up in the SLD.
As we show now, the exponential postprocessing cost can always be avoided.
Later in Sec.~\ref{sec:sample_complexity}, we will turn back to the more subtle issue about sample complexity.

\section{On triplet probability distributions}
\label{app:moments_and_bounds_of_tpds}

In this appendix, we translate results from the sector length literature into the picture of triplet probabilities.
First, we express moments of triplet probability distributions (TPDs) in terms of few-body Pauli correlators, 
in {\add App.}~\ref{app:moments_tpd}.
The question which probability distributions  may or may not arise in Bell sampling experiments is  addressed in {\add App.}~\ref{app:attainable_tpds}.

\subsection{Moments of triplet probability distributions}
\label{app:moments_tpd}

Consider an $n$-qubit state $\rho$ and denote the $i$-th moment of its TPD ${\mathbf{\tilde a}}[\rho]  = (\tilde a_x[\rho])_{x=0}^n$ by
\begin{align}
     \langle \tilde{x}^i \rangle 
    = \sum_{x=0}^n \tilde a_x[\rho] x^i \, .
\end{align}
Simplifying the $i$-th component of 
$\mathbf{a}[\rho] = \tilde T ^{-1} \mathbf{\tilde a}[\rho]$
and denoting by $A_i[\rho] = 2^n a_i[\rho]$ the unnormalized version of Shor--Laflamme QWEs,
we find
\begin{align} 
\label{eq:a1_as_tpd_moments}
A_1[\rho] 
&=  4 \langle \tilde{x} \rangle  - 3n \,,
\\
\label{eq:a2_as_tpd_moments}
A_2[\rho] 
&= 8  \langle \tilde{x}^2\rangle 
- (12n-4)  \langle \tilde{x}\rangle 
+ 9 \binom{n}{2} \,,
\\
\label{eq:a3_as_tpd_moments}
\text{and} \hspace{2mm}
A_3[\rho] &= 
\frac{32}{3} \langle \tilde{x}^3 \rangle  
- (24n-16)  \langle \tilde{x}^2 \rangle  \\
&  + \frac{{54n^2-90n+28}}{3}   \langle \tilde{x} \rangle  
-  27 \binom{n}{3} \,.
\end{align}
Rearranging Eq.~\eqref{eq:a1_as_tpd_moments} yields the TPD's mean
\begin{align} \label{eq:tpd_mean}
     \langle \tilde{x}\rangle & = \frac{3n+ A_1[\rho]}{4}\, .
\end{align}
Similarly, by inserting Eq.~\eqref{eq:tpd_mean} into Eq.~\eqref{eq:a2_as_tpd_moments} we obtain
the second moment of the TPD,
\begin{align}
     \langle \tilde{x}^2\rangle & = \frac{A_2[\rho] + (3n-1)(3n+A_1[\rho]) - 9\binom{n}{2} }{8} 
\end{align}
as well as its variance  
\begin{align} \label{eq:tpd_variance_via_sld}
     \langle \tilde{x}^2\rangle -  \langle \tilde{x} \rangle^2 = \frac{2A_2[\rho] - A_1[\rho]^2  -2A_1[\rho]  +3n}{16} \,. 
\end{align}
Note that, for pure graph states \cite{hein_multiparticle_entanglement_2004}, $A_1[\rho]$ is equal to the number of isolated vertices, while $A_2[\rho]$ counts leaves and twin pairs of the underlying graph~\cite{miller_shor_laflamme_2023}.

In principle, we could go on and express the $i$-th moment $ \langle \tilde{x} ^i  \rangle$ of the TPD as a linear combination of $A_0[\rho]$, \ldots, $A_i[\rho]$.
When 
\begin{align}
    \rho = \frac{1}{|\mathcal{S}| } \sum_{S\in\mathcal{S}} S
\end{align}
is a pure stabilizer state or, more generally, the maximally mixed state of a stabilizer QECC with stabilizer group $\mathcal{S}$,
we can  compute $A_i[\rho] = |\{S\in\mathcal{S} \ \vert \ \wt(S) = i\}|$
in classical runtime $O(n^i)$,
which is efficient if $i $ is constant.
Similar is true for matrix product states with a low bond dimension.

\subsection{Bounds on triplet probability distributions}
\label{app:attainable_tpds}

For every $n$-qubit state $\rho$, 
the TPD $\mathbf{\tilde a}[\rho]  $ is a discrete probability distribution on the finite set $ \{0, \ldots, n \}$.
However, not every mathematically well-defined probability distribution on $ \{0, \ldots, n \}$ can be physically realized as a TPD arising in a two-copy Bell measurement.
For example, from Eq.~\eqref{def:rains_qwe} 
{\add of the main text},
we know that
the zero-triplet probability $\tilde{a}_0[\rho] = \tfrac{\Tr[\rho \tilde \rho]}{2^n} $ of any state $\rho$ is upper bounded by $2^{-n}$. 
This bound is tight and attained iff {\add $\rho$ is pure and} $\rho= \tilde\rho$, e.g., for half-filled Dicke states and GHZ states, assuming $n$ even.
At the other extreme of the  distribution,
however, 
we can have probabilities as large as $\tilde a_n[\ket{0}\!\bra{0}^{\otimes n} ] = 1 $.
These observations are manifestations of an even more general asymmetry in TPDs that we are going to unfold now.

Recall from Eq.~\eqref{eq:tpd_mean} that the mean $ \langle \tilde{x}\rangle  $ of any TPD
can be expressed in terms of the (unnormalized) 1-body sector length $A_1[\rho] = 2^n a_1[\rho]$.
From its definition [Eq.~\eqref{def:sld}] it is clear that $A_1[\rho]$ cannot be negative.
Hence, we can lower bound the mean of the TPD via 
\begin{align} \label{eq:tpd_mean_bound}
 \langle \tilde{x} \rangle \ge \frac{3n}{4} \,,
\end{align}
which implies that one will always observe a fairly large number of triplets in two-copy Bell sampling experiments.
The bound in Ineq.~\eqref{eq:tpd_mean_bound} is tight and attained for the maximally mixed state whose triplet probabilities
\begin{align}
    \tilde a_i \left[ \frac{\mathbbm 1}{2^n}\right] = \binom{n}{i} \frac{3^i}{4^n}
\end{align}
are binomially distributed since every Bell measurement randomly (with equal probability $25\%$) results in one of three triplets or in a singlet.
{\add This proves Cor.~\ref{cor:triplet_mean_bound} in the main text.}
Conversely, we recover the well-known bound $A_1[\rho] \le n$ by inserting the trivial bound $
 \langle \tilde{x} \rangle \le n$ into Eq.~\eqref{eq:tpd_mean}.
More generally, it has been shown~\cite{wyderka_characterizing_quantum_2020} that
\begin{align} \label{eq:bounds_on_low_body_sector_lengths}
0 \le A_i[\rho] \le \begin{cases}
        \binom{n}{1} &  \text{ if $i=1$ and $n\ge 1$}\, , \\
        \binom{n}{2} &  \text{ if $i=2$ and $n\ge 3$}\, , \\
        \binom{n}{3} &  \text{ if $i=3$ and $n\ge 5$}\, , \\
    \end{cases}
\end{align}
which directly translates into bounds on linear combinations of mean, variance, and skewness of admissible TPDs via Eqs.~\eqref{eq:a1_as_tpd_moments}--\eqref{eq:tpd_variance_via_sld}.
Unfortunately, 
there cannot exist a non-trivial lower bound on the TPD's variance as the example of any pure product state shows.
Nonetheless, yet another bound on linear combinations of sector lengths (Cor.~9 in Ref.~\cite{wyderka_characterizing_quantum_2020}) translates into
\begin{align} \label{eq:tpd_ssa_bound}
    \sum_{i=0}^n \binom{n-i}{2}(2i+2-n) \tilde{a}_i[\rho] \ge 0
\end{align}
for all $n$-qubit states $\rho$.
For example, inserting $n=3$ into Ineq.~\eqref{eq:tpd_ssa_bound} yields $\tilde a_1[\rho] \ge 3\tilde a_0[\rho]$ which is a
{\add non-trivial result} 
about two-copy Bell sampling experiments.
Finally, the fact that the GHZ state maximizes $n$-body correlations~\cite{tran_correlations_between_2016, eltschka_maximum_nbody_2020},
\begin{align}
0 \le A_n[\rho] \le A_n[\Psi_\text{GHZ}] = 2^{n-1} + \delta_{n, \text{even}}
\end{align}
also translates into a bound on linear combinations of triplet probabilities.

\begin{figure}
    \centering
    \includegraphics{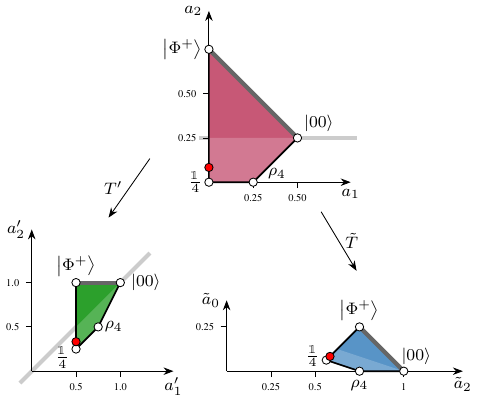}
    \caption{2D {\add crosssections through}   3D polytopes of admissible QWEs for $n=2$ qubits.
    SLDs (pink) were characterized in Ref.~\cite{wyderka_characterizing_quantum_2020} and the admissible APDs (green) and TPDs (blue) follow via the transforms $T'$ and $\tilde T $ from Fig.~\ref{fig:macwilliams}.
    Each polytope is the convex hull of four vectors of QWEs for the states $\ket{\Phi^+}\!\bra{\Phi^+}$, $\ket{0,0}\!\bra{0,0}$, $\mathbbm 1 /4$, and
    $\rho_4 = {\add \tfrac{1}{2}}(\ket{0,0}\!\bra{0,0} + \ket{0,1}\!\bra{0,1})$.
    The thickened 
    {\add gray} 
    boundary lines precisely correspond to pure states. 
    In darker regions, all states are entangled. 
    However, also in lighter regions, entangled states do exist, e.g., the Werner state~\cite{werner_quantum_states_1989} $(\tfrac{1}{3}+ \varepsilon) \ket{\Phi^+ }\!\bra{\Phi^+} + (\tfrac{2}{3}-\varepsilon) \tfrac{\mathbbm 1}{4}$  is NPT entangled for every $\varepsilon > 0$  (red dot at $\varepsilon = 0$)~\cite{peres_separability_criterion_1996}.} 
    \label{fig:polytopes2qubits}
\end{figure}

{\add Next, let us discuss} the polytopes of admissible QWEs for $n=2$ qubits, see Fig.~\ref{fig:polytopes2qubits}.
It was shown in Ref.~\cite{wyderka_characterizing_quantum_2020} that a putative SLD $\mathbf{a}=(a_0,a_1,a_2)$ can be physically realized (i.e., $\mathbf{a}=\mathbf{a}[\rho]$ for some $2$-qubit state $\rho$) iff $\mathbf{a}\in \RR^3$ is contained in the convex hull of  
\begin{align}
   \hspace{-12mm}  & \mathbf{a}\left[\ket{\Phi^+}\!\bra{\Phi^+}\right]                      \hspace{-15mm} &= \hspace{3mm} \tfrac{1}{4}(1,0,3)\,,   \\
   \hspace{-12mm}  & \mathbf{a}\left[\ket{0,0}\!\bra{0,0}\right]                              \hspace{-15mm} &= \hspace{3mm} \tfrac{1}{4}(1,2,1)\,,    \\
  \hspace{-12mm} & \mathbf{a}\left[\ket{0}\!\bra{0} \otimes \tfrac{\mathbbm 1}{2}\right]  \hspace{-15mm} &= \hspace{3mm} \tfrac{1}{4}(1,1,0)  \, ,\\
 \hspace{-12mm} \text{and } \hspace{2mm}
     & \mathbf{a}\left[\tfrac{\mathbbm1}{4}\right]                            \hspace{-15mm} &= \hspace{3mm} \tfrac{1}{4}(1,0,0) \,. 
\end{align}
In Fig.~\ref{fig:polytopes2qubits}, we depict (without any loss of information a two-dimensional 
{\add crosssection through}) 
the resulting sector length polytope (pink).
By applying the basis change matrices $T'$ and $\tilde T $ from Fig.~\ref{fig:macwilliams},
we  convert the SLD polytope into the corresponding polytopes of admissible $2$-qubit averaged purity distributions (green)
and triplet probability distributions (blue).
In each basis, we have available one entanglement criterion 
[recall Eqs.~\eqref{eq:nbody_criterion}, \eqref{eq:purity_criterion}, and~\eqref{eq:concurrence_criterion}],
namely
\begin{itemize}
    \item $a_2[\rho] > \tfrac{1}{4} \ \Longrightarrow \ \rho$ entangled ($n$-body sector length),
    \item $a'_2[\rho] > a'_1[\rho] \ \Longrightarrow \ \rho$ entangled (purity), and
    \item $\tilde a_2[\rho] > 1-3\tilde a_0[\rho]  \ \Longrightarrow \ \rho$ entangled (concurrence).
\end{itemize}
We highlight the regions of definitively-entangled states with a darker haze in Fig.~\ref{fig:polytopes2qubits}.
Coincidentally, all three criteria are equivalent in the present case of $n=2 $ qubits.
For $n\ge3$, on the other hand, Fig.~\ref{fig:dicke_robustness} from the main text clearly shows that the three entanglement criteria are no longer equivalent.
Note that also for $n=3$ qubits, the set of admissible QWEs is known to be a polytope (the convex hull of seven four-dimensional vectors)~\cite{wyderka_characterizing_quantum_2020}
while it is a grand open challenge to fully characterize the admissible sets of SLDs, APDs, and TPDs in the general case of $n\ge 4 $ qubits.

{\add 
Finally, 
let us remark that---for small qubit numbers---QWEs are strong enough to detect multiparite entanglement.
For example, every 3-qubit state $\rho$ and every 4-qubit state $\sigma$ with $a_3[\rho]>3/8$ and  $a_3[\sigma] > 7/16$, respectively,  must be genuinely multipartite entangled~\cite{wyderka_characterizing_quantum_2020}.}

{\add
\section{Error mitigation for learning weight enumerators}
\label{app:error_mitigation_qwes}

In this appendix, we introduce two new error mitigation techniques designed to heuristically reduce noise in experimentally measured \emph{quantum weight enumerators}  (QWEs).
The first technique addresses both \emph{state preparation and Bell measurement}  (SPABM) errors, as it is designed to reproduce the QWEs of a pure state. 
The second technique targets only errors in the Bell measurement circuit, which are primarily due to noisy $\Cnot$ gates.
Both techniques share the general limitation of error mitigation strategies in that they are not scalable. 
In our case, the obstacle is that \emph{Shor–Laflamme distributions} (SLDs) cannot always be measured efficiently.
In Sec.~\ref{sec:experiment_states} of the main text, however, we report on a two-copy Bell sampling experiment with $n=6$ qubits per copy, where this restriction is not yet prohibitive. 
While both mitigation strategies could be applied in principle,
we only apply the second one to the experimental data produced in this work.
As this choice aims to remove only the readout noise from the Bell measurements,
it allows us to investigate the entanglement structure of the noisy states produced by our quantum computer, which is the focus of our experimental demonstration.

To remove all SPABM errors, our first mitigation strategy makes the heuristic assumption that all noise processes can be effectively combined into a local depolarizing error channel, $\mathcal{E}^{\otimes n}_p$, 
with a single unknown error parameter $p\in[0,1]$.
By Eq.~\eqref{eq:sector_length_decay} of the main text, 
such noise reduces the $i$-th Shor--Laflamme QWE by a factor of $(1-p)^{2i}$ for all $i\in\{0,\ldots,n\}$. 
We therefore determine the parameter $p$ such that
\begin{align}
    \sum_{i=0}^n a_i^\text{miti}[\rho_\text{exp}] = \Tr[\rho_\text{targ} ^2]
\end{align}
under the condition $(1-p)^{2i} a_i^\text{miti} [\rho_\text{exp}] = a_i^\text{raw}[\rho_\text{exp}]$,
where $\rho_{\text{targ}}$ and $\rho_{\text{exp}}$ denote the target state and its experimental realization, respectively. For most applications, one will work with $\Tr[\rho_{\text{targ}}^2] = 1$.

The second mitigation strategy removes only Bell measurement errors, leaving state-preparation noise intact. 
This is appropriate for characterizing the noisy states generated by current prototypes of quantum computers.
To this end, we adopt a weaker heuristic assumption than in the previous case, namely that the effective noise (when errors are backpropagated from the Bell measurements onto the state) acts diagonally on the SLD.
In other words, different-body sector lengths do not interfere with one another.
Unlike the local depolarizing model, where each $i$-body sector length is reduced by the factor $(1-p)^{2i}$, we instead allow each to be independently reduced by its own parameter $\lambda_i$.
To calibrate the $\lambda_i$ values, we use the reference state $\Psi_1 =  \ket{0}\!\bra{0}^{\otimes n}$, for which preparation errors can be neglected. 
Its measured QWEs, $a_i^{\text{raw}}[\rho_1]$, are smaller than the ideal values $a_i[\Psi_1]= \binom{n}{i}/2^n$ due to Bell measurement errors.
This motivates the definition
\begin{align} \label{eq:migation_strategy}
    \lambda _i 
    =  \frac{ a^\text{raw}_i[\rho_1] }{\binom ni / 2^n  }
\end{align}
for all $i\in\{0,\ldots, n\}$.
For all other states $\rho_j$ under investigation,
we reuse these damping factors
to define the mitigated SLDs via
\begin{align}
    a^\text{miti}_i[\rho_j] 
    = \lambda_i^{-1}
    a^\text{raw}_i[\rho_j] 
\end{align}
By definition, we thus have ${a}_i^\text{miti}[\rho_1] = {a}_i[\Psi_1]$.
For the other states, we generally expect 
$a^\text{raw}_i[\rho_j] \le a^\text{miti}_i[\rho_j] \le a _i[\Psi_j]$.
Further applying the transforms $T'$ and $\tilde T $ from App.~\ref{app:notation} to the vector $\mathbf{a}^\text{miti}[\rho_j]$,
we can also obtain mitigated values for Rains' unitary enumerators and Rains' shadow enumerators.

In the experiment reported in Sec.~\ref{sec:experiment_states} of the main text, the application of this second mitigation strategy yields the damping vector
$\boldsymbol{\lambda} =  (1, 0.957 , 0.919 ,  0.885  , 0.852 , 0.819  , 0.785)$,
which is in excellent agreement with the local depolarizing noise model, i.e., $\lambda_i \approx (1-p)^{2i}$, corresponding to an effective per-qubit error rate of $p \approx 0.02$.

}

\section{Preparing 6-qubit AME states with only 7 CNOTs}
\label{app:ame_circuit}

\begin{figure}
    \centering
    \includegraphics[width=.8\columnwidth]{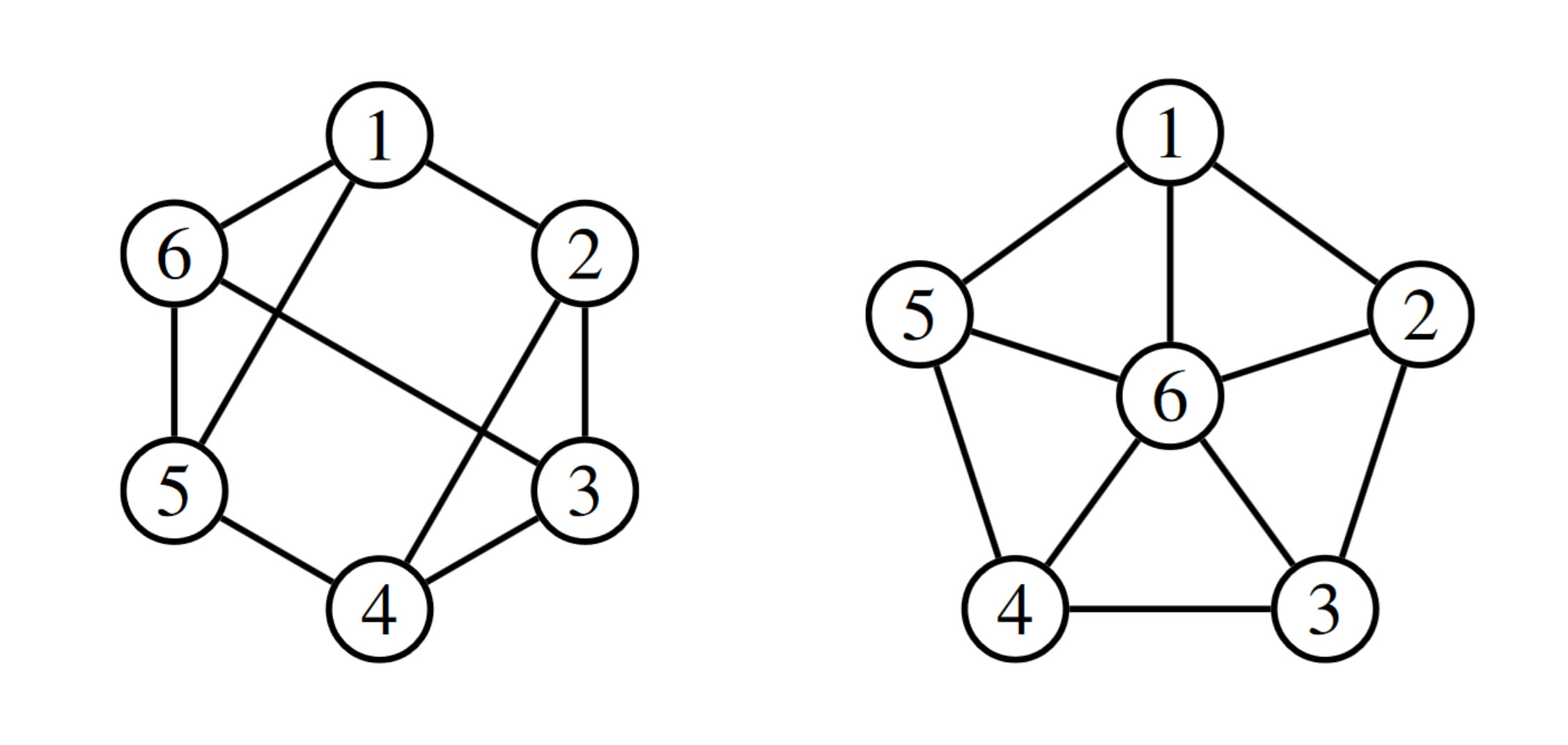}
    \caption{Realizations of six-qubit AME states as graph states with 9 (left) and 10 (right) edges, respectively.}
    \label{fig:ame_graph_states}
\end{figure}

\begin{figure}
    \centering 
    
    \includegraphics[width=.8\columnwidth]{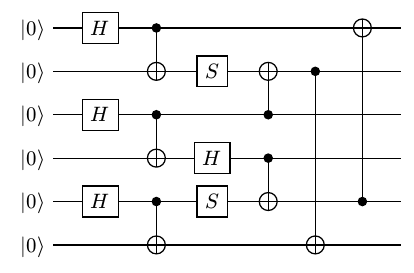}
    \caption{Preparation circuit for a six-qubit absolutely maximally entangled state. This circuit requires fewer \textsc{Cnot} gates than the naive graph state preparation circuits from Fig.~\ref{fig:ame_graph_states}.}
    \label{fig:ame_circuit}
\end{figure}

{\add 
In this appendix, we present a new preparation circuit 
for a six-qubit state that is \emph{absolutely maximally entangled} (AME).
Recall that a pure state $\Psi = \ket{\psi}\!\bra{\psi}$ is called \emph{$m$-uniform} 
iff all of its $m$-body marginals are maximally mixed, i.e., iff 
\begin{align}\label{eq:sld_m_uniform}
a_1[\Psi] = \ldots = a_m[\Psi] = 0 \, .
\end{align}
By the Schmidt decomposition, the largest value of  $m$ for which
there is any hope for Eq.~\eqref{eq:sld_m_uniform} to hold
is $m=\lfloor \tfrac{n}{2}\rfloor$.
In this case, $\Psi$ is called an AME state.
For qubits, such AME states only exist for $n\in\{2,3,5,6\}$ parties~\cite{huber_absolutely_maximally_2017}.

A six-qubit AME state can be realized in form of a \href{https://graphvis.uber.space/?graph=6_66A7}{cylinder graph state}
or a \href{https://graphvis.uber.space/?graph=6_4E6F}{wheel graph state}}
whose graphs are displayed on the left and right of Fig.~\ref{fig:ame_graph_states}, respectively.
However, in the graph state paradigm, the number of edges (here 9 and 10) is precisely equal to the number of two-qubit gates in the corresponding preparation circuit.
We can do better by leaving the graph state paradigm.
Via a randomized search over low-depth Clifford circuits,  we discover an AME state that can be prepared with only seven two-qubit gates, see Fig.~\ref{fig:ame_circuit}.
To our knowledge this reduces the $\Cnot$ cost of the best AME state preparation circuit from 9 to 7.
This is also the circuit that we used 
in our experimental demonstration presented in Sec.~\ref{sec:experiment_states} from the main text.

\section{QWEs for superpositions versus mixtures}
\label{app:superposition_vs_mixtures}

\begin{figure*}
    \centering
    \includegraphics{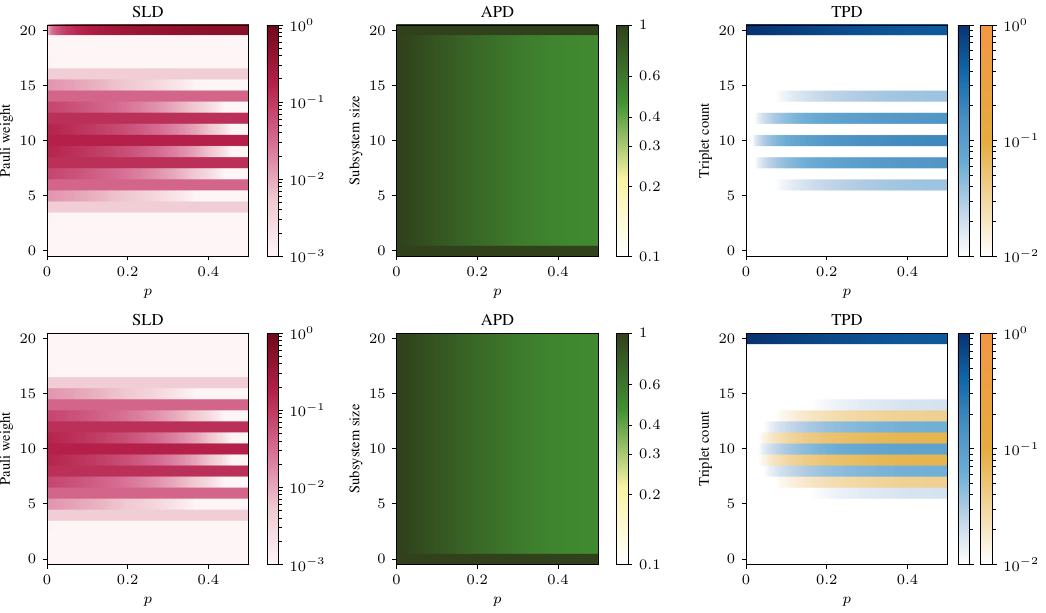}
    \caption{QWEs for a parameterized superposition (top)  and mixture (bottom) of  $\ket{0}^{\otimes n}$ and $\ket{1}^{\otimes n}$ for $n=20 $ qubits.
    See Eqs.~\eqref{eq:parameterized_superposition} and~\eqref{eq:parameterized_mixture} for precise definitions. 
    Vertical slices through the plots (rotated {\add clockwise} by $90^\circ$) correspond to QWE distributions of a given state, similar to Fig.~\ref{fig:experiment_states} from the main text. The three distributions show the same information in different vector space bases of $\RR^{n+1}$ as discussed in Fig.~\ref{fig:macwilliams} of the main text.
    }
    \label{fig:superposition_vs_mixtures}
\end{figure*}

Here
we showcase how superpositions and mixtures may differently manifest themselves in the three distributions of 
{\add \emph{quantum weight enumerators}}
(QWEs).
For concreteness, we consider the following two families of parameterized $n$-qubit states.
First, a superposition state ${\Psi_\text{sup}(p)}= \ket{\psi(p)} \! \bra{\psi(p)}$ with state vector 
\begin{align} \label{eq:parameterized_superposition}
    \ket{\psi(p)} = \sqrt{p} \ket{0}^{\otimes n} + \sqrt{1-p} \ket{1}^{\otimes n}
\end{align}
and second, the corresponding incoherent mixture  
\begin{align} 
\label{eq:parameterized_mixture}
\rho_\text{mix}(p) = p \ket{0}\!\bra{0}^{\otimes n} + (1-p) \ket{1}\!\bra{1}^{\otimes n} \,.
\end{align}

Note that $\ket{\psi(1-2/n)}$ recently appeared as a logical state of an $\llbracket n,1,2\rrbracket$ code that was used to construct a family of $\llbracket n, k, d\rrbracket$ QECCs (for every $k,d=\text{const}$), which have logical states with vanishingly small geometric entanglement in the limit of large $n$~\cite{bravyi_how_much_2025}.
This is somewhat surprising because, under very mild assumptions, logical states are highly entangled~\cite{bravyi_how_much_2025}.

By symmetry (QWEs are local-unitary invariants), we can assume $0\le p \le 0.5 $ without loss of generality.
Then, ${\Psi_\text{sup}(p)}$ is entangled for all values of $p\neq 0$ whereas $
\rho_\text{mix}(p)$ is always fully separable.
This form of entanglement is a global property as both states share exactly the same marginals and, in consequence,
the same (averaged) subsystem purities
\begin{align} \label{eq:apd_superpos_mix}
    a'_i[\Psi_\text{sup}(p)] = 
    a'_i[\rho_\text{mix}(p)] =   1-2p(1-p)
\end{align}
for $0<i<n$.
{\add For $i=0$, on the other hand, we have}  $a'_0[\Psi_\text{sup}(p)] = a'_0[\rho_\text{mix}(p)] = 1$
{\add  by normalization}.
The only difference that is thus captured by the APD are the global purities
$a'_n[\Psi_\text{sup}(p)] = 1$ and
$a'_n[\rho_\text{mix}(p)] = 1-2p(1-p)$, respectively.
As discussed in the main text,
we can convert APDs into SLDs and TPDs.
We plot all of them for both parameterized states on $n=20$ qubits in Fig.~\ref{fig:superposition_vs_mixtures}.
The differences are subtle but clearly visible.

Let us first discuss the sector length distributions  (pink), which are displayed in  the left panel of Fig.~\ref{fig:superposition_vs_mixtures}. 
They look mostly the same for $\Psi_\text{sup}(p)$ and $\rho_\text{mix}(p)$.
In fact, they precisely coincide in all entries except for the last---a property inherited from their APDs [Eq.~\eqref{eq:apd_superpos_mix}] because
$T'^{-1}$ [Eq.~\eqref{eq:trafo_apd_to_sld}]
is a lower-triangular matrix.
For $p=0$, we still have $\Psi_\text{sup}(0)=\rho_\text{mix}(0)$  and the SLD is given by the symmetrical binomial distribution---the SLD unique to pure product states [Eq.~\eqref{eq:sld_pure_product}].
As we increase $p$, we see that
$\mathbf{a}[\Psi_\text{sup}(p)]$ continuously changes until the SLD 
\begin{align} \label{eq:ghz_sld}
    a_i [\Psi_\text{GHZ}] &= \frac{1}{2^n}\binom{n}{i}\delta_{i,\text{even}}  + \frac{1}{2}\delta_{i,n} 
\end{align}
of the GHZ state $\Psi_\text{GHZ} = \Psi_\text{sup}(p=0.5)$ is reached~\cite{aschauer_local_invariants_2004}.
Similarly, the odd components of $\mathbf{a}[\rho_\text{mix}(p)]$ converge to zero until the SLD
\begin{align}
  a_i[\rho_\text{rep} ] = \frac{1}{2^n}\binom{n}{i}\delta_{i,\text{even}}
\end{align}
of the maximally mixed state 
$\rho_\text{rep} =  \tfrac{1}{2} \ket{0}\!\bra{0}^{\otimes n} + \tfrac{1}{2} \ket{1}\!\bra{1}^{\otimes n}$ within the $n$-qubit repetition code space is reached.
Note that $\Psi_\text{sup}(p)$ lies within this subspace too.
We analytically find  (assuming $n$ is even)
\begin{align}
    a_n[\Psi_\text{sup}(p)]  &=  \frac{1}{2^n} + 2p(1-p) \hspace{2mm} \text{ and} \\
    a_n[\rho_\text{mix}(p)]  &=  \frac{1}{2^n} \, .
\end{align}
Therefore, the $n$-body sector length criterion [Eq.~\eqref{eq:nbody_criterion}] identifies $\Psi_\text{sup}(p)$ as entangled for all values of $p\in (0, 0.5]$ while $a_n[\rho_\text{mix}(p)]$ is as large as possible given the fact that $\rho_\text{mix}(p)$ is a separable state. 
Finally, note that the  GHZ state maximizes the $n$-body sector length among all $n$-qubit states~\cite{tran_correlations_between_2016, eltschka_maximum_nbody_2020}.

Next, consider the central panel of Fig.~\ref{fig:superposition_vs_mixtures},
where we have plotted the averaged purity distributions (APDs).
As mentioned above, the APDs (green) coincide except for $a'_n[\Psi_\text{sup}(p)]=1$
and
$a'_n[\rho_\text{mix}(p)]=1-2p(1-p)$.
Hence, also the purity criterion [Eq.~\eqref{eq:purity_criterion}] detects entanglement in $\Psi_\text{sup}(p)$ for all values of $p \in (0, 0.5]$ 
while $\rho_\text{mix}(p)$ again only meets the separability bound without surpassing it.

Last but not least, the triplet probability distributions 
{\add TPDs}
are shown in the right panel of Fig.~\ref{fig:superposition_vs_mixtures}.
Here, we use blue and yellow colors to distinguish the cases of even and odd singlets, respectively.
In other words, we are---strictly speaking---depicting the dual TPD $\mathbf{\tilde{b}}$ rather than the TPD $\mathbf{\tilde a}$ itself.
Recall that the two are related via $\tilde b_i[\rho] = (-1)^{n-i} \tilde a_i[\rho]$ for all $n$-qubit states $\rho$, i.e.,
by Eq.~\eqref{eq:purity_from_tpd}
{\add in the main text,}
we can always write $\Tr[\rho^2]=\sum_i \tilde b_i[\rho]$.
Consistently, we only find yellow stripes emerging in the plot of $\rho_\text{mix}(p)$ but not for $ \Psi_\text{sup}(p)$ 
as the former becomes increasingly mixed as $p$ grows whereas the latter remains pure.
Also here (in the TPD picture), the QWEs of the two parameterized states otherwise show similar features.
For $p=0$, both TPDs are completely concentrated in the all-triplet bin, i.e., $\tilde a_i[\Psi_\text{sup}(0)]= \tilde a_i[\rho_\text{mix}(0)] =\delta_{i,n}$ 
as is the hallmark of pure product states.
As $p$ is increased, 
this probability is redistributed into the other bins---indicating mixedness.
For $\Psi_\text{sup}(p)$, this mixedness is only found in the marginals (but not in the global state), thus certifying entanglement.
For $\rho_\text{mix}(p)$, the mixedness has nothing to do with entanglement.
Indeed, we analytically find (assuming $n$ is even)
\begin{align}
    \tilde a_n[\Psi_\text{sup}(p)] 
    &=  1-2p(1-p)\left(1-\tfrac{2}{2^n}\right) \,,
\end{align}
yielding the  
concurrence $C[\Psi_\text{sup}(p)] = 1- \tilde a_n[\Psi_\text{sup}(p)]$,
as well as
\begin{align}
    \tilde a_n[\rho_\text{mix}(p)]
    &=  1-2p(1-p)\left(1-\tfrac{1}{2^n}\right)\,.
\end{align}
From this follows that the concurrence $C[\Psi_\text{sup}(p)] >0$ certifies entanglement for $p\in(0,0.5]$ while the lower bound on $C[\rho_\text{mix}(p)]$ 
from Eq.~\eqref{eq:lower_bound_concurrence}
is identically equal to zero.
Similarly, we find  (if $n$ is even)
\begin{align}
    \tilde a_0[\Psi_\text{sup}(p)]
    =  \frac{4p(1-p)}{2^n}\,,
\end{align}
which yields the $n$-tangle  
$\text{Tr}[\Psi_\text{sup}(p) \tilde{\Psi}_\text{sup}(p) ] = 4p(1-p)$.
Unsurprisingly, the latter reaches its maximal possible value  for $p=0.5$.
For completeness, we remark 
\begin{align} \label{eq:no_triplets_prob_mixture}
    \tilde a_0[\rho_\text{mix}(p)]
    =  \frac{2p(1-p)}{2^n}\,,
\end{align}
however, as the state is mixed,  Eq.~\eqref{eq:no_triplets_prob_mixture} does not carry the interpretation of the rescaled $n$-tangle here.

\section{QWEs of many-qubit states}
\label{app:manybody_examples}

\begin{figure*}
    \centering
    \includegraphics{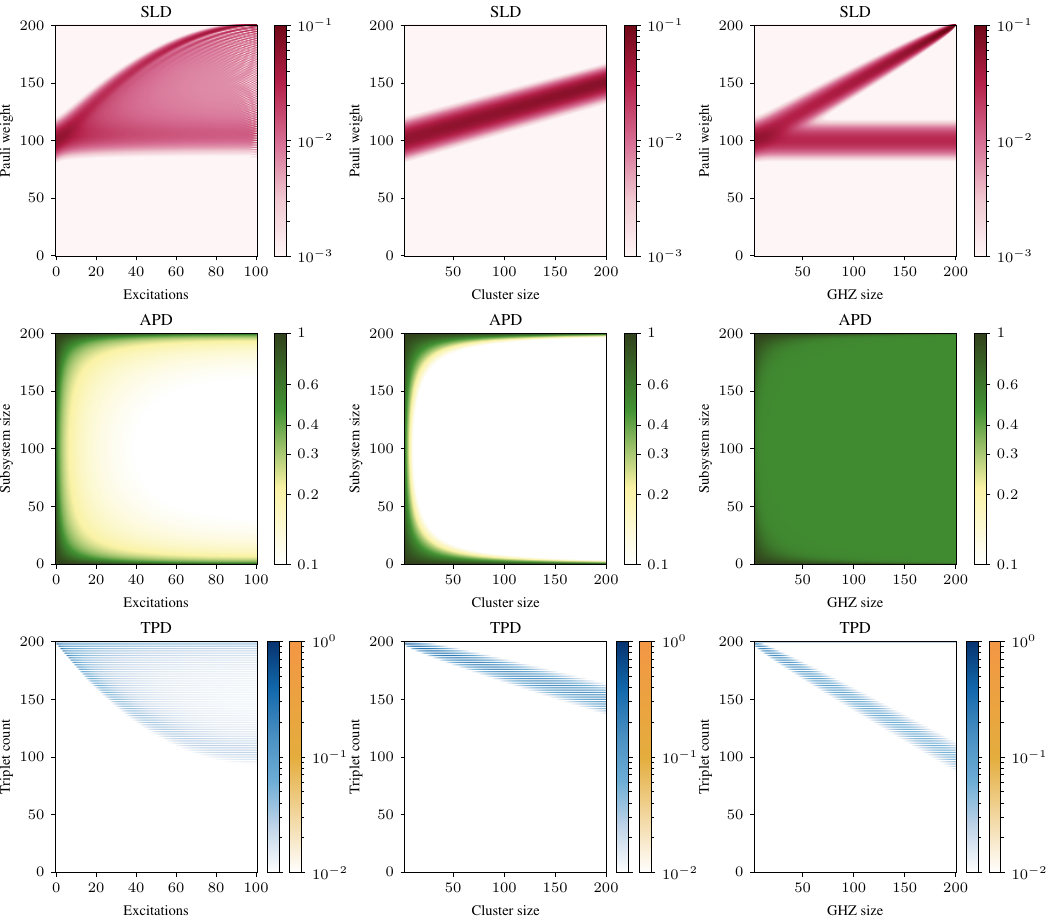}
    \caption{\justifying {Quantum weight enumerators of noiseless parameterized states defined in Eqs.~\eqref{eq:dicke_app}---\eqref{eq:cluster_app} on $n=200$ qubits}.
    Shown are Shor--Laflamme distributions (top, pink),  
    averaged purity distributions (center, green), and
    triplet probability distributions (bottom, blue).
    On the $y$-axis of each panel, the index $i$ in the entry of the respective distributions [recall Defs.~\eqref{def:sld}, \eqref{def:apd}, and~\eqref{def:rains_qwe}] is varied.
    On the $x$-axis, we vary the parameter $e$ from Eqs.~\eqref{eq:dicke_app}---\eqref{eq:cluster_app}.
    This figure illustrates that QWEs can capture various distinct entanglement features of QWEs.
    }
    \label{fig:manybody_noiseless}
\end{figure*}

\begin{figure*}
    \centering
    \includegraphics{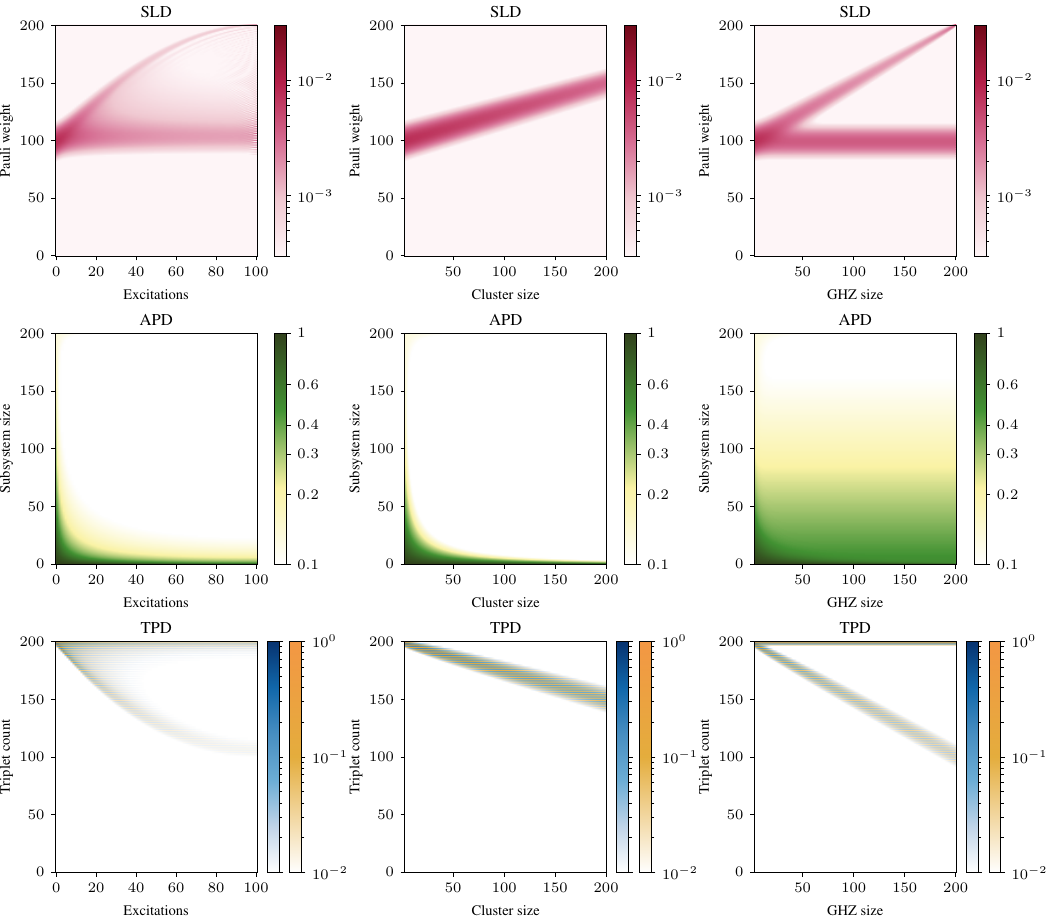}
    \caption{Same as Fig.~\ref{fig:manybody_noiseless} but with local depolarizing noise of strength $p=0.01$ on every qubit. 
    Due to the noise, we find non-zero triplet probabilities  $\tilde a_{n-j}[\rho]>0$ [Eq.~\eqref{eq:tpd}] for odd numbers $j$ of singlets (yellow).
    This figure illustrates that many QWE entanglement features observed in Fig.~\ref{fig:manybody_noiseless} are stable under small noise.
    }
    \label{fig:manybody_noisy}
\end{figure*}

In this appendix, we present and discuss various illustrative examples for the three distributions of {quantum weight enumerators} (QWEs). 
These are:
(i) Shor--Laflamme QWEs aka sector lengths, which stratify the purity by Pauli weight;
(ii) Rains' unitary QWEs aka averaged purities;
and (iii) Rains' shadow QWEs aka triplet probabilities.
Specifically, we consider the following three families of quantum state vectors
\begin{align} \label{eq:dicke_app}
    \ket{D^n_e} &= \tfrac{1}{\sqrt{\binom{n}{e}}} \sum_{ \vert\mathbf{x}\vert = e} \ket{\mathbf{x}}\,, \\
    \label{eq:ghz_app}
    \ket{\text{GHZ}^e, \textbf{0}^{ n-e} } &= \left(\tfrac{\ket{0}^{\otimes e} + \ket{1}^{\otimes e}}{\sqrt{2}} \right) \otimes \ket{0}^{\otimes n-e} \,,\\
  \text{and }  \ket{\text{Cycle}^e , \textbf{+}^{n-e}} & = \prod_{s=1}^e \CZ_{s, s+1 \text{ mod } e}  \ket{+}^{\otimes n}\, .
  \label{eq:cluster_app}
\end{align}
Each state family is parameterized by an integer $e$. 
For Dicke state vectors $\ket{D^n_e}$, this integer is the number of excitations, which we vary from $e=0$ to $e=n/2$.
Further increasing $e$ would not lead to any new QWEs as $\ket{D^n_e} $ is local unitary equivalent to $\ket{D^n_{n-e}} = X^{\otimes n } \ket{D^n_e}$.
From GHZ and cycle graph states with $e$ entangled qubits, meanwhile, we construct families of $n$-qubit states by simply appending $n-e$ qubits in a pure product state.

For $\ket{D^n_e}$ and $\ket{\text{GHZ}^e}$, we already know the QWEs from Eqs.~\eqref{eq:apd_dicke_states} and~\eqref{eq:ghz_sld}, respectively.
Similarly, those of $\ket{\text{Cycle}^e}$ were computed in Eq.~(21) of Ref.~\cite{miller_shor_laflamme_2023}.
Leveraging the convolution~\cite{wyderka_characterizing_quantum_2020}
\begin{align} \label{eq:sld_convolution}
    a_j[\rho \otimes \sigma] = \sum_{i=0}^j a_i[\rho] \, a_{j-i}[\sigma] \,,
\end{align}
together with Eq.~\eqref{eq:sld_pure_product} from the main text, we can thus compute the SLDs of 
$\ket{\text{GHZ}^e, \textbf{0}^{ n-e} } $ and $\ket{\text{Cycle}^e , \textbf{+}^{n-e}} $.
This then allows us to calculate the respective APDs and TPDs by applying the transforms $T'$ and $\tilde T$ from App.~\ref{app:notation}.

In Fig.~\ref{fig:manybody_noiseless}, we display all QWEs for the three discussed state families in the case of $n=200$ qubits.
In the (pink)  picture of \emph{sector length distributions} (SLDs), we see that 
increasing $e$ causes a movement of the SLDs mean value from $n/2$ towards sectors of larger Pauli weight.
Thereby, distinct features manifest themselves, depending on the state family.
This observation is 
consistent with the fact that, for pure states, the mean value of the SLD must lie in the range between $n/2$ and $3n/4$~\cite{miller_shor_laflamme_2023}.

For Dicke states, we see in the top left panel of  Fig.~\ref{fig:manybody_noiseless}  that a significant proportion of the SLD remains around $n/2$ for all values of $e$.
Meanwhile, a second peak emerges.
By increasing $e$, this peak gives rise to a curved ridge that deeply extends into the high-weight sector until full Pauli weight is reached around ${ \add  e \lesssim n/2}$.
The region between this curved ridge and the one at Pauli weight $n/2$ is   significantly occupied by the SLD.
Towards larger values of $e$, an interference-like pattern arises in this region.
Precisely at half-filling, the odd-weight sector lengths ``destructively interfere'' in the sense that we observe $a_i[|{D^n_{n/2}}\rangle\! \langle|{D^n_{n/2}}|]=0$ for all odd values of $i$.
We explain the latter observation as follows:
the half-filled Dicke state is invariant under the state inversion map [Eq.~\eqref{eq:spin_flipped_state}]. 
Therefore, TPD [Eq.~\eqref{def:rains_qwe}] and dual SLD [Eq.~\eqref{def:sld_dual}] coincide with each other and, since the state is pure, also with the SLD.
This also explains the subtle feature that, for $e=n/2$, SLD (pink) and TPD (blue) of the Dicke state are exactly the same.

The second state family, cycle graph states $ \ket{\text{Cycle}^e , \textbf{+}^{n-e}}$, shows a different behavior in the top central panel of Fig.~\ref{fig:manybody_noiseless}.
Here, the SLD only features a single peak, whose center moves from $n/2$ for $e=0$ entangled qubits to $3n/4$ for $e=200$.
At the same time, the width of the peak slightly decreases.
This is explained as follows: 
for $e=0$, we have a pure product state, whose SLD is a symmetric binomial distribution [Eq.~\eqref{eq:sld_pure_product}].
For $e>0$, the SLD of $\ket{\text{Cycle}^e }$ is strikingly similar to that of a random state~\cite{miller_shor_laflamme_2023}, which is very close to an asymmetric binomial distribution centered at $3e/4$ [Eq.~\eqref{eq:sld_2design}].
Note that, for graph states, mean and variance of the SLD can always be expressed in terms of the numbers of isolated vertices $I$, leaves $L$, and twin pairs $T$ of the corresponding graph, see Cor.~3 of Ref.~\cite{miller_shor_laflamme_2023}.
Here, these invariants are given by $I=n-e$, $L =0$, and $T=\binom{n-e}{2}+2\delta_{e,4}$, which implies that mean and variance of the SLD are given by $(2n+e)/4$ and $(4n-e+4\delta_{e,4})/16$, respectively.
This explains the position and width of the ray in the of central panel of Fig.~\ref{fig:manybody_noiseless} as a function of $e$.

Finally, consider in the top right panel of Fig.~\ref{fig:manybody_noiseless} the SLD for the third state family: GHZ states in a tensor product with pure product states.
Here, we observe qualitatively different features compared to the already-discussed state families.
Similar to the case of Dicke states, we see two ridges in the SLD: again a horizontal one at Pauli weight $n/2$, and a second one that increases from weight $n/2$ at $e=0$ to weight $n$ at $e=n$.
This time, however, the second ridge is straight (not curved) and there are no interference pattern between the two ridges.
The reason for the observed pattern is this:
from Eq.~\eqref{eq:ghz_sld}, we know that the SLD of $\ket{\text{GHZ}^e}$ arises from the symmetrical binomial distribution $\binom{e}{i}/2^e$ after redistributing all odd-weight sector lengths into the weight-$e$ bin.
Convoluting with the ordinary binomial distribution $\binom{n-e}{j}/2^{n-e}$ via Eq.~\eqref{eq:sld_convolution} smeers out both the centered distribution as well as the peak at $e$, resulting in the observed pattern.
We also point out that at the very right of the SLD (just as in the case of Dicke states), all odd-weight sector lengths vanish, 
which again causes the curious feature that that, for $e=n$, SLD (pink) and TPD (blue) of $\ket{\text{GHZ}^n}$ precisely coincide.

Let us next turn our attention to the (green) averaged purity distributions (APDs) in Fig.~\ref{fig:manybody_noiseless}.
In general, we observe the following pattern: 
when the size $|S|$ of the subsystems $S$ whose purities $\Tr[\rho_S^2]$ are averaged [Eq.~\eqref{def:apd}] is decreased from $|S|=n$, the value of the APD decreases from $1$ until, at $|S| = n/2$, it starts to increase again, which results in a symmetric pattern with a horizontal symmetry axis.
The symmetry is easily explained by the fact that---by virtue of the Schmidt decomposition or, alternatively, by MacWilliams' identity---for pure states, a marginal has the same purity as its complement.
The decrease of the APD value with decreasing $|S|$ reflects that the global state is pure and (if $e>0$) entangled.
Therefore, some of its marginals are mixed, which causes the drop in the APD.
Quantitatively, however, we can easily distinguish the three state families:
the GHZ state (center right) has fairly pure marginals, which are only mixtures of $\ket{0}^{\add \otimes e}$ and $\ket{1}^{ \add \otimes e}$; 
taking the tensor product with the pure state $\ket{0}^{\otimes n-e}$ only increases the subsystem purity, which is visible as a dark green shade in the top and bottom left corner of the APD panel of the GHZ state.
For Dicke states (center left), the values of the APD become significantly smaller than for the GHZ states,
showcasing in which sense Dicke states can be regarded as more entangled than GHZ states.
For cycle graph states (center) the APD drops to even smaller values, which is unsurprising because generic states have exponentially small subsystem purities.

Last but not least, consider the (blue) triplet probability distributions (TPDs) in bottom of Fig.~\ref{fig:manybody_noiseless}.
These are the QWEs that can be straightforwardly measured in two-copy Bell sampling experiments.
For all three state families, we see that for $e=0$ the zero-singlet probability is equal to $100\%$.
When $e$ is increased, the probabilities of measuring more than zero singlets increases, however, only for even singlet counts. 
{\add In accordance with Cor.~\ref{cor:triplet_mean_bound} of the main text,}
all TPDs concentrate in the upper half (low singlet count) of their respective panels.
The fact that odd-singlet counts are never observed is easily attributed to the fact that the global state is always pure [Eq.~\eqref{eq:purity_from_tpd}].
Moreover, the occurrence of non-zero singlet counts is due to entanglement in the states for $e>0$,
{\add which} can be quantified in terms of the concurrence [Eq.~\eqref{eq:concurrence}].

Looking more closely at the TPDs in the bottom of Fig.~\ref{fig:manybody_noiseless},
we see that Dicke states have non-zero probabilities for all 
{\add allowed} 
triplet counts above $n-e$.
For large $e$, the TPD of Dicke states (bottom left) has a heavy tail in the center of the distribution at $n/2$ triplets.
Similarly, for $\ket{\text{GHZ}^e, \textbf{0}^{ n-e} } $ (bottom right), we observe one ray that extends down to $n/2$ triplets; a second (less visible) horizontal ray (supporting half the TPD) remains in the zero-singlet bin.
On the other hand, for $\ket{\text{Cycle}^e, \textbf{+}^{n-e}}$, the TPD panel (bottom center) only features a single ray that connects $n$ triplets for $e=0$  and $3n/4$ triplets for $e=n$.
The observed fact that the TPDs of the Dicke and GHZ states exhibit heavy tails around $n/2$ triplets, while those of the cluster state do not, explains the sample requirements discussed 
in Fig.~\ref{fig:sample_requirements} of the main text.

Besides Fig.~\ref{fig:manybody_noiseless}, where we show the QWEs of ideal states, we also show those of slightly-perturbed states in Fig.~\ref{fig:manybody_noisy}. 
This time, we apply $p=0.01$ of local depolarizing noise per qubit. 
The QWEs of the noise states are computed as explained in Sec.~\ref{sec:dicke_entanglement_stability} of the main text.
We see that many of the entanglement features that we just discussed survive in the presence of 
{\add such} 
small noise.
In particular, the SLD features (pink) qualitatively remain the same.
This is unsurprising because local depolarizing noise acts diagonally on SLDs [Eq.~\eqref{eq:sector_length_decay}].

\section{\add QWEs of an approximate QECC from a CFT}
\label{app:qwes_of_cft_code}

\begin{figure*}
    \centering 
    \includegraphics{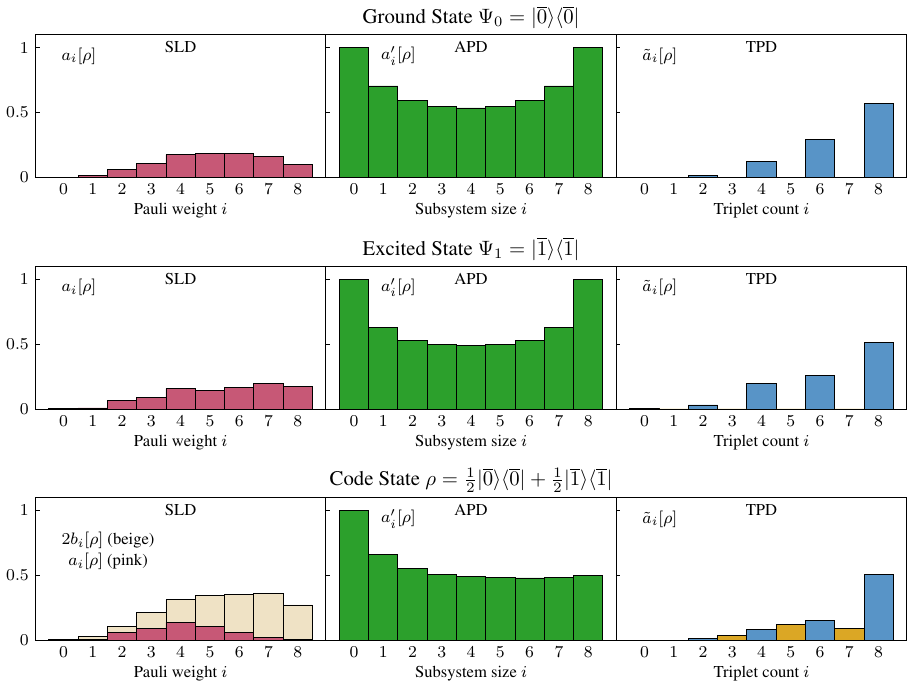}
    \caption{\add Quantum weight enumerators of the
    logical basis states (top, center) and the code state (bottom) for a physically-inspired approximate quantum error correcting code with $n=8$ physical qubits. This code is defined in terms of a critical transverse-field Ising model~\cite{sang_approximate_quantum_2024} and was recently demonstrated experimentally~\cite{zhang_probing_mixed_2025}.
    As discussed in App.~\ref{app:qwes_of_cft_code}, their weight enumerators reveal that these three states are entangled.
     }
    \label{fig:qwes_of_cft_code}
\end{figure*}

{\add 
Here,
we investigate   
\emph{quantum weight enumerators} (QWEs) 
of an approximate \emph{quantum error-correcting code} (QECC)
whose logical basis states are derived from a 
\emph{conformal field theory} (CFT).
More precisely, we consider the
approximate QECC that encodes  $k \ge \Omega(\log\log n)$ logical qubits
into the low-energy sector of
the critical transverse field Ising model,
$H = -\sum_{i=1}^n (Z_i + X_iX_{i+1\text{ mod }n})$.
In the limit of $n\rightarrow \infty$, 
this code family has a finite threshold against $Y$- and $Z$- but not against $X$-errors. 
For any finite value of $n$, however, 
this code can only approximately correct $X$- and $Z$-errors~\cite{sang_approximate_quantum_2024}

We restrict ourselves to the case of $n=8$ logical and $k=1$ physical qubits,
as this corresponds to the largest instance of a recent experimental
demonstration~\cite{zhang_probing_mixed_2025}.
Via numerical diagonalization, we compute the logical basis vectors 
$\ket{\bar 0}$ and $\ket{\bar 1}$
with eigenenergies approximately
$-10.25$ and
$-10.05$, respectively.
As expected, these are not stabilizer states.
Indeed, their stabilizer $2$-Rényi entropies \cite{leone_stabilizer_renyi_2022},
\begin{align}
    M_2(\ket{\psi}) = - \log_2\left( \sum_{P \in \{I,X,Y,Z\}^{\otimes n}} \frac{ \vert  \langle \psi \vert P \vert \psi \rangle \vert ^4  }{2^n}\right) \, ,
\end{align}
are found to be $M_2(\ket{\bar0}) \approx 3.05$ and  $M_2(\ket{\bar1}) \approx 2.50$. 
Next, we compute the QWEs of $\Psi_0 =\ket{\bar0}\bra{\bar 0}$, of  $\Psi_1 =\ket{\bar 1}\bra{\bar 1}$, and of  $\rho = \rho_\text{QECC} = \tfrac{1}{2} (\ket{\bar0}\bra{\bar 0}+\ket{\bar1}\bra{\bar 1})$,  the maximally mixed state inside this code.
The results are shown  
in Fig.~\ref{fig:qwes_of_cft_code}.

First, we observe $\mathbf{a}[\Psi_0] \neq \mathbf{a}[\Psi_1]$.
Due to the LU invariance of QWEs, this implies that the logical $X$-gate of the investigated code cannot be implemented transversally.
Next, 
we can infer---from all three QWE distributions---that the logical basis states are entangled:
(i) their $n$-body sector lengths (pink),
$a_8[\Psi_0] \approx 0.104 $
and
$a_8[\Psi_1] = 0.173$,
clearly exceed the classical bound,
$2^{-n} \approx 0.004$;
(ii) their subsystem purities (green),
$a'_i[\Psi_0]$
and $a'_i[\Psi_1]$ are smaller than  1 (their global purity);
and (iii) from their triplet probabilities (blue), we find that 
their concurrences $C[\Psi_0] = 1- \tilde a_n[\Psi_0] \approx 0.431$ 
and 
$C[\Psi_1]  \approx 0.486$ 
as well as their $n$-tangles 
$2^n\tilde{a}_0 [\Psi_1] = 0.0625$ and 
$2^n\tilde{a}_0 [\Psi_1] = 0.125$ 
are strictly positive.
Applying Eq.~\eqref{eq:lower_bound_concurrence} from the main text to the TPD of 
$\rho$, the mixed code state,
unfortunately results in a trivial lower bound on its concurrency, $C[\rho ]$.
Nevertheless, also this state is entangled:
for one its $n$-body sector length $a_8[\rho] \approx 0.0055$
still exceeds the classical bound of $2^{-n}$
and, second, we observe
$a'_i[\rho]< a'_n[\rho]=\tfrac{1}{2}$, which
implies the existence of an entangled partition of size $i$ vs $n-i$ 
for all $i\in\{4,5,6,7\}$.

Having discussed the entanglement structure of our approximate QECC,
let us next turn to its error-correcting capabilities.
To this end, we are plotting (beige) the rescaled dual SLD, $2^k \mathbf{b}[\rho]$, 
in the background of the code's SLD (pink).
Recall 
from Eq.~\eqref{eq:distance} in the main text 
that the code distance can be inferred as the smallest index $i$ for which the inequality
$a_i[\rho] \le 2^k b_i[\rho]$
is strict.
In the bottom-left panel of Fig.~\ref{fig:qwes_of_cft_code}
we see that this is already the case for $i=1$.
This is unsurprising because---as an approximate QECC---the exact distance of the code is trivially given by $d=1$.
To understand from which type of single-qubit errors the code suffers most from,
we compute the different contributions to $A_1 = 2^na_1[\rho]$  and $B_1 = 2^{n+k}b_1[\rho]$.
We find that each qubit $j$ contributes $\Tr[\rho X_j ]^2=\Tr[ \rho Y_j ]^2=0$
and $\Tr[ \rho Z_j]^2 \approx0.327$ 
to $A_1$.
Similarly, the contributions to $B_1$ are given by
$2\Tr[(\rho X_j)^2] \approx 0.607$,
$2\Tr[ (\rho Y_j)^2] \approx 0.006$, 
and
$2\Tr[(\rho Z_j)^2 ] \approx 0.332$.
Hence, $X$-errors are (as expected) the main cause of problems.
The contribution of $Y$- and $Z$-errors, on the other hand,
is much smaller albeit also non-zero, reflecting the fact that $Y$- and  $Z$-errors can only be corrected approximately.

Last but not least,
we point out that the average fidelity and the entanglement fidelity under the local depolarizing channel can be computed from the QWEs of any code~\cite{shor_quantum_analog_1997}.

}
\end{document}